\documentclass[article]{IEEEtran} 
\usepackage{amsmath,amssymb,latexsym,cite}
\usepackage{epsf, pstricks,pgf, xcolor}
\usepackage{graphicx}
\usepackage{epstopdf}
\usepackage{epsfig}
\usepackage{calc}

\def\argmax{\operatornamewithlimits{arg\,max}}
\def\argmin{\operatornamewithlimits{arg\,min}}

\newcommand{\snr}{\text{$\mathsf{SNR}$}}
\newcommand{\inr}{\text{$\mathsf{INR}$}}

\DeclareMathOperator{\real}{Re}
\DeclareMathOperator{\imag}{Im}
\def\mrt{{M_{\text{\textnormal{T}}}}}
\def\mrr{{M_{\text{\textnormal{R}}}}}
\def\mct{{N_{\text{\textnormal{T}}}}}
\def\mcr{{N_{\text{\textnormal{R}}}}}
\def\mri{M}
\def\c{{\text{\textnormal{C}}}}
\def\tx{\text{\textnormal{TX}}}
\def\comp{{\text{\textnormal{comp}}}}
\def\iftext{{\text{\textnormal{IF}}}}

\begin{document}
\newtheorem{theorem}{Theorem}
\newtheorem{corollary}{Corollary}
\newtheorem{remark}{Remark}
\newtheorem{example}{Example}
\newtheorem{definition}{Definition}
\newtheorem{lemma}{Lemma}

\title{Integer-Forcing Linear Receivers}

\author{Jiening Zhan, Bobak Nazer, \IEEEmembership{Member, IEEE}, Uri Erez, \IEEEmembership{Member, IEEE}, and Michael Gastpar, \IEEEmembership{Member, IEEE}
\thanks{J. Zhan, B. Nazer and M. Gastpar were supported in part by NSF grants CCR-0347298, CNS-0627024, and CCF-0830428, and by an Okawa Foundation Research Grant. J. Zhan was supported by an NSF Graduate Fellowship. B. Nazer was supported by an NSF CAREER grant CCF-1253918. M. Gastpar was supported by 3TU.CeDICT: Centre for Dependable ICT Systems, The Netherlands, and by a European ERC Starting Grant 259530-ComCom. U. Erez was supported by the U.S. - Israel Binational Science Foundation under grant 2008455
and by the Israel Science Foundation under grant 1557/11. The material in this paper was presented in part at the IEEE International Symposium on Information Theory, Austin, TX, June 2010 and at the 72nd IEEE Vehicular Technology Conference, Ottawa, Canada, September 2010.}
\thanks{J. Zhan was with the Department of Electrical Engineering and Computer Sciences, University of California, Berkeley, Berkeley, CA, 94720-1770, USA (email: jiening@eecs.berkeley.edu) and is now with Google, Inc. B. Nazer is with the Department of Electrical and Computer Engineering, Boston University, Boston, MA, 02215, USA (email: bobak@bu.edu). U. Erez is with the Department of Electrical Engineering - Systems, Tel Aviv University, Ramat Aviv, Israel (email: uri@eng.tau.ac.il). M. Gastpar is with the School of Computer and Communication Sciences, Ecole Polytechnique F\'ed\'erale, Lausanne, Switzerland (e-mail: michael.gastpar@epfl.ch).}}

\markboth{IEEE Trans Info Theory, to appear}{~}

\maketitle

\begin{abstract}
Linear receivers are often used to reduce the implementation complexity of multiple-antenna systems. In a traditional linear receiver architecture, the receive antennas are used to separate out the codewords sent by each transmit antenna, which can then be decoded individually. Although easy to implement, this approach can be highly suboptimal when the channel matrix is near singular. This paper develops a new linear receiver architecture that uses the receive antennas to create an effective channel matrix with integer-valued entries. Rather than attempting to recover transmitted codewords directly, the decoder recovers integer combinations of the codewords according to the entries of the effective channel matrix. The codewords are all generated using the same linear code which guarantees that these integer combinations are themselves codewords. Provided that the effective channel is full rank, these integer combinations can then be digitally solved for the original codewords. This paper focuses on the special case where there is no coding across transmit antennas and no channel state information at the transmitter(s), which corresponds either to a multi-user uplink scenario or to single-user V-BLAST encoding. In this setting, the proposed integer-forcing linear receiver significantly outperforms conventional linear architectures such as the zero-forcing and linear MMSE receiver. In the high SNR regime, the proposed receiver attains the optimal diversity-multiplexing tradeoff for the standard MIMO channel with no coding across transmit antennas. It is further shown that in an extended MIMO model with interference, the integer-forcing linear receiver achieves the optimal generalized degrees-of-freedom.

\end{abstract}


\section{Introduction}
\label{sec:intro}
It is by now well known that increasing the number of antennas in a wireless system can significantly increase its capacity. Since the seminal papers of Foschini \cite{foschini96}, Foschini and Gans \cite{fmg98}, and Telatar \cite{telatar99}, multiple-input multiple-output (MIMO) channels have been thoroughly investigated in theory (see \cite{gjjv00} for a survey) and implemented in practice \cite{802.11n}. A significant challenge encountered in MIMO systems is that channel knowledge at the transmitter is often quite limited. The focus of the present work is on a quasi-static block fading model where the channel remains constant throughout the transmission of a codeword but the transmitter has only statistical knowledge of the channel realization (sometimes referred to as slow fading). In other words, channel state information (CSI) is only available at the receiver.

An enormous body of work has strived to develop MIMO receiver architectures that can attain the promised capacity gains with an implementation complexity similar to that of single-antenna systems. The vast majority of these architectures fall into one of the following two categories:

\noindent\textbf{Joint Maximum Likelihood (ML) Receivers:} Clearly, the ML decision rule is optimal and thus yields the best possible rates and probability of error. However, if the transmitter employs a capacity-approaching channel code, finding the ML estimate directly is prohibitively complex as the size of the search space is exponential in the product of the code's blocklength and the number of antennas. As a result, most joint ML decoding architectures are geared towards MIMO systems that either employ uncoded constellations
or where ML detection is performed on a symbol-level basis. This includes the vast literature on space-time codes \cite{alamouti98,tsc98,tjc99,hh02,ed03,srs03,ecd04,jafarkhani,orbv06,tv06,ekpkl06}, which are known to be optimal in the high signal-to-noise ratio (SNR) regime, both in terms of multiplexing (i.e., rate) and diversity (i.e., probability of error).  The complexity of joint ML detection can be significantly reduced through the use of sphere decoding algorithms \cite{vb99,aevz02,dec03,hv05,jo05,bbwzfb05,sej12,je12}. Further savings are possible by employing lattice-based constellations and exploiting this structure at the receiver via lattice-aided reduction \cite{yw02,tmk07,tmk07lll,glm09,je10}. Both of these approaches can achieve high SNR optimality in terms of the diversity-multiplexing tradeoff (DMT) \cite{zt03}.

The finite SNR performance can be enhanced by coupling the space-time symbols with an outer channel code while still maintaining some separation between detection and decoding. One approach is for the receiver to feed the soft outputs from its symbol-level joint detection to a decoder for the outer code. This approach can be enhanced by iterating between detection and decoding using an iterative decoder \cite{lwn02,ht03,tka04,lyw04,sps05}. While it can be shown numerically that these approaches improve upon the performance of uncoded systems, it is difficult to argue that they can operate close to the MIMO capacity at practically-relevant SNRs. Overall, this class of architectures is well-suited to the high SNR regime as well as scenarios where diversity is far more important than high data rates.

\begin{figure*}[t!]
\begin{center}
\psset{unit=0.76mm}
\begin{pspicture}(-40,-5)(197,52)
\footnotesize

\rput(-35,27){Info}
\rput(-35,23){Bits}

\rput(15,40){
\psline{->}(-45,-13)(-35,0)(-25,0) \rput(-31.5,3){$\mathbf{w}_1$}
\psframe(-25,-5)(0,5) \rput(-12.5,0){Linear Code}
\psline(0,0)(8,0)
\psline(8,0)(8,6)(5,11)(11,11)(8,6)
}

\rput(15,25){
\psline{->}(-45,0)(-25,0) \rput(-31.5,3){$\mathbf{w}_2$}
\psframe(-25,-5)(0,5) \rput(-12.5,0){Linear Code}
\psline(0,0)(8,0)
\psline(8,0)(8,6)(5,11)(11,11)(8,6)
}

\rput(2.5,16.5){\Large{$\vdots$}}

\rput(15,5){
\psline{->}(-45,18)(-35,0)(-25,0) \rput(-30.25,3){$\mathbf{w}_\mrt$}
\psframe(-25,-5)(0,5) \rput(-12.5,0){Linear Code}
\psline(0,0)(8,0)
\psline(8,0)(8,6)(5,11)(11,11)(8,6)
}

\rput(56,35){
\rput(-3,0){
\psline(-12,5)(-12,11)(-15,16)(-9,16)(-12,11)
\psline{-}(-12,5)(-4,5) 
}
\psline{->}(13,5)(25,5)
\psframe(25,0)(55,10) \rput(40,5){SISO Decoder}
\psline{->}(55,5)(77,5)
\rput(62.5,8){$\mathbf{\hat{u}}_1$}
\psline{->}(105,5)(115,5)(123.5,-4.5)
\rput(110.5,8){$\mathbf{\hat{w}}_1$}
}

\rput(56,20){
\rput(-3,0){
\psline(-12,5)(-12,11)(-15,16)(-9,16)(-12,11)
\psline{-}(-12,5)(-4,5) 
}
\psline{->}(13,5)(25,5)
\psframe(25,0)(55,10) \rput(40,5){SISO Decoder}
\psline{->}(55,5)(77,5)
\rput(62.5,8){$\mathbf{\hat{u}}_2$}
\psline{->}(105,5)(122.25,5)
\rput(110.5,8){$\mathbf{\hat{w}}_2$}
}

\rput(96,16.5){\Large{$\vdots$}}

\rput(56,0){
\rput(-3,0){
\psline(-12,5)(-12,11)(-15,16)(-9,16)(-12,11)
\psline{-}(-12,5)(-4,5) 
}
\psline{->}(13,5)(25,5)
\psframe(25,0)(55,10) \rput(40,5){SISO Decoder}
\psline{->}(55,5)(77,5)
\rput(63.5,8){$\mathbf{\hat{u}}_\mrt$}
\psline{->}(105,5)(115,5)(123.5,19.5)
\rput(111.5,8){$\mathbf{\hat{w}}_\mrt$}
}

\rput(-5,0){
\psframe[linecolor=blue,linestyle=dashed](78,-18)(132,49)
\rput(105,-4){\textcolor{blue}{Decode Linear Combinations}}
\rput(99,-12){$\mathbf{u}_m = \Big[ {\displaystyle \sum_\ell a_{m,\ell} \mathbf{w}_\ell} \Big]$}\rput(121.25,-11){$\mod{p}$}
}

\psframe(49,0)(69,45) \rput(59,28){Linear} \rput(59,22){Equalizer}

\psframe(133,0)(161,45) \rput(147,28){Solve Linear} \rput(147,22){Combinations}
\rput(187,28){Estimated}
\rput(187,22){Info Bits}

\end{pspicture}
\end{center}
\caption{Block diagram of the integer-forcing architecture.} \label{f:highlevel}
\end{figure*}
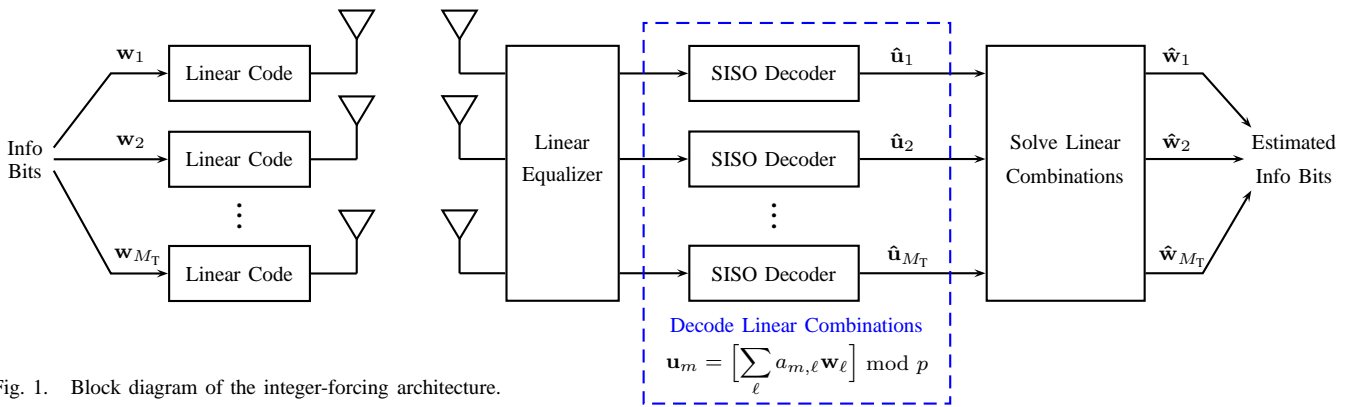
\noindent\textbf{Zero-Forcing and MMSE Linear Receivers:} The added complexity in a MIMO receiver stems from the fact that the transmitted data streams are coupled in space (i.e., across antennas) by the channel. Conventional linear receivers attempt to remove this coupling by first passing the received signals through a linear front-end \cite{verdu,lv89,mh94}. For instance, consider a MIMO system in which each transmit antenna encodes an independent data stream (i.e., codeword). The zero-forcing receiver (or decorrelator) first inverts the channel matrix so that each data stream can be recovered via a single-input single-output (SISO) decoding algorithm (i.e., a single-user decoder). The minimum mean-squared error (MMSE) receiver does the same except with a regularized channel inverse that accounts for possible noise amplification. Both of these architectures permit the use of powerful channel codes that can achieve high data rates at practically-relevant SNRs. The associated achievable rates can be written down in closed form and approached closely using capacity-approaching codes developed for the single-antenna setting \cite{fc07}.

Unfortunately, by inverting the channel matrix, these architectures distribute the noise unequally across data streams. If the channel realization is not known to the transmitter, this can lead to a significant degradation in the outage rate. In particular, conventional linear receivers exhibit a suboptimal DMT \cite{zt03,hn07,kcm09}. Their performance can be significantly improved via successive interference cancellation (SIC) \cite{vg97,wfgv98} but this technique is insufficient to obtain the optimal DMT \cite{kcm09,jvl11}. Overall, this class of architectures is well-suited to scenarios where high data rates are far more important than diversity.

In this paper, we propose a novel class of receiver architectures for quasi-static MIMO channels that exhibits qualitatively and quantitatively distinct behavior from both of the classes described above.

\noindent\textbf{Integer-Forcing Linear Receivers:} The architectures discussed above operate on the implicit assumption that the decoding algorithm is limited to recovering a subset of the data streams while treating the rest as noise. Recent work on compute-and-forward \cite{ng11IT} has revealed a new possibility: the decoder can directly recover a linear combination of interfering data streams. Specifically, consider the scenario where each data stream is drawn from the same linear or lattice codebook. The codebook structure ensures that any \textit{integer combination} of codewords is itself a codeword, and thus decodable at high rates. The integer-forcing linear receiver exploits this property to flip the usual decoding process: it first eliminates noise and only then eliminates interference between data streams in the digital domain. That is, first the linear front-end is used to create an effective integer-valued matrix while amplifying the noise as little as possible. Then, the resulting equalized channel outputs are fed into SISO decoders which recover linear combinations of the data streams. Finally, these linear combinations are solved for the original data streams. See Figure~\ref{f:highlevel} for an illustration. The achievable rates of this architecture can be written down in closed form and approached closely using either nested lattice codes incorporating shaping \cite{tenBrink05,fsk11} (suitable for low SNR) or linear codes with no shaping (sufficient for high SNR), such as a low-density parity-check (LDPC) code combined with quadrature amplitude modulation (QAM) \cite{ozegn11,hc11,hc12IT}. More generally, any low-complexity coding framework for compute-and-forward~\cite{fsk11,oe12,hc11,hc12IT,she12,ylhyc12,bl12,tnp13,hnt14} can also be used to implement an integer-forcing linear receiver.

The key step underpinning this approach is the selection of an integer matrix $\mathbf{A}$ to approximate the channel matrix $\mathbf{H}$. As we will show, if $\mathbf{A}$ is properly chosen, this architecture can operate quite close to the MIMO outage capacity without incurring the complexity of joint ML decoding. Moreover, by setting $\mathbf{A}$ to be the identity matrix, it can be shown that integer-forcing includes the performance of zero-forcing and linear MMSE receivers as a special case. Its complexity is nearly the same as that of a conventional linear receiver: it employs linear equalization and single-user decoding followed by inverting $\mathbf{A}$ in the digital domain. The additional complexity is mainly due to the search for the appropriate $\mathbf{A}$. Although finding the optimal $\mathbf{A}$ has a worst-case complexity that is exponential in the number of antennas, this search only needs to be performed once per coherence interval. In practice, efficient approximation algorithms, such as the Lenstra-Lenstra-Lovasz (LLL) algorithm \cite{lll82}, can be used to find near-optimal $\mathbf{A}$ in polynomial time.

Like conventional linear receiver architectures, the integer-forcing receiver first equalizes the channel and then feeds the result into several SISO decoders. However, conventional linear receivers attempt to isolate the data streams by creating an effective identity matrix, which can significantly amplify the noise. Since the integer-forcing linear receiver can equalize the channel to any full-rank integer matrix, it can optimize over the choice of this matrix to reduce the noise amplification. For example, consider the following $2 \times 2$ MIMO system: \begin{align*}
 \begin{bmatrix} \mathbf{y}_1 \\ \mathbf{y}_2 \end{bmatrix} = \begin{bmatrix}
  2 & 1 \\
  1 & 1 \\
 \end{bmatrix} \begin{bmatrix} \mathbf{x}_1 \\ \mathbf{x}_2 \end{bmatrix}   +  \begin{bmatrix} \mathbf{z}_1 \\ \mathbf{z}_2 \end{bmatrix}. 
\end{align*} The zero-forcing receiver applies the matrix inverse to the received signal,
 \begin{align*}
\begin{bmatrix}
  1 & -1 \\
  -1 & 2 \\
 \end{bmatrix} \begin{bmatrix} \mathbf{y}_1 \\ \mathbf{y}_2 \end{bmatrix} =  \begin{bmatrix} \mathbf{x}_1 \\ \mathbf{x}_2 \end{bmatrix}   +  \begin{bmatrix} \mathbf{z}_1 - \mathbf{z}_2  \\ -\mathbf{z}_1 + 2 \mathbf{z}_2 \end{bmatrix}, 
\end{align*} which enlarges the effective noise variances by factors of $2$ and $5$, respectively. On the other hand, the integer-forcing receiver can directly decode the linear combinations $2\mathbf{x}_1 + \mathbf{x}_2$ and $\mathbf{x}_1 + \mathbf{x}_2$ while leaving the noise variances unchanged. These linear combinations can then be digitally solved for the original data streams.

In this paper, we will focus on the important special case of MIMO systems where each transmit antenna encodes an independent data stream (i.e., there is no coding across the transmit antennas\footnote{It is worth noting that the use of the integer-forcing receiver does not preclude space-time encoding at the transmitter. Very recent work has examined this possibility \cite{de12} and shown that integer-forcing achieves the full diversity-multiplexing tradeoff \cite{oe13}. In fact, it has been shown in \cite{oe13} that it attains the capacity of any MIMO channel up to a constant gap.}). This could correspond to a multiple-access or uplink scenario where each user has a single antenna and the basestation has multiple antennas. It also describes the V-BLAST approach for operating single-user MIMO channels at very high data rates \cite{wfgv98}. We will argue, through a combination of analytical results and outage plots, that the integer-forcing linear receiver can very closely approach the performance of the optimal joint ML receiver for the entire SNR range. In particular, we are able to show that integer-forcing achieves the optimal DMT, which has remained out of reach for conventional linear receivers \cite{zt03,hn07,kcm09,jvl11}. 

We will also consider MIMO channel models that include external interference \cite{bfhy02,wsg94} and argue that the integer-forcing receiver architecture is an attractive approach to the problem of interference mitigation. By selecting the integer coefficients in a direction that depends on both the interference space and the channel matrix, the proposed architecture reduces the impact of interference and attains a non-trivial gain over traditional linear receivers. Furthermore, we show that the integer-forcing receiver achieves the same generalized degrees-of-freedom as the joint ML decoder.

In the remainder of the paper, we begin with a formal problem statement in Section~\ref{sec:problem}, and then overview conventional MIMO receiver architectures and their achievable rates in Section~\ref{sec:existingarchitecture}. In Section~\ref{sec:architecture}, we present the integer-forcing receiver architecture and a basic performance analysis. Through a series of examples, we explore the performance of integer-forcing compared to conventional architectures in Section~\ref{sec:fixedchannels}. We study the outage performance of the integer-forcing linear receiver under a slow fading channel model in Section~\ref{sec:outage}. We show that in the case where each antenna encodes an independent data stream, our architecture achieves the same diversity-multiplexing tradeoff as that of the optimal joint decoder. In Section~\ref{sec:interference}, we consider the MIMO channel with interference and show that the integer-forcing receiver can be used to effectively mitigate interference. We characterize the generalized degrees-of-freedom for the integer-forcing receiver and find that it is the same as for the joint ML decoder.

\section{Problem Statement}
\label{sec:problem}

Throughout the paper, we will use boldface lowercase to refer to vectors, e.g., $\mathbf{a} \in \mathbb{Z}^M$, and boldface uppercase to refer to matrices, e.g., $\mathbf{H} \in \mathbb{R}^{M \times M}$. Let $\mathbf{H}^T$ denote the transpose of a matrix $\mathbf{H}$ and $| \mathbf{H}| $ denote its determinant. Also, let $\mathbf{H}^{-1}$ denote the inverse of $\mathbf{H}$, $\mathbf{H}^{-T}$ denote the transpose of $\mathbf{H}^{-1}$, and $\mathbf{H}^\dagger \triangleq (\mathbf{H}^T \mathbf{H})^{-1} \mathbf{H}^T $ denote the pseudoinverse. The notation $\| \mathbf{a} \| \triangleq \sqrt{ \sum_{i} a_i^2}$ will refer to the $\ell_2$-norm of the vector $\mathbf{a}$ while $\| \mathbf{a} \|_\infty \triangleq \max_i{|a_i|}$ will refer to the $\ell_\infty$-norm. We will use $\lambda_{\text{max}}(\mathbf{H})$ and $\lambda_{\text{min}}(\mathbf{H})$ to refer to the maximum and minimum singular values of the matrix $\mathbf{H}$. Finally, let $\mathbf{I}$ denote the identity matrix and $\mathbf{0}$ denote the all-zeros matrix (where the size will be clear from the context).

The baseband representation of a MIMO channel usually takes values over the complex field. For notational convenience, we will work with the real-valued representation of these complex matrices. Recall that any equation of the form $\mathbf{Y}_\c = \mathbf{H}_\c \mathbf{X}_\c + \mathbf{Z}_\c$ over the complex field can be expressed via its real-valued representation,
\begin{align*}
 \begin{bmatrix}
  \real(\mathbf{Y}_\c) \\
  \imag(\mathbf{Y}_\c) \\
 \end{bmatrix}
 =
 \begin{bmatrix}
  \real(\mathbf{H}_\c) & \!\!\! -\imag(\mathbf{H}_\c)\\
  \imag(\mathbf{H}_\c) & \real(\mathbf{H}_\c) \\
 \end{bmatrix}
 \begin{bmatrix}
  \real(\mathbf{X}_\c) \\
  \imag(\mathbf{X}_\c) \\
 \end{bmatrix}
+
 \begin{bmatrix}
  \real(\mathbf{Z}_\c) \\
  \imag(\mathbf{Z}_\c) \\
 \end{bmatrix} 
\end{align*}

We will denote the number of transmit and receive antennas in the complex domain by $\mct$ and $\mcr$, respectively. The corresponding real-valued representation of this $\mct \times \mcr$ channel has $\mrt = 2\mct$ effective transmit antennas, each with a real-valued input, and $\mrr = 2 \mcr$ effective receive antennas, each with a real-valued observation. We will use $\mrt$ single-user encoders for the resulting real-valued channel.\footnote{The implementation complexity of our scheme can be decreased slightly by specializing it to the complex field using the techniques in \cite[Section~II.B]{ng11IT}. For notational convenience, we will focus solely on the real-valued representation.}

\begin{definition}[Messages]
\label{def:messages}
There are $\mrt$ \textit{data streams} (or messages) $\mathbf{w}_1,\ldots,\mathbf{w}_\mrt$ of length $k$, which are each drawn independently and uniformly from $\mathbb{Z}_p^k$ where $p$ is prime. (Recall that $\mathbb{Z}_p$ refers to the integers $\mathbb{Z}$ modulo $p$.)
\end{definition}

\begin{remark} The messages are represented over a $p$-ary alphabet in order to make the connection to the compute-and-forward framework~\cite{ng11IT} explicit. 
\end{remark}

\begin{definition}[Encoders]
\label{def:encoders}
For $\ell = 1,\ldots,\mrt$, the $\ell^{\text{th}}$ data stream $\mathbf{w}_\ell$ is mapped onto a length $n$ channel input $\mathbf{x}_\ell \in \mathbb{R}^{n}$ by the $\ell^{\text{th}}$ \textit{encoder},
\begin{align*}
\mathcal{E}_\ell: \mathbb{Z}_p^k  \rightarrow \mathbb{R}^{n} \ .
\end{align*} An equal \textit{power allocation} is assumed across transmit antennas
\begin{align*}
\frac{1}{n}\| \mathbf{x}_\ell \|^2  \leq \snr \ .
\end{align*}
\end{definition}

While we formally impose a separate power constraint on each antenna, we note that the performance at high SNR (in terms
of the diversity-multiplexing tradeoff) remains unchanged if this is replaced by a sum power constraint over all antennas instead.

\begin{definition}[Rate]
\label{def:rate}
Each of the $\mrt$ encoders transmits at the same\footnote{Since the transmitters do not have knowledge of the channel matrix, we focus on the case where the $\mrt$ data streams are transmitted at equal rates. We will compare the integer-forcing receiver against successive cancellation V-BLAST schemes with asymmetric rates in Section \ref{subsec:rateallocation}.} \text{rate}
\begin{align*}
 R_\tx = \frac{k}{n} \log_2{p} \ .
 \end{align*} The \textit{total rate} of the MIMO system is just the number of transmit antennas times the rate, $\mrt \cdot R_\tx$.
\end{definition}

\begin{definition}[Channel]
\label{def:channel}
Let $\mathbf{X} \triangleq \mathbb{R}^{\mrt \times n}$ be the matrix of transmitted vectors, \begin{align*}
\mathbf{X} =
\left[
\begin{array}{c}
  \mathbf{x}^T_1 \\
   \vdots \\
   \mathbf{x}^T_{\mrt}
\end{array}
\right] \ .
\end{align*} The \textit{MIMO channel} takes $\mathbf{X}$ as an input, multiplies it by the \textit{channel matrix} $\mathbf{H} \in \mathbb{R}^{\mrr \times \mrt}$ and adds noise $\mathbf{Z} \in \mathbb{R}^{\mrr \times n}$ whose entries are independent and identically distributed (i.i.d.)~Gaussian random variables with zero mean and unit variance. The signal $\mathbf{Y} \in \mathbb{R}^{\mrr \times n}$ observed across the $\mrr$ receive antennas over $n$ channel uses can be written as\begin{equation}
\label{eq:observation}
\mathbf{Y} = \mathbf{H}\mathbf{X} + \mathbf{Z} \ .
\end{equation} We assume that the channel realization $\mathbf{H}$ is known to the receiver but unknown to the transmitter and remains constant throughout the transmission block of length $n$. In other words, CSI is only available to the receiver. Let $h_{m,\ell}$ denote the channel coefficient occupying the $m^{\text{th}}$ row and $\ell^{\text{th}}$ column of $\mathbf{H}$.
\end{definition}

\begin{remark} We will begin by investigating the performance of various schemes for fixed channel matrices $\mathbf{H}$. In Section~\ref{sec:outage}, we will consider the slow fading scenario, i.e., when the channel matrix $\mathbf{H}$ is generated randomly according to some distribution and held fixed over the blocklength of the code. 
\end{remark}

\begin{definition}[Decoder]
\label{def:decoder}
At the receiver, a \textit{decoder} makes an estimate of the messages
\begin{align*}
&\mathcal{D}: \mathbb{R}^{\mrr \times n}\rightarrow \mathbb{Z}_p^{\mrt\times k}  \\
&(\mathbf{\hat{w}}_1, \ldots,\mathbf{\hat{w}}_{\mrt}) = \mathcal{D}(\mathbf{Y}).
\end{align*}
\end{definition}

\begin{definition}[Achievable Rates]
\label{def:achrate}
We say that sum rate $R(\mathbf{H})$ is \textit{achievable} if for any $\epsilon > 0$ and $n$ large enough, there exist encoders and  a decoder such that reliable decoding is possible
\begin{align*}
\mathbb{P} \left( (\mathbf{\hat{w}}_1, \ldots,\mathbf{\hat{w}}_{\mrt}) \neq (\mathbf{w}_1, \ldots,\mathbf{w}_{\mrt}) \right) \leq \epsilon
\end{align*} so long as the total rate does not exceed $R(\mathbf{H})$,
\begin{align*}
\mrt \cdot R_{\text{TX}} \leq R(\mathbf{H}).
\end{align*}
\end{definition}

\section{Conventional Receiver Architectures}
\label{sec:existingarchitecture}
Many approaches to MIMO decoding have been studied in the literature. We now provide a brief overview of the rates achievable via the joint ML, zero-forcing, linear MMSE, and linear MMSE-SIC receivers.
\subsection{Joint ML Receiver}
\begin{figure}[ht]
\begin{center}
\psset{unit=0.65mm}
\begin{pspicture}(-21,-23)(110,47)
\rput(-15,28){
\rput(-2.5,0){$\mathbf{w}_1$} \psline{->}(1,0)(7,0) \psframe(7,-5)(21,5)
\rput(14,0){$\mathcal{E}_1$} \rput(28.5,3.5){$\mathbf{x}_1$}
\psline{->}(21,0)(36,0)
}

\rput(-1,10){$\vdots$}

\rput(-15,-13.5){
\rput(-2.5,0){$\mathbf{w}_\mrt$} \psline{->}(2,0)(7,0) \psframe(7,-5)(21,5)
\rput(14,0){$\mathcal{E}_\mrt$} \rput(28.5,3.5){$\mathbf{x}_\mrt$}
\psline{->}(21,0)(36,0)
}

\rput(20, -22.5){
\psframe(1,0)(28,60)
\rput(14.5, 32){\Large{$\mathbf{H}$}}
}
\rput(52,32){
\psline{->}(-4,0)(5,0)
\rput(8,0){
\pscircle(0,0){3}
\psline(-1.5,0)(1.5,0)
\psline(0,-1.5)(0,1.5)
\psline{->}(0,8)(0,3)
\rput(0,11.5){$\mathbf{z}_1$}
\psline{->}(3,0)(18,0)
}

\rput(18.5,4){$\mathbf{y}_1$}
}

\rput(60,10){$\vdots$}
\rput(52,-16){
\psline{->}(-4,0)(5,0)
\rput(8,0){
\pscircle(0,0){3}
\psline(-1.5,0)(1.5,0)
\psline(0,-1.5)(0,1.5)
\psline{->}(0,8)(0,3)
\rput(1,11.5){$\mathbf{z}_{\mrr}$}
\psline{->}(3,0)(18,0)
}
\rput(18.5,4){$\mathbf{y}_{\mrr}$}
}

\rput(78,-22.5){
\psframe(0,0)(18,60)
\rput(9,30.5){\large{$\mathcal{D}$}}
}

\rput(84,8){
\psline{->}(12,0)(18,0)
\rput(22,15){$\mathbf{\hat{w}}_1$}
\rput(22,-15){$\mathbf{\hat{w}}_\mrt$}
\rput(22,2){$\vdots$}
}

\end{pspicture}
\end{center}
\caption{MIMO channel with single stream encoding.} \label{f:mimo}
\end{figure}
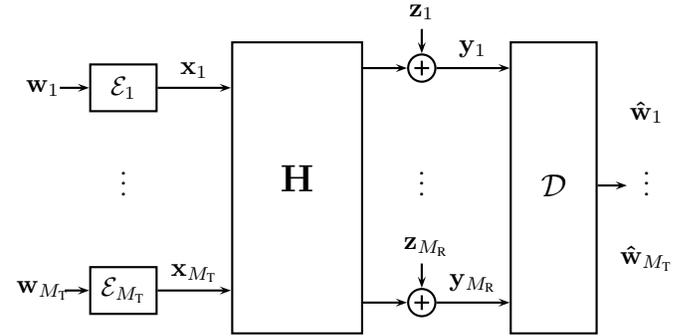

Clearly, the best performance is attainable by joint ML decoding across all $\mrr$ receive antennas and $n$ time slots. This situation is illustrated in Figure~\ref{f:mimo}. Let $\mathbf{H}_\mathcal{S}$ denote the submatrix of $\mathbf{H}$ formed by taking the columns with indices in $\mathcal{S} \subseteq \{1,2,\ldots,\mrt\}$. If we use a \textit{joint ML receiver} that searches for the most likely set of transmitted message vectors $\mathbf{\hat{w}}_1, \ldots, \mathbf{\hat{w}}_{\mrt}$, then the following rate is achievable\footnote{With joint encoding (across the transmit antennas) and joint ML decoding, a rate of $\frac{1}{2}\log\det{\left(\mathbf{I} + \snr\  \mathbf{HH}^T\right)}$ is achievable.} (if the transmitters employ i.i.d~Gaussian codebooks):
\begin{align}
\label{eq:rjoint}
R_{\text{ML}}(\mathbf{H}) = \mrt \cdot \min_{\mathcal{S} \subseteq \{1,\ldots,\mrt\}}\frac{1}{2|\mathcal{S}|} \log \det \big( \mathbf{I} + \snr\ \mathbf{H}_\mathcal{S}\mathbf{H}^T_{\mathcal{S}}\big) 
\end{align} Note that this is also the capacity of the channel subject to equal rate constraints per transmit antenna. The worst-case complexity of this approach is exponential in the product of the blocklength $n$ and the number of antennas $\mrt$. As discussed in Section~\ref{sec:intro}, the complexity of the joint ML receiver can be reduced in practice via sphere decoding.

\subsection{Conventional Linear Receivers} \label{s:convlinear}
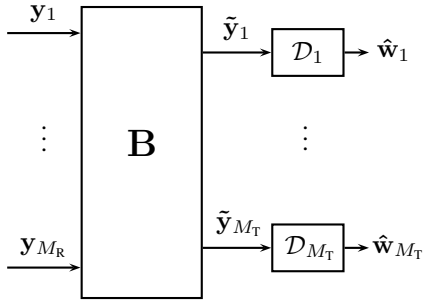
\begin{figure}[ht]
\begin{center}
\psset{unit=0.65mm}
\begin{pspicture}(35,-25)(122,40)
\rput(25,32){
\rput(8,0){
\psline{->}(2,0)(17,0)
}
\rput(17.5,4){$\mathbf{y}_1$}
}

\rput(55,-22.5){
\psframe(-5,0)(20,60)
\rput(7.5, 32){\Large{$\mathbf{B}$}}
}

\rput(64,28){
\psline{->}(11,0)(25,0)
\rput(18,5){$\mathbf{\tilde{y}}_1$}
\psframe(25,5)(40,-5)
\rput(32.5,0){$\mathcal{D}_1$}
\psline{->}(40,0)(45,0)
\rput(50,.5){$\mathbf{\hat{w}}_1$}
}

\rput(42.5, 12){$\vdots$}

\rput(25,-16){
\rput(8,0){
\psline{->}(2,0)(17,0)
}
\rput(17.5,4){${\mathbf{y}}_{\mrr}$}
}

\rput(96, 12){$\vdots$}

\rput(64,-12){
\psline{->}(11,0)(25,0)
\rput(18.5,5){$\mathbf{\tilde{y}}_\mrt$}
\psframe(25,5)(40,-5)
\rput(33,0){$\mathcal{D}_\mrt$}
\psline{->}(40,0)(45,0)
\rput(51,.25){$\mathbf{\hat{w}}_\mrt$}
}
\end{pspicture}
\end{center}
\caption{A conventional linear receiver. Each of the individual message vectors is decoded directly from the projected channel output. The goal of the linear equalizer is to approximately invert the channel and cancel the interference from other streams.} \label{f:trad}
\end{figure}

Rather than processing all the observed signals from the antennas jointly, one simple approach is to separate out the transmitted data streams using linear equalization and then decode each data stream individually, as shown in Figure~\ref{f:trad}. Given the channel output $\mathbf{Y}$ from $\eqref{eq:observation}$, the receiver applies the equalization matrix $\mathbf{B} \in \mathbb{R}^{\mrt \times \mrr}$ to obtain 
\begin{align*}
\mathbf{\tilde{Y}} &= \mathbf{B} \mathbf{Y} \\
&=\mathbf{B}\mathbf{H}\mathbf{X} + \mathbf{B}\mathbf{Z} \ .
\end{align*} Each row $\mathbf{\tilde{y}}^T_m$ of $\mathbf{\tilde{Y}}$ is treated as an estimate of $\mathbf{x}^T_m$,
\begin{equation*}
\mathbf{\tilde{y}}_m^T = \mathbf{b}_m^T \mathbf{h}_m \mathbf{x}_m^T + \underbrace{\sum_{i\neq m} \mathbf{b}_m^T \mathbf{h}_i \mathbf{x}_i^T + \mathbf{b}_m^T \mathbf{Z}}_{\text{effective noise}}  
\end{equation*} where $\mathbf{b}^T_m$ is the $m^{\text{th}}$ row of $\mathbf{B}$ and  $\mathbf{h}_m$ is the $m^{\text{th}}$ column of $\mathbf{H}$. This estimate is fed into a single-user decoder $\mathcal{D}_m: \mathbb{R}^n \rightarrow \mathbb{Z}_p^k$ to decode the $m^{\text{th}}$ data stream, $\mathbf{\hat{w}}_m = \mathcal{D}_m(\mathbf{\tilde{y}}_m)$. 

The following rate is achievable for the $m^{\text{th}}$ data stream using a conventional linear receiver with i.i.d.~Gaussian codebooks:
\begin{align*}
&R_{\text{linear},m}(\mathbf{H},\mathbf{b}_m) \\&~~~=\frac{1}{2} \log \Bigg( 1 + \frac{\snr \big(\mathbf{b}^T_m \mathbf{h}_m \big)^2}{\|\mathbf{b}_m\|^2 + \snr \sum\limits_{i\neq m}{\big(\mathbf{b}_m^T\mathbf{h}_i \big)^2}}\Bigg) \ .
\end{align*}
Since we focus on the case where each data stream is encoded at the same rate, the achievable sum rate is dictated by the worst stream,
\begin{align}
\label{eq:rlinear}
R_{\text{linear}}(\mathbf{H},\mathbf{B}) =  \mrt \cdot \min_{m=1,\ldots,\mrt}  R_{\text{linear},m}(\mathbf{H},\mathbf{b}_m)\ .
\end{align}

\begin{remark}
The complexity of a linear receiver architecture is dictated primarily by the choice of decoding algorithm for the individual data streams. In the worst case (when ML decoding is used for each data stream), the complexity is exponential in the blocklength of the data stream. In practice, one can employ modern codes with iterative decoders, such as low-density parity-check (LDPC) codes~\cite{rsu01}, to approach rates close to the capacity with linear complexity. \end{remark}

For the \textit{zero-forcing receiver}, we choose the equalization matrix to be the pseudoinverse of the channel matrix, 
\begin{align}
\mathbf{B}_{\text{ZF}} &= \mathbf{H}^\dag  \label{e:zfprojection} \ ,
\end{align} which leads to the zero-forcing sum rate 
\begin{align} \label{e:rzeroforcing}
R_{\text{ZF}}(\mathbf{H}) &= R_{\text{linear},m}(\mathbf{H},\mathbf{B}_{\text{ZF}}) \ .
\end{align} In the case where $\mrr \geq \mrt$ and $\mathbf{H}$ is full rank, the resulting channel outputs $\mathbf{\tilde{y}}_1,\ldots,\mathbf{\tilde{y}}_\mrt$ are interference free. If $\mathbf{H}$ is orthogonal, the zero-forcing receiver attains the performance of a joint ML decoder. However, as the condition number of $\mathbf{H}$ increases, the performance gap between the zero-forcing receiver and the joint ML receiver increases due to noise amplification (see the example in Section \ref{ex:2}). 

The \textit{linear MMSE receiver} applies the MMSE equalization matrix
\begin{equation}
\mathbf{B}_{\text{MMSE}} =  \snr \ \mathbf{H}^T\big( \mathbf{I} + \snr \ \mathbf{HH}^T \big)^{-1} \label{e:mmseprojection} \ , 
\end{equation} which maximizes the rate expression~\eqref{eq:rlinear} and yields the linear MMSE sum rate
\begin{equation} \label{e:rmmse}
R_{\text{MMSE}}(\mathbf{H}) = R_{\text{linear}}(\mathbf{H},\mathbf{B}_{\text{MMSE}}) \ . 
\end{equation} In the low SNR regime, this receiver can significantly outperform zero-forcing.

\subsection{Successive Interference Cancellation}
\label{sec:sic}

The performance of linear receivers can be improved using
\textit{successive interference cancellation} (SIC) \cite{vg97,wfgv98}. That is, after a codeword is
decoded, it may be subtracted from the observed vector prior to decoding
the next codeword, which increases the effective signal-to-noise ratio. Let
$\Pi$ denote the set of all permutations of $\left\{1, 2, \ldots, \mrt
\right\}$. For a fixed decoding order $\pi \in \Pi$, let $\pi_m =
 \left\{\pi(m), \pi(m + 1), \ldots, \pi(\mrt) \right\}$ denote the indices of the data
streams that have not yet been decoded. Let $\mathbf{h}_{\pi(m)}$ denote
the $\pi(m)^{\text{th}}$ column vector of $\mathbf{H}$ and let $\mathbf{H}_{\pi_m}$ be
the submatrix consisting of the columns with indices $\pi_m$, i.e., $\mathbf{H}_{\pi_m} = \big[\mathbf{h}_{\pi(m)} \cdots
\mathbf{h}_{\pi(\mrt)}\big]$. The following rate is achievable for the
$\pi(m)^{\text{th}}$ stream using SIC:
\begin{align}
\label{eq:sicstream}
&R_{\text{SIC},\pi(m)}(\mathbf{H},\mathbf{b}_m) \\&~~~=\frac{1}{2} \log \Bigg( 1 + \frac{\snr
\big(\mathbf{b}^T_m \mathbf{h}_{\pi(m)} \big)^2}{\|\mathbf{b}_m\|^2 + \snr
\sum\limits_{i > m}{\big( \mathbf{b}_m^T\mathbf{h}_{\pi(i)} \big)^2}}\Bigg) \ . \nonumber
\end{align}
where $\mathbf{b}_m$ is the equalization vector
for decoding the $\pi(m)^{\text{th}}$ stream after canceling interference
due to streams $\pi(1)$ through $\pi(m-1)$. The (optimal) MMSE equalization vector is
\begin{equation}
\mathbf{b}_{\text{MMSE-SIC},m}^T = \snr \  \mathbf{h}_{\pi(m)}^T \big(\mathbf{I} + \snr \ \mathbf{H}_{\pi_m}\mathbf{H}_{\pi_m}^T\big)^{-1} \ . \label{e:mmsesicprojection}
\end{equation}  Using this equalization vector and a fixed decoding
order $\pi$, we obtain the SIC scheme often referred to as V-BLAST I,
\begin{align}
\label{eq:rmmsesic1}
&R_{\text{V-BLAST I}}(\mathbf{H}) \nonumber \\
&~=~ \mrt \cdot \min_{m=1,\ldots,\mrt} R_{\text{SIC},\pi(m)}(\mathbf{H},\mathbf{b}_{\text{MMSE-SIC},m}) \ .
\end{align}

The sum rate can be further improved by choosing the decoding order that maximizes the rate of the worst stream,
\begin{align}
\label{eq:rmmsesic2}
&R_{\text{V-BLAST II}}(\mathbf{H}) \nonumber \\
&~=~ \mrt \cdot \max_{\pi \in \Pi} \min_{m=1,\ldots,\mrt}  R_{\text{SIC},\pi(m)}(\mathbf{H},\mathbf{b}_{\text{MMSE-SIC},m}) \ ,
\end{align} which is known as V-BLAST II. We postpone the discussion of V-BLAST III (which permits asymmetric rate allocation) to Section \ref{sec:outage}.

The complexity of the linear MMSE-SIC architecture is determined by the decoding algorithm used for the individual data streams. Note that, unlike the zero-forcing and linear MMSE receivers, not all $\mrt$ streams can be decoded in parallel and delay is incurred as later streams have to wait for earlier streams to finish decoding.

\begin{figure}
\begin{center}
\psset{unit=0.65mm}
\begin{pspicture}(57,-25)(181,42)
\rput(46,32){
\rput(8,0){
\psline{->}(3,0)(18,0)
}
\rput(19,4){$\mathbf{y}_1$}
}

\rput(67,-22.5){
\psframe(5,0)(30,60)
\rput(17.5, 32){\Large{$\mathbf{B}$}}
}

\rput(87,28){
\psline{->}(10,0)(25,0)
\rput(17.5,5){$\mathbf{\tilde{y}}_1$}
\psframe(25,5)(40,-5)
\rput(32.5,0){$\mathcal{D}_1$}
\psline{->}(40,0)(55,0)
\rput(47.5,4.5){$\mathbf{\hat{u}}_1$}
}

\rput(65, 12){$\vdots$}

\rput(46,-16){
\rput(8,0){
\psline{->}(3,0)(18,0)
}
\rput(19,4){${\mathbf{y}}_{\mrr}$}
}

\rput(119, 12){$\vdots$}

\rput(87,-12){
\psline{->}(10,0)(25,0)
\rput(17.5,5){$\mathbf{\tilde{y}}_\mrt$}
\psframe(25,5)(40,-5)
\rput(33,0){$\mathcal{D}_\mrt$}
\psline{->}(40,0)(55,0)
\rput(47.5,4.5){$\mathbf{\hat{u}}_\mrt$}
}

\rput(143,-22.5){
\psframe(-1,3)(24,57)
\rput(13,32){\Large{$\mathbf{A}_p^{-1}$}}
}
\psline{->}(167,10)(173,10)
\rput(176, 24){$\mathbf{\hat{w}}_1$}
\rput(176, 12){$\vdots$}
\rput(176, -6){$\mathbf{\hat{w}}_\mrt$}
\end{pspicture}
\end{center}\caption{The proposed integer-forcing linear receiver. The goal of the linear equalizer is to create a full-rank, integer-valued effective channel matrix $\mathbf{A}$ that minimizes the effective noise seen at the decoders. After the equalization step, each decoder recovers a linear combination $\mathbf{u}_m = \big[ \sum_\ell a_{m,\ell} \mathbf{w}_m \big]\hspace{-0.05in} \mod{p}$ of the transmitted messages. These linear combinations are then collected and solved for the individual messages. In the figure, $\mathbf{A}_p^{-1}$ represents the inverse of $[\mathbf{A}]\bmod{p}$ over $\mathbb{Z}_p$.} \label{f:latticedemod}
\end{figure}
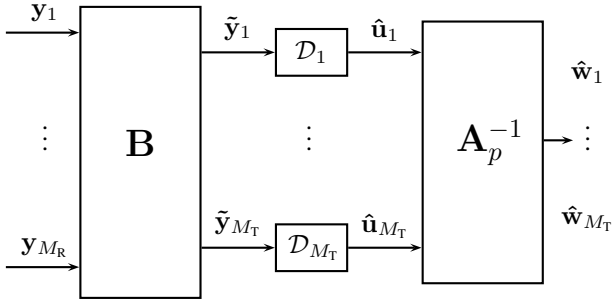

\section{Proposed Receiver Architecture}
\label{sec:architecture}
\subsection{Architecture Overview}
Linear receivers such as the zero-forcing or linear MMSE receiver directly decode the data streams after the equalization step. In other words, they use the equalization matrix $\mathbf{B}$ to invert the channel matrix at the cost of amplifying the noise. Although low in complexity, these approaches are far from optimal when the channel matrix is ill-conditioned. In the integer-forcing architecture, each encoder uses the same linear code and the receiver exploits the code-level linearity to recover linear combinations of the transmitted messages. Instead of inverting the channel, the scheme uses $\mathbf{B}$ to force the effective channel to a full-rank integer matrix $\mathbf{A}$. As in the case of traditional linear receivers, each row of the effective output is then sent to a separate decoder. Since each encoder uses the same linear codebook, each integer combination is itself a codeword and can be decoded reliably up to the codebook's noise tolerance. The integer-forcing receiver is free to optimize over all full-rank integer matrices $\mathbf{A}$ in order to minimize the effective noise seen at each decoder, and the transmitter is agnostic to the choice of $\mathbf{A}$. Using the compute-and-forward framework~\cite{ng11IT}, each integer combination can be mapped to a linear combination of the messages. See Figure~\ref{f:latticedemod} for a block diagram. Finally, these linear combinations are solved for the original messages. Before discussing the details of the architecture and its achievable rates, we give a brief overview of the compute-and-forward framework below.

\subsection{Compute-and-Forward Preliminaries}

Earlier work on \textit{compute-and-forward} \cite{ng11IT} demonstrated that it is possible for a receiver to decode linear combinations of the transmitted messages without recovering the messages individually. We now briefly review the basic framework of compute-and-forward as it will serve as a building block for our integer-forcing architecture. 

\begin{definition}[Lattice] A \emph{lattice} $\Lambda$ is a discrete additive subgroup of $\mathbb{R}^n$ that is closed under additive and reflection. We call $\mathbf{G}$ a \emph{generator matrix} for $\Lambda$ if
\begin{align}
\Lambda = \left\{\mathbf{G}\mathbf{d} : \mathbf{d} \in \mathbb{Z}^{n} \right\} \ . \nonumber
\end{align}
\end{definition} As a consequence of this definition, the zero vector $\mathbf{0}$ is always an element of $\Lambda$. 

A lattice codebook is simply a code whose codewords are elements of a lattice \cite{zamirbook}. Lattice codebooks can be formed by combining a regular (i.e., low-dimensional lattice) constellation (e.g., PAM, QAM) with a linear code (e.g., an LDPC code) or via more intricate nested lattice constructions \cite{ez04}. Note that any integer-linear combination of lattice points is itself a lattice point. That is, if $\mathbf{x}_1,\ldots, \mathbf{x}_{\mrt} \in \Lambda$, then $\sum_{\ell} a_{m,\ell} \mathbf{x}_\ell \in \Lambda$ for any $a_{m,\ell} \in \mathbb{Z}$. Both compute-and-forward and integer-forcing require that all transmitted codewords are drawn from the same lattice codebook.

Assume that the channel is not equalized, $\mathbf{B} = \mathbf{I}$ and consider the channel output observed at the $m^{\text{th}}$ receive antenna,
\begin{align} \label{e:cfchannel}
\mathbf{y}_m^T &= \mathbf{h}^T_m\mathbf{X} + \mathbf{z}_m^T \ .
\end{align} The decoder's goal is to recover the linear combination $\mathbf{u}_m = \big[ \sum_{\ell} a_{m,\ell} \mathbf{w}_\ell \big] \bmod{p}$ for some $a_{m,\ell} \in \mathbb{Z}$ directly from $\mathbf{y}_m$. Let $\mathbf{a}_m = [a_{m,1} ~\cdots ~ a_{m,\mrt}]^T$ denote the integer coefficient vector. After scaling by $\beta_m \in \mathbb{R}$, we can rewrite the channel output as follows:
\begin{align}
\beta_m \mathbf{y}_m^T = \underbrace{\mathbf{a}^T_m\mathbf{X}}_{\text{lattice codeword}} +~~~ \underbrace{(\beta_m \mathbf{h}^T_m - \mathbf{a}^T_m) \mathbf{X} + \beta_m \mathbf{z}_m^T}_{\text{effective noise}} \ ,\label{e:compeffecchannel}
\end{align} The effective noise variance is 
\begin{align*}
\sigma_{\text{eff},m}^2 &= \frac{1}{n} \mathbb{E} \big\| (\beta_m \mathbf{h}^T_m - \mathbf{a}^T_m) \mathbf{X} + \beta_m \mathbf{z}_m^T \big\|^2 \\
&= \beta_m^2  + \snr \| \beta_m \mathbf{h}_m - \mathbf{a}_m \|^2 
\end{align*} where we have used the fact that the data streams $\mathbf{x}_\ell$ are independent of one another and assumed that $\frac{1}{n} \mathbb{E}\| \mathbf{x}_\ell \|^2 = \snr$. In the above expression, the leading $\beta_m^2$ term corresponds to the variance of the additive noise after scaling by $\beta_m$. The more interesting term $\snr \|\beta_m \mathbf{h}_m - \mathbf{a}_m\|^2$ corresponds to a ``non-integer" penalty due to any mismatch between the effective channel vector $\beta_m \mathbf{h}_m$ and the integer coefficient vector $\mathbf{a}_m$.

Roughly speaking, if the rate of the lattice codebook is $\frac{1}{2} \log\big(\snr / \sigma_{\text{eff},m}^2\big)$, then it can tolerate effective noise of variance $\sigma_{\text{eff},m}^2$ or higher, and the decoder can recover the integer-linear combination $\mathbf{a}_m^T \mathbf{X}$ by decoding to the closest lattice codeword to $\mathbf{y}_m^T$ in Euclidean distance. It can then map this integer-linear combination to the desired linear combination of the messages. The following theorem from~\cite{ng11IT} makes this precise. Define $\log^{+}(x) \triangleq \max \left\{ \log(x), 0\right\}$.

\begin{theorem}[{\cite[Theorem 1]{ng11IT}}]
\label{thm:rate1}
For any $\epsilon > 0$ and $n, p$ large enough, there exist encoders and decoders, $\mathcal{E}_1, \ldots, \mathcal{E}_{\mrt}, \mathcal{D}_1, \ldots ,\mathcal{D}_{\mrt}$, such that the decoders can recover the linear combinations $\mathbf{u}_m = \big[ \sum_{\ell} a_{m,\ell} \mathbf{w}_\ell \big] \bmod{p}$ from the channel outputs~\eqref{e:cfchannel} with total probability of error at most $\epsilon$ for any choice of integer coefficient vectors $\mathbf{a}_1,\ldots, \mathbf{a}_{\mrt} \in \mathbb{Z}^{\mrt}$ satisfying
\begin{align}
R_{\text{TX}} &< \min_{m = 1, \ldots , \mrt} R_{\comp}(\mathbf{H},\mathbf{a}_m,\beta_m) \label{eqn:comprate} \\
R_{\comp}(\mathbf{H},\mathbf{a}_m,\beta_m)& = \frac{1}{2}\log^{+} \left( \frac{\snr}{\beta_m^2 +  \snr \|\beta_m \mathbf{h}_m - \mathbf{a}_m \|^2} \right)  \nonumber 
\end{align} for some $\beta_1,\ldots,\beta_{\mrt} \in \mathbb{R}$. 
\end{theorem}

\begin{remark}
Note that the decoders in Theorem \ref{thm:rate1} are free to choose any integer coefficients that satisfy (\ref{eqn:comprate}) using their knowledge of $\mathbf{H}$ and $\snr$. The encoders can operate without knowledge of $\mathbf{H}$ (and hence the choice of integer coefficients) by allowing for some probability of outage, as discussed in Section \ref{sec:outage}.
\end{remark}

\subsection{Integer-Forcing Achievable Rates}
\label{decoding_equations}

We now describe the integer-forcing linear receiver and its achievable rates in detail. See Figure~\ref{f:latticedemod} for a block diagram. Throughout, we assume that the $\ell^{\text{th}}$ channel input $\mathbf{x}_\ell$ is the result of mapping the message $\mathbf{w}_\ell$ onto a codeword from a lattice codebook that is shared across the transmitters.

Upon observing $\mathbf{Y}$, the receiver applies an equalization matrix $\mathbf{B} \in \mathbb{R}^{\mrt \times \mrr}$ to obtain the effective channel output 
\begin{align*}
\mathbf{\tilde{Y}} &= \mathbf{B}\mathbf{Y} \\
&=\mathbf{B}\mathbf{H} \mathbf{X} + \mathbf{B}\mathbf{Z} \\
&= \mathbf{A} \mathbf{X} + ( \mathbf{BH} - \mathbf{A} ) \mathbf{X} + \mathbf{BZ}
\end{align*} where $\mathbf{A} = \{a_{m,\ell}\} \in \mathbb{Z}^{\mrt \times \mrt}$ is the matrix of desired integer coefficients. Let $\mathbf{\tilde{y}}_m^T$, $\mathbf{b}_m^T$, and $\mathbf{a}_m^T$ denote the $m^{\text{th}}$ rows of $\mathbf{\tilde{Y}}$, $\mathbf{B}$, and $\mathbf{A}$, respectively. We can write $\mathbf{\tilde{y}}_m^T$ as the sum of a lattice codeword plus some effective noise, 
\begin{align}
\mathbf{\tilde{y}}_m^T &= \mathbf{b}^T_m \mathbf{H} \mathbf{X} + \mathbf{b}^T_m \mathbf{Z} \nonumber \\
&=  \underbrace{\mathbf{a}^T_m \mathbf{X}}_{\text{lattice codeword}} + \underbrace{\big( \mathbf{b}^T_m \mathbf{H} - \mathbf{a}^T_M \big) \mathbf{X} + \mathbf{b}^T_m \mathbf{Z}}_{\text{effective noise}} \label{e:ifeffecchannel} \ .
\end{align} The effective noise variance is
\begin{align}
\sigma_{\text{eff},m}^2 &= \frac{1}{n} \mathbb{E} \Big\|  \big( \mathbf{b}^T_m \mathbf{H} - \mathbf{a}^T_M \big) \mathbf{X} + \mathbf{b}^T_m \mathbf{Z} \Big\|^2 \nonumber \\
&= \| \mathbf{b}_m \|^2 + \snr \big\| \mathbf{H}^T \mathbf{b}_m - \mathbf{a}_m \big\|^2 \ . \label{e:ifeffecnoise}
\end{align}

The $m^{\text{th}}$ effective channel output $\mathbf{\tilde{y}}_m$ and the desired integer coefficient vector $\mathbf{a}_m$ are fed into a SISO decoder $\mathcal{D}_m: \mathbb{R}^{n} \times \mathbb{Z}^{\mrt} \rightarrow \mathbb{Z}_p^k$. This decoder makes an estimate $\mathbf{\hat{u}}_m = \mathcal{D}_m( \mathbf{\tilde{y}}_m, \mathbf{a}_m)$ of the linear combination 
\begin{align*}
\mathbf{u}_m = \left[\sum_{\ell =1}^{\mrt}{a_{m,\ell} \mathbf{w}_\ell}\right]\bmod{p} \ .
\end{align*} 

As shown in the following lemma, the desired linear combinations can be recovered reliably if the lattice codebook can tolerate noise of effective variance $\max_{m} \sigma_{\text{eff},m}^2$.
\begin{lemma} \label{l:mimocomp}
For any $\epsilon > 0$ and $n, p$ large enough, there exist encoders and decoders, $\mathcal{E}_1, \ldots, \mathcal{E}_{\mrt}, \mathcal{D}_1, \ldots ,\mathcal{D}_{\mrt}$, such that the decoders can recover the linear combinations $\mathbf{u}_m = \big[ \sum_{\ell} a_{m,\ell} \mathbf{w}_\ell \big] \bmod{p}$ from the channel outputs~\eqref{e:ifeffecchannel} with probability of error at most $\epsilon$ for any choice of integer matrix $\mathbf{A} = [\mathbf{a}_1~\cdots ~\mathbf{a}_{\mrt}]^T \in \mathbb{Z}^{\mrt \times \mrt}$ satisfying
\begin{align}
&\qquad \qquad \qquad R_{\text{TX}} <  R_{\comp}(\mathbf{H},\mathbf{A},\mathbf{B}) \label{eqn:comprate} \\
&R_{\comp}(\mathbf{H},\mathbf{A},\mathbf{B})  \label{eqn-rateexp-1} \\
&~~ = \min_{m=1,\ldots,\mrt} \frac{1}{2}\log^{+} \left( \frac{\snr}{ \| \mathbf{b}_m \|^2 + \snr \big\| \mathbf{H}^T \mathbf{b}_m - \mathbf{a}_m \big\|^2} \right)  \nonumber 
\end{align} for some $\mathbf{B} = [\mathbf{b}_1~\cdots ~\mathbf{b}_{\mrt}]^T \in \mathbb{R}^{\mrt \times \mrr}$.
\end{lemma} This lemma follows directly from Theorem~\ref{thm:rate1} by noting that the effective channel output from~\eqref{e:ifeffecchannel} has the same form as~\eqref{e:compeffecchannel}, and by substituting in the effective noise variance from~\eqref{e:ifeffecnoise}.

Assuming the linear combinations have been decoded correctly, they can be solved (in the digital domain) for the desired messages. Let $\mathbf{W} =  [\mathbf{w}_1~\cdots~\mathbf{w}_{\mrt}]^T$ denote the matrix of message vectors, $\mathbf{U} = [\mathbf{u}_1~\cdots~\mathbf{u}_{\mrt}]^T$ denote the matrix of their linear combinations, and $\mathbf{A}_p  = [ \mathbf{A} ] \bmod{p}$ denote the desired integer matrix modulo $p$. If $\mathbf{A}_p$ is full-rank over $\mathbb{Z}_p$, the message vectors can be recovered from the linear combinations, 
\begin{align}
\mathbf{W} = \big[\mathbf{A}_p^{-1} \mathbf{U} \big] \bmod{p}
\end{align} where the inverse $\mathbf{A}_p^{-1}$ is taken over $\mathbb{Z}_p$.

In summary, by optimizing over the equalization matrix $\mathbf{B}$ and the integer matrix $\mathbf{A}$, we obtain the following sum rate.
\begin{theorem}
\label{thm:rate}
Under the integer-forcing architecture, the following sum rate is achievable:
\begin{align}
R_{\iftext}(\mathbf{H}) &= \mrt \cdot \max_{\substack{\mathbf{A} \in \mathbb{Z}^{\mrt \times \mrt}\\  \mathrm{rank}(\mathbf{A}) = \mrt}} \max_{\mathbf{B} \in \mathbb{R}^{\mrt \times \mrr}} R_{\text{comp}}(\mathbf{H},\mathbf{A},\mathbf{B}) \label{e:totalrate}
\end{align}
\end{theorem}

Note that the optimization is over all integer matrices $\mathbf{A} \in \mathbb{Z}^{\mrt \times \mrt}$ that are full rank over the reals, rather than over $\mathbb{Z}_p$. To show that this condition suffices, we will first need to establish an upper bound on the magnitudes of the elements in $\mathbf{A}$.

\begin{lemma}\label{lm:asearch}
The rate expression~$R_{\comp}(\mathbf{H},\mathbf{A},\mathbf{B})$ from Lemma~\ref{l:mimocomp} is $0$ for any integer matrix $\mathbf{A}$ such that, for some $m \in \{1,2,\ldots,\mrt\}$,
\begin{align}
\| \mathbf{a}_m \| \geq 1 + \sqrt{\snr} \   \lambda_{\text{max}}(\mathbf{H}) \ . \label{e:normbound}
\end{align} 
\end{lemma}
\begin{IEEEproof}
First, note that if the denominator in~\eqref{eqn-rateexp-1} is larger than $\snr$, the rate expression is equal to $0$. Therefore, we must have that 
\begin{align}
\| \mathbf{b}_m \|^2 &\leq \snr \label{e:projectionnorm} \\
\big\| \mathbf{H}^T \mathbf{b}_m - \mathbf{a}_m \big\| &\leq 1 \label{e:nonintegernorm} \ .
\end{align} By applying the reverse triangle inequality to~\eqref{e:nonintegernorm}, we obtain
\begin{align*}
\| \mathbf{a}_m \| &\leq 1 + \| \mathbf{H}^T \mathbf{b}_m \| \\
& \leq 1 + \lambda_{\text{max}}(\mathbf{H}) \| \mathbf{b}_m\| \\
& \leq 1 + \sqrt{\snr} \ \lambda_{\text{max}}(\mathbf{H}) 
\end{align*} where the last step is due to~\eqref{e:projectionnorm}.
\end{IEEEproof}

\begin{IEEEproof}[Proof of Theorem~\ref{thm:rate}]
Applying Lemma~\ref{l:mimocomp}, it follows that the decoders can reliably recover the linear combinations $\mathbf{u}_1, \ldots, \mathbf{u}_{\mrt}$ with integer coefficient matrix $\mathbf{A}$ using equalization matrix $\mathbf{B}$ so long as the codebook rate $R_{\tx}$ does not exceed $R_{\comp}(\mathbf{H},\mathbf{A},\mathbf{B})$. To recover the messages, we need that $[\mathbf{A}]\bmod{p}$ is full rank over $\mathbb{Z}_p$. As shown in~\cite[Theorem 11]{ng11IT}, if the magnitudes of the elements of $\mathbf{A}$ are upper bounded by a constant, it suffices to check whether $\mathbf{A}$ is full rank over the reals, rather than $\mathbb{Z}_p$. From Lemma~\ref{lm:asearch}, the rate will be positive only if $\| \mathbf{a}_m \| < 1 + \sqrt{\snr} \ \lambda_{\text{max}}(\mathbf{H})$ for $m=1,2,\ldots,\mrt$, which in turn bounds the elements of $\mathbf{A}$.
\end{IEEEproof}

\begin{remark} In Section~\ref{sec:outage}, we will examine the outage performance of integer-forcing, i.e., $\mathbf{H}$ will be drawn according to some distribution and its realization will be unknown to the transmitter(s). In this setting, $\lambda_{\text{max}}(\mathbf{H})$ will not be known (and $\lambda_{\text{max}}(\mathbf{H})$ may be not upper bounded by an absolute constant). However, it follows from~\cite[Remark 10]{ng11IT} that, if $\mathbf{H}$ is drawn from a distribution such that $\mathbb{P}( \lambda_{\text{\textnormal{max}}}(\mathbf{H}) > \gamma) \rightarrow 0$ as $\gamma \rightarrow \infty$, then it still suffices to check the rank of  $\mathbf{A}$ over the reals. 
\end{remark}

In the following subsections, we will demonstrate that integer-forcing can match the performance of a conventional linear receiver by setting the integer matrix $\mathbf{A} = \mathbf{I}$, derive the optimal equalization matrix $\mathbf{B}$ for a given $\mathbf{A}$, discuss how to select $\mathbf{A}$ to maximize the rate, and explore the implementation complexity.

\subsection{Conventional Linear Receivers as a Special Case of Integer-Forcing with $\mathbf{A} = \mathbf{I}$} \label{s:specialcase}

The following lemma establishes that integer-forcing can match the achievable rate of any conventional linear receiver.

\begin{lemma} \label{l:ifidentity}
For any channel matrix $\mathbf{H}$, the achievable sum rate $R_{\text{linear}}(\mathbf{H},\mathbf{B})$ for a conventional linear receiver with equalization matrix $\mathbf{B}$ is also achievable via integer-forcing by setting the integer matrix $\mathbf{A}$ to be the identity matrix and using equalization matrix $\mathbf{\tilde{B}}$ whose $m^{\text{th}}$ row is $\mathbf{\tilde{b}}_m^T = \alpha_m \mathbf{b}_m^T$ where 
\begin{align*}
\alpha_m = \frac{\snr \ \mathbf{b}_m^T \mathbf{h}_m}{\| \mathbf{b}_m \|^2 + \snr \sum_{i=1}^{\mrt} \big( \mathbf{b}_m^T \mathbf{h}_i \big)^2} \ .
\end{align*} That is, $R_{\comp}(\mathbf{H},\mathbf{I},\mathbf{\tilde{B}}) = R_{\text{linear}}(\mathbf{H},\mathbf{B})$.
\end{lemma} 
\begin{IEEEproof}
From~\eqref{e:ifeffecnoise}, the effective noise variance is 
\begin{align*}
\sigma_{\text{eff},m}^2 &= \alpha_m^2 \| \mathbf{b}_m \|^2 + \snr \ ( \alpha_m \mathbf{b}_m^T \mathbf{h}_m - 1)^2  \\ 
&~~~~+~\snr \sum_{i \neq m} (\alpha_m \mathbf{b}_m^T \mathbf{h}_i )^2 \\
&= \snr \left( \frac{\| \mathbf{b}_m \|^2 + \snr \sum_{i\neq m} (\mathbf{b}_m^T \mathbf{h}_i)^2 }{\| \mathbf{b}_m \|^2 + \snr \sum_{i=1}^{\mrt} (\mathbf{b}_m^T \mathbf{h}_i)^2 }\right) \ .
\end{align*} It follows that the $m^{\text{th}}$ data stream can be decoded successfully up to rate
\begin{align*}
\frac{1}{2} \log\bigg(\frac{\snr}{\sigma_{\text{eff},m}^2}\bigg) &= \frac{1}{2} \log\bigg(\frac{\| \mathbf{b}_m \|^2 + \snr \sum_{i=1}^{\mrt} (\mathbf{b}_m^T \mathbf{h}_i)^2 }{\| \mathbf{b}_m \|^2 + \snr \sum\limits_{i\neq m} (\mathbf{b}_m^T \mathbf{h}_i)^2 }\bigg) \\
&= R_{\text{linear},m}(\mathbf{H},\mathbf{b}_m) \ .
\end{align*} 
\end{IEEEproof}
\begin{remark}
Readers familiar with~\cite{ez04} will recognize $\alpha_m$ as the MMSE scaling coefficient for estimating $\mathbf{x}_m$ from $\mathbf{b}_m^T \mathbf{Y}$. It can be shown that it suffices to set $\alpha_m = 1$ for the special case of the MMSE equalization matrix $\mathbf{B}_{\text{MMSE}}$.
\end{remark}

\begin{remark} It is possible to develop an integer-forcing receiver that employs successive cancellation and show that it includes the performance of conventional V-BLAST architectures as a special case. See~\cite{oen13} for more details. \end{remark}

\subsection{Optimizing the Equalization Matrix $\mathbf{B}$}

Theorem \ref{thm:rate} states an achievable integer-forcing rate for any choice of equalization matrix $\mathbf{B}$ and full-rank integer matrix $\mathbf{A}.$ (Recall that these choices do not need to be revealed to the transmitters, only the target rate $R_{\tx}$.) The remaining task is to select these matrices in such a way as to {\em maximize} the rate~\eqref{e:totalrate}. This turns out to be a non-trivial task. We consider it in two steps. In particular, we first observe that for a fixed integer matrix $\mathbf{A}$ it is straightforward to characterize the optimal equalization matrix $\mathbf{B}.$ In the next subsection, we will discuss the harder problem of selecting the integer matrix $\mathbf{A}.$

To start, consider the special case when the rank of $\mathbf{H}$ is equal to the number of transmit antenna, $\mathrm{rank}(\mathbf{H}) = \mrt$. For a fixed integer matrix $\mathbf{A}$, a simple choice for the equalization matrix is \begin{align}
\mathbf{B}_{\text{exact}} = \mathbf{A}\mathbf{H}^{\dag}  \ . \label{Eq-Bexact}
\end{align} We call this scheme ``exact'' integer-forcing since the effective channel matrix after equalization is simply the full-rank integer matrix $\mathbf{A}$.
We also note that by setting $\mathbf{A} = \mathbf{I}$ we recover the zero-forcing receiver from~\eqref{e:zfprojection}. More generally, the performance of exact integer-forcing is summarized in the following corollary.
\begin{corollary}
\label{cor:ratesub}
Assume that $\mathrm{rank}(\mathbf{H}) = \mrt$. The following sum rate is achievable via integer-forcing with $\mathbf{B}_{\text{exact}}$: \begin{align}
&R_{\iftext,\text{exact}}(\mathbf{H}) = \mrt \cdot  \max_{\substack{\mathbf{A} \in \mathbb{Z}^{\mrt \times \mrt} \\ \mathrm{rank}(\mathbf{A}) = \mrt}} R_{\comp}(\mathbf{H}, \mathbf{A}, \mathbf{B}_{\text{exact}}) \label{eq:exactrate}\\
&R_{\comp}(\mathbf{H}, \mathbf{A}, \mathbf{B}_{\text{exact}}) = \min_{m = 1,\ldots,\mrt} \frac{1}{2} \log^+ \left( \frac{\snr}{\|(\mathbf{H}{^T})^{\dag}\mathbf{a}_m \|^2 } \right)  \ .  \label{eq:ratesub}
\end{align}
\end{corollary}
Note that the achievable rate in (\ref{eq:exactrate}) is determined by the largest effective noise variance,
\begin{align}
\label{eq:effectivenoise1}
\sigma^2_{\text{exact}} = \max_{m=1,\ldots,\mrt} \big\|(\mathbf{H}{^T})^{\dag}\mathbf{a}_m \big\|^2 \ .
\end{align}Hence, the goal is to choose linearly independent integer vectors $\mathbf{a}_1, \mathbf{a}_2, \ldots, \mathbf{a}_{\mrt}$ to minimize $\sigma^2_{\text{exact}}$. As we will discuss in Section~\ref{sec:equations}, this problem corresponds to finding the shortest basis for the lattice induced by $(\mathbf{H}{^T})^{\dag}$.

We now characterize the optimal equalization matrix $\mathbf{B}$ for a fixed integer matrix $\mathbf{A}$ and provide an equivalent rate expression for Theorem \ref{thm:rate} that depends only on $\mathbf{H}$, $\mathbf{A}$, and $\snr$. In Section \ref{sec:equations}, we will use this expression to provide insight on selecting the optimal integer matrix $\mathbf{A}$.

\begin{theorem}
\label{thm:bopt}
The optimal equalization matrix for a fixed integer matrix $\mathbf{A}$ is 
\begin{align}
\label{eq:bopt}
\mathbf{B}_\text{{opt}} = \snr \ \mathbf{A}\mathbf{H}^T \left(\mathbf{I} + \snr \ \mathbf{H}\mathbf{H}^T \right)^{-1} \ ,
\end{align} which leads to the following expression for the achievable integer-forcing sum rate:
\begin{align}
&R_{\iftext}(\mathbf{H}) = \mrt \cdot  \max_{\substack{\mathbf{A} \in \mathbb{Z}^{\mrt \times \mrt} \\ \mathrm{rank}(\mathbf{A}) = \mrt}} R_{\comp}(\mathbf{H}, \mathbf{A}, \mathbf{B}_{\text{opt}}) \label{eq:optrate}\\
&R_{\comp}(\mathbf{H}, \mathbf{A}, \mathbf{B}_{\text{opt}}) = \min_{m = 1,\ldots,\mrt} \frac{1}{2} \log^+ \left( \frac{\snr}{\sigma^2_{\text{opt},m}} \right)  \\
&\sigma_{\text{opt},m}^2 = \snr \ \mathbf{a}_m^T \Big(\mathbf{I} - \snr \ \mathbf{H}^T \big( \mathbf{I} + \snr \  \mathbf{H}\mathbf{H}^T \big)^{-1} \mathbf{H} \Big) \mathbf{a}_m \ . \nonumber
\end{align}
\end{theorem}
\begin{IEEEproof}Let $\mathbf{B} = [\mathbf{b}_1~\cdots~\mathbf{b}_{\mrt}]^T$. We solve for each $\mathbf{b}_m$ separately to maximize the achievable rate in Theorem \ref{thm:rate},
\begin{align}
\mathbf{b}_m &= \argmax_{\mathbf{b}_m}  \frac{1}{2}\log^+ \left( \frac{\snr}{\| \mathbf{b}_m \|^2 +  \snr \| \mathbf{H}^T \mathbf{b}_m - \mathbf{a}_m \|^2} \right) \nonumber\\
&=\argmin_{\mathbf{b}_m} \sigma_{\text{eff},m}^2 \nonumber
\end{align} where $\sigma_{\text{eff},m}^2$ is given in~\eqref{e:ifeffecnoise}. Expanding, we find that $\sigma_{\text{eff},m}^2$ is equal to
\begin{align*}
&\mathbf{b}_m^T\mathbf{b}_m + \snr (\mathbf{H}^T\mathbf{b}_m - \mathbf{a}_m)^T(\mathbf{H}^T\mathbf{b}_m - \mathbf{a}_m)\\
&= \mathbf{b}_m^T\mathbf{b}_m + \snr \big( \mathbf{b}^T_m \mathbf{H}\mathbf{H}^T\mathbf{b}_m - 2 \mathbf{b}_m^T\mathbf{H}\mathbf{a}_m + \mathbf{a}_m^T\mathbf{a}_m \big) \\
&= \mathbf{b}^T_m \left( \mathbf{I} + \snr \ \mathbf{H}\mathbf{H}^T \right) \mathbf{b}_m -  \snr \  2\mathbf{b}_m^T\mathbf{H} \mathbf{a}_m + \snr \ \mathbf{a}_m^T \mathbf{a}_m
\end{align*}
We then take the derivative with respect to $\mathbf{b}_m$, 
\begin{align}
\frac{d \sigma_{\text{eff},m}^2}{d\mathbf{b}_m} = 2\left( \mathbf{I} + \snr \ \mathbf{H}\mathbf{H}^T \right)\mathbf{b}_m - \snr \ 2\mathbf{H}\mathbf{a}_m \ , \nonumber
\end{align}
and set it equal to zero to get
\begin{align}
\mathbf{b}^T_{\text{opt},m} = \snr \  \mathbf{a}^T_m\mathbf{H}^T\left( \mathbf{I}  + \snr \  \mathbf{H}\mathbf{H}^T \right)^{-1} \ . \nonumber
\end{align} Plugging back in, we find that $\sigma_{\text{eff},m}^2 = \sigma_{\text{opt},m}^2$ as desired.
\end{IEEEproof}
\begin{remark} If $\mathrm{rank}(\mathbf{H}) = \mrt$, the optimal equalization matrix converges to exact integer-forcing as the SNR tends to infinity, $\lim_{\snr \rightarrow \infty} \mathbf{B}_\text{opt}=\mathbf{B}_\text{exact}$. \end{remark}

We now derive an alternate expression for the achievable rate in Theorem \ref{thm:bopt}. Recall that any real, symmetric matrix $\mathbf{S}$ can be written in terms of its \emph{eigendecomposition} $\mathbf{S} = \mathbf{V}\mathbf{D} \mathbf{V}^T$ where $\mathbf{V}$ is an orthogonal matrix whose columns contain the (real) eigenvectors of $\mathbf{S}$ and $\mathbf{D}$ is a diagonal matrix whose entries are the (real) eigenvalues of $\mathbf{S}$. 

\begin{theorem}
\label{thm:alterrate}
Let $\mathbf{V}\mathbf{D}\mathbf{V}^T$ be the eigendecomposition of the symmetric matrix \mbox{$\mathbf{I} + \snr \ \mathbf{H}^T \mathbf{H}$}. The achievable integer-forcing sum rate from Theorem~\ref{thm:bopt} can be equivalently written as 
\begin{align}
&R_{\iftext}(\mathbf{H}) = \mrt \cdot  \max_{\substack{\mathbf{A} \in \mathbb{Z}^{\mrt \times \mrt} \\ \mathrm{rank}(\mathbf{A}) = \mrt}}  \min_{m = 1,\ldots,\mrt} R_{\comp}(\mathbf{H},\mathbf{a}_m) \label{eq:alternaterate}\\
&R_{\comp}(\mathbf{H}, \mathbf{a}_m) = \max\bigg(-\frac{1}{2} \log\Big( \big\| \mathbf{D}^{-1/2} \mathbf{V}^T \mathbf{a}_m\big\|^2 \Big) \ , \ 0 \bigg) \ . \nonumber
\end{align}
\end{theorem}

\begin{IEEEproof}
Using the Matrix Inversion Lemma~\cite{hs81}, it can be shown that
\begin{align*}
\mathbf{I} - \snr \ \mathbf{H}^T \big( \mathbf{I} + \snr \  \mathbf{H}\mathbf{H}^T \big)^{-1} \mathbf{H} \ = \ \big(\mathbf{I} + \snr \ \mathbf{H}^T \mathbf{H} \big)^{-1}, 
\end{align*} which enables us to write the effective noise variance from Theorem~\ref{thm:bopt} as
\begin{align*}
\sigma^2_{\text{opt},m} = \snr \ \mathbf{a}_m^T \big(\mathbf{I} + \snr \ \mathbf{H}^T \mathbf{H} \big)^{-1} \mathbf{a}_m \ .
\end{align*} Next, we express $\big(\mathbf{I} + \snr \ \mathbf{H}^T \mathbf{H} \big)^{-1}$ in terms of the eigendecomposition $\mathbf{VDV}^T$ of its inverse $\mathbf{I} + \snr \ \mathbf{H}^T \mathbf{H}$ to get
\begin{align*}
\sigma^2_{\text{opt},m} &= \snr \ \mathbf{a}_m^T \mathbf{V} \mathbf{D}^{-1} \mathbf{V}^T \mathbf{a}_m \\
&= \snr \ \big\| \mathbf{D}^{-1/2} \mathbf{V}^T \mathbf{a}_m \big\|^2 \ . 
\end{align*} Finally, we plug into $\frac{1}{2} \log^+(\snr/\sigma^2_{\text{opt},m})$ to get the desired result.
\end{IEEEproof}

\begin{remark} Note that we can express the entries of the diagonal matrix $\mathbf{D}$ in Theorem~\ref{thm:alterrate} as 
\begin{equation}
\mathbf{D}_{i,i} = \left\{
\begin{array}{c l}
1 + \lambda^2_i \ \snr & i  \leq \mbox{rank}(\mathbf{H})\\
  1 & i > \mbox{rank} (\mathbf{H})
\end{array}
\right. \label{eq:singularvalue}
\end{equation}
where $\lambda_i$ is the $i^{\text{th}}$ singular value of $\mathbf{H}$ (indexed in decreasing order). Furthermore, we can express the columns of $\mathbf{V}$ as the right singular vectors of $\mathbf{H}$.
\end{remark}

\begin{remark}
Note that the rate expression from Theorem~\ref{thm:alterrate} should be used in simulations, rather than the expression from Theorem~\ref{thm:rate}, owing to its superior numerical stability.
\end{remark}

\subsection{Selecting the Integer Matrix $\mathbf{A}$}
\label{sec:equations}

In the previous subsection, we characterized the optimal $\mathbf{B}$ for a fixed full-rank integer matrix $\mathbf{A}$. Now, we discuss how to select $\mathbf{A}$ to maximize the achievable rate. In general, for a fixed $\snr$ and channel matrix $\mathbf{H}$, finding the best $\mathbf{A}$ is a combinatorial problem, and seems to require an exhaustive search. In fact, finding the optimal $\mathbf{A}$ is linked to the hard combinatorial problem of finding the shortest set of linearly independent lattice vectors \cite{mg02}. Fortunately, the size of the search space is bounded in terms of the $\snr$ and the number of transmit antennas $\mrt$ and the search for $\mathbf{A}$ must only be performed once per coherence interval. Moreover, powerful approximation algorithms, such as the LLL algorithm can provide near-optimal solutions in polynomial time \cite{lll82}.

The following corollary of Lemma~\ref{lm:asearch} roughly characterizes the search space (for a naive exhaustive search).
\begin{corollary}
\label{c:searchspace}
To optimize the achievable rate in Theorem \ref{thm:alterrate} (or, equivalently, in Theorem~\ref{thm:rate} or~\ref{thm:bopt}), it is sufficient to check the space of integer matrices $\mathbf{A} \in \mathbb{Z}^{\mrt \times \mrt}$ whose rows $\mathbf{a}_m^T$ satisfy 
\begin{align*}
\| \mathbf{a}_m \| < 1 + \sqrt{\snr} \ \lambda_\text{max}(\mathbf{H}) \ .
\end{align*}
\end{corollary}
Thus, an exhaustive search only needs to check roughly $\snr^{\mrt}$ possibilities.

An initially tempting choice for $\mathbf{A}$ might be $\mathbf{A} = \mathbf{I}.$
As discussed in Section~\ref{s:specialcase}, this choice matches the performance of a conventional linear receiver. We now explicitly show how and why the choice $\mathbf{A} = \mathbf{I}$ is suboptimal. 

\begin{figure}
\begin{center}
\psset{unit=0.6mm}
\begin{pspicture}(0,-75)(120,120)
\rput(0,0){
\rput(60,112){\textbf{Zero-Forcing}}
\rput(15,0){
\psdots[dotscale=0.5](0,30)(0,45)(0,60)(0,75)(0,90)(0,105)
}

\rput(15,0){
\psdots[dotscale=0.5](0,30)(0,45)(0,60)(0,75)(0,90)(0,105)
}

\rput(30,0){
\psdots[dotscale=0.5](0,30)(0,45)(0,60)(0,75)(0,90)(0,105)
}

\rput(45,0){
\psdots[dotscale=0.5](0,30)(0,45)(0,60)(0,75)(0,90)(0,105)
}

\rput(60,0){
\psdots[dotscale=0.5](0,30)(0,45)(0,60)(0,75)(0,90)(0,105)
}

\rput(75,0){
\psdots[dotscale=0.5](0,30)(0,45)(0,60)(0,75)(0,90)(0,105)
}

\rput(90,0){
\psdots[dotscale=0.5](0,30)(0,45)(0,60)(0,75)(0,90)(0,105)
}

\rput(105,0){
\psdots[dotscale=0.5](0,30)(0,45)(0,60)(0,75)(0,90)(0,105)
}


\psline[linewidth=1.2]{->}(60,45)(72.5,55)
\psline[linewidth=1.2]{->}(60,45)(20,90)
\rput(87,51){$\lambda_\text{max}^{-1} \mathbf{v}_\text{max}$}
\rput(28,98){$\lambda_\text{min}^{-1} \mathbf{v}_\text{min}$}
\psline[linestyle=dashed,linewidth=0.6]{->}(60,45)(75,45)
\psline[linestyle=dashed,linewidth=0.6]{->}(60,45)(60,60)
\rput(60,65){$\mathbf{a}_1$}
\rput(81,43){$\mathbf{a}_2$}
}

\rput(0,-100){
\rput(60,112){\textbf{Integer-Forcing}}
\rput(15,0){
\psdots[dotscale=0.5](0,30)(0,45)(0,60)(0,75)(0,90)(0,105)
}

\rput(15,0){
\psdots[dotscale=0.5](0,30)(0,45)(0,60)(0,75)(0,90)(0,105)
}

\rput(30,0){
\psdots[dotscale=0.5](0,30)(0,45)(0,60)(0,75)(0,90)(0,105)
}

\rput(45,0){
\psdots[dotscale=0.5](0,30)(0,45)(0,60)(0,75)(0,90)(0,105)
}

\rput(60,0){
\psdots[dotscale=0.5](0,30)(0,45)(0,60)(0,75)(0,90)(0,105)
}

\rput(75,0){
\psdots[dotscale=0.5](0,30)(0,45)(0,60)(0,75)(0,90)(0,105)
}

\rput(90,0){
\psdots[dotscale=0.5](0,30)(0,45)(0,60)(0,75)(0,90)(0,105)
}

\rput(105,0){
\psdots[dotscale=0.5](0,30)(0,45)(0,60)(0,75)(0,90)(0,105)
}

\psline[linewidth=1.2]{->}(60,45)(72.5,55)
\psline[linewidth=1.2]{->}(60,45)(20,90)
\rput(87,51){$\lambda_\text{max}^{-1} \mathbf{v}_\text{max} $}
\rput(28,98){$\lambda_\text{min}^{-1}{\mathbf{v}_\text{min}}$}
\psline[linestyle=dashed,linewidth=0.6]{->}(60,45)(105,75)
\psline[linestyle=dashed,linewidth=0.6]{->}(60,45)(75,60)
\rput(112,75){$\mathbf{a}_2$}
\rput(75,65){$\mathbf{a}_1$}
}

\end{pspicture}
\end{center}\caption{The zero-forcing linear receiver (top) is equivalent to an integer-forcing linear receiver with integer vectors fixed to $\mathbf{a}_1 = [1~~0]^T$ and $\mathbf{a}_2 = [0~~1]^T$. By optimizing over all linearly independent integer vectors, the integer-forcing linear receiver can attain significantly higher rates. These vectors should be chosen in the direction of $\mathbf{v}_{\text{max}}$ to avoid noise amplification by $\lambda_{\text{min}}^{-1}$.}
\label{fig:equations}
\end{figure}
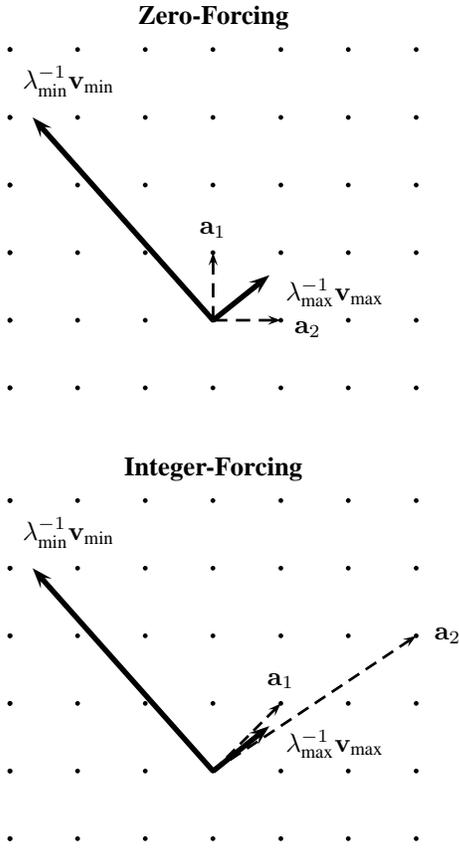

The rate expression from Theorem~\ref{thm:alterrate} suggests that we should select the integer vectors $\mathbf{a}_1, \ldots,\mathbf{a}_{\mrt}$ to be short and in the direction of the maximum eigenvector of $\mathbf{H}^T\mathbf{H}$. To make this concrete, consider a $2 \times 2$ real MIMO channel for which $\mathbf{H}^T\mathbf{H}$ has eigenvalues $\lambda_{\text{max}} > \lambda_{\text{min}} > 0$, with corresponding eigenvectors $\mathbf{v}_\text{min}$ and $\mathbf{v}_\text{max}$, as illustrated in Figure \ref{fig:equations}. Here, decoders 1 and 2 recover linear combinations according to integer vectors $\mathbf{a}_1 = [a_{1,1}~ a_{1,2}]^T$ and $\mathbf{a}_2 = [a_{2,1}~a_{2,2}]^T$, respectively. From Theorem~\ref{thm:alterrate}, the following rate is achievable:
\begin{align*}
&R_{\iftext}(\mathbf{H}) = \min_{m = 1,2} \log^+ \left(\frac{\snr}{\sigma^2_{\text{opt},m}}\right) \\
&\sigma^2_{\text{opt},m} = \frac{1}{1 + \lambda^2_{\text{min}}\snr} |\mathbf{v}_\text{min}^T\mathbf{a}_m|^2 + \frac{1}{1+\lambda^2_\text{max}\snr}|\mathbf{v}_\text{max}^T\mathbf{a}_m|^2 \ .
\end{align*} As argued above, the linear MMSE receiver is equivalent to setting $\mathbf{a}_1 = [1~~0]^T$ and $\mathbf{a}_2 = [0~~1]^T$. As a result, the noise variance in at least one of the streams will be heavily amplified by $(1 + \lambda^2_{\text{min}}  \snr)^{-1}$ and the rate will be limited by the minimum singular value of the channel matrix. With integer-forcing, we are free to choose \textit{any linearly independent} $\mathbf{a}_1$ and $\mathbf{a}_2$ since we only require that $\mathbf{A}$ be invertible. By choosing $\mathbf{a}_1$ and $\mathbf{a}_2$ in the direction $\mathbf{v}_{\text{\tiny{MAX}}}$, we can significantly reduce noise amplification in the case of near-singular channel matrices.

As the number of antennas or the SNR increases, an exhaustive search for the optimal integer matrix $\mathbf{A}$ rapidly becomes infeasible. The optimization problem from~\eqref{eq:alternaterate} can be written as
\begin{align}
\argmin_{\substack{\mathbf{A} \in \mathbb{Z}^{\mrt \times \mrt} \\ \mathrm{rank}(\mathbf{A}) = \mrt}}\max_m \big\| \mathbf{D}^{-1/2} \mathbf{V}^T \mathbf{a}_m \big\|^2 \ .\nonumber
\end{align} Thus, the search for the optimal $\mathbf{A}$ is equivalent to the search for the shortest set of linearly independent vectors in the lattice generated by $\mathbf{D}^{-1/2} \mathbf{V}^T$.\footnote{For exact integer-forcing, we can instead search over the lattice generated by $(\mathbf{H}^T)^\dag$.} This is known in the computer science literature as the Shortest Independent Vector Problem (SIVP) \cite{mg02}. Although SIVP is suspected to be NP-hard \cite{regev09}, several polynomial-time approximation algorithms have been developed, such as the LLL algorithm \cite{lll82}. Very recent work has examined the connection between SIVP and compute-and-forward \cite[Section VIII]{fsk11} as well as integer-forcing \cite{hc12IT,she12,shv13} and we refer the interested reader to these papers for more details and specialized algorithms.

\subsection{Implementation Issues}

One appealing feature of the integer-forcing architecture is that it can operate using similar codes and constellations as those used in conventional architectures. As discussed earlier, the main requirement placed on the coding scheme is that any integer combination of codewords is itself a codeword, i.e., the codewords are drawn from a lattice. While it may initially seem that lattice encoding and decoding is quite complex, one can select lattices that enable very efficient implementations.

As a starting point, we can construct a simple nested lattice pair by coupling $q$-ary pulse amplitude modulation (PAM) with any $q$-ary linear code and employing one-dimensional modulo operations. This coding scheme can operate quite close to the Gaussian capacity at high SNR, with a loss of no more than $0.255$ bits per dimension \cite{tenBrink05} owing to the lack of shaping. Furthermore, through the use of modern coding techniques (such as LDPC codes) and iterative decoding algorithms, these lattices can be efficiently encoded and decoded. Recent work has examined the performance of this approach for both compute-and-forward as well as integer-forcing \cite{ozegn11,oe12,hc11,hc12IT,ylhyc12,hn13,tnp13}.

When operating at low rates, the loss incurred by one-dimensional modulo operations (see, e.g., \cite{tenBrink05}) becomes significant and we may wish to include some form of shaping. It was demonstrated in \cite{ez04} that nested lattice codes are able to achieve the Gaussian capacity. In this framework, the codebook is comprised of the elements of the fine lattice (i.e., the inner code) that fall within the fundamental Voronoi region of the coarse lattice (i.e., the outer code). Most of the shaping gain can be attained by using a simple coarse lattice, such as one generated via a convolutional code with a small number of states. For instance, a $4$-state rate-$1/2$ binary convolutional code suffices to reduce the shaping penalty to $0.094$ bits per real dimension.  An implementation of a nested lattice dirty-paper coding scheme was proposed in \cite{et05}, which could also be used as a foundation for integer-forcing.  Recent work by Feng, Silva, and Kschischang \cite{fsk11} has taken an algebraic approach to compute-and-forward, which provides an excellent framework for selecting good codes and constellations.

\begin{remark}In the case of uncoded transmission, the integer-forcing linear receiver reduces to lattice reduction without the constraint that the integer matrix $\mathbf{A}$ is unimodular.\footnote{Recall that a matrix is \textit{unimodular} if has integer entries and its inverse has integer entries.} In Appendix \ref{ap:latticereduction}, we provide a detailed discussion of this connection.
\end{remark}

\section{Fixed Channel Matrices}
\label{sec:fixedchannels}
In this section, we explore the behavior of integer-forcing through a series of three examples that have been chosen to highlight the differences between zero-forcing, integer-forcing, and joint ML decoding. Later, in Section \ref{subsec:outage}, we will compare the average performance under Rayleigh fading. In Example 1, we show that, in order to attain the highest rates, the integer matrix must change as the SNR increases. In Example 2, we demonstrate that, integer-forcing can sometimes achieve arbitrarily higher rates than zero-forcing, i.e., integer-forcing does not merely yield a power gain. In Example 3, we demonstrate that the gap between the integer-forcing rate and joint ML rate can be arbitrarily large, i.e., integer-forcing does not achieve the capacity in general.

\begin{remark}
Recent work by Ordentlich and Erez has demonstrated that, when combined with an appropriate space-time code, the integer-forcing receiver can attain the capacity of any MIMO channel up to a constant gap \cite{oe13}.
\end{remark}

\subsection{Example 1: SNR Dependence of the Integer Coefficients}
\label{ex:1}
In this example, we show that the choice of the optimal integer matrix $\mathbf{A}$ depends on the $\snr$ (even if the channel matrix is fixed). Consider the $2 \times 2$ real-valued MIMO channel with channel matrix
\newline
\begin{align}
 \mathbf{H} = \begin{bmatrix}
  0.7 & 1.3 \\
  0.8 & 1.5 \\
 \end{bmatrix}. \label{e:fixedmatrixEx1}
\end{align}\begin{figure}[ht]
\centering
\includegraphics[width=3.4in]{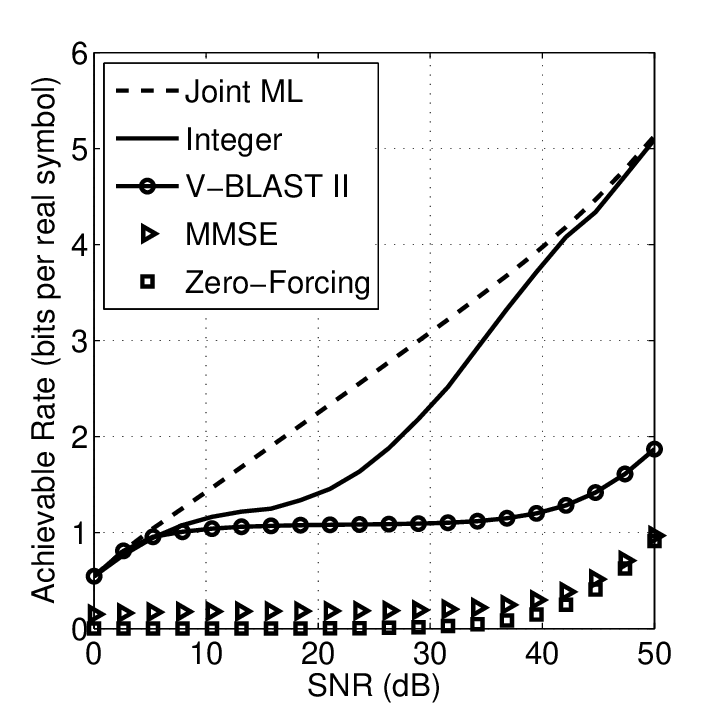}
\caption{Achievable rates for the $2 \times 2$ real-valued MIMO channel from~\eqref{e:fixedmatrixEx1}.}
\label{f:fixedchannel}
\end{figure}
In Figure \ref{f:fixedchannel}, we have plotted the performance of the joint ML~\eqref{eq:rjoint}, integer-forcing~\eqref{eq:alternaterate}, V-BLAST II~\eqref{eq:rmmsesic2}, linear MMSE~\eqref{e:rmmse}, and zero-forcing~\eqref{e:rzeroforcing} receivers. (Recall that we assume equal-rate data streams on both transmit antennas, as in Definition \ref{def:rate}.) Conventional linear receivers perform poorly since the columns of the channel matrix are far from orthogonal. The integer-forcing receiver attempts to exploit the interference by decoding two linearly independent integer vectors in the direction of the maximum eigenvector $\mathbf{v}_\text{max} \approx  [0.47~~0.88]^T$ of $\mathbf{H}^T\mathbf{H}$. Recall that the best integer vectors correspond to finding the shortest basis for the lattice generated by $\mathbf{D}^{-1/2} \mathbf{V}^T$, which is itself a function of the channel matrix $\mathbf{H}$ and the $\snr$. For example, at $\snr = 30$dB the optimal integer vectors are $\mathbf{a}_1 = [1 ~~2]^T$ and $\mathbf{a}_2 = [6~~11]^T,$ while for $\snr = 40$dB they are $\mathbf{a}_1 = [1 ~~7]^T$ and $\mathbf{a}_2 = [2~~13]^T.$ Thus, as the $\snr$ increases, we may have to adjust our choice of integer vectors. 

\subsection{Example 2: The Gap Between Integer-Forcing and Zero-Forcing}
We now show that the integer-forcing rate can be arbitrarily larger than the zero-forcing rate.
\label{ex:2}
Consider the $2 \times 2$ real-valued MIMO channel with channel matrix
 \begin{align}
 \mathbf{H} = \begin{bmatrix}
  1 & 1+ \sqrt{\epsilon} \\
  0 &  \epsilon \\
 \end{bmatrix} \nonumber 
 \end{align}
where we assume $0 < \epsilon \ll 1$, $\frac{1}{\sqrt{\epsilon}}$ is an integer and $\snr \gg 1$. We first note that the channel inverse is
 \begin{align}
 \mathbf{H}^{-1} =  \frac{1}{\epsilon}  \begin{bmatrix}
  \epsilon & -(1 + \sqrt{\epsilon}) \\
  0 &  1 \\
 \end{bmatrix} \ .  \nonumber
 \end{align}
From~\eqref{e:rzeroforcing}, the zero-forcing rate is
\begin{align*}
R_\text{ZF}(\mathbf{H}) &= 2 \min \Bigg( \frac{1}{2} \log \bigg( 1 + \frac{\epsilon^2 \snr}{\epsilon^2 + \epsilon + 2 \sqrt{\epsilon}  + 1}\bigg)\ , \\ &\qquad\qquad~~~~ \frac{1}{2} \log( 1 + \epsilon^2 \snr) \Bigg) \nonumber\\
&\leq \log( 1 + \epsilon^2 \snr) \ .  \nonumber
\end{align*}
Since we have assumed $\frac{1}{\sqrt{\epsilon}}$ is an integer, we can set the integer vectors to be \begin{align}
\mathbf{a}_1^T = [1~~1] \qquad \qquad
\mathbf{a}_2^T = \left[\frac{1}{\sqrt{\epsilon}}~~\frac{1}{\sqrt{\epsilon}} + 1\right] \ . \nonumber
\end{align}
From~\eqref{eq:ratesub}, the exact integer-forcing rate for $\mathbf{A}  = [\mathbf{a}_1~\mathbf{a}_2]^T$ is \begin{align*}
R_{\iftext,\text{exact}}(\mathbf{H}) &= 2 \min_{m = 1,2} \frac{1}{2} \log \left( \frac{\snr}{\big\|\mathbf{H}{^{-T}}\mathbf{a}_m \big\|^2 } \right). \\
&=2\min \left(\frac{1}{2} \log\left(\frac{\snr}{1 + \frac{1}{\epsilon}} \right),\ \frac{1}{2} \log \left(\frac{\snr}{\frac{1}{\epsilon}} \right) \right)\\
&= \log\left(\frac{\snr}{1 + \frac{1}{\epsilon}} \right)\\
&\geq \log\left(\frac{\snr}{\frac{2}{\epsilon}} \right)\\
&= \log\left(\frac{\epsilon \snr}{2} \right)
\end{align*}
where the inequality follows since $0 < \epsilon \ll 1$. From~\eqref{eq:rjoint}, the rate of joint ML decoding is upper bounded by
\begin{align*}
&R_\text{ML}(\mathbf{H}) \\&\leq \frac{1}{2} \log \det \left(\mathbf{I} + \snr \  \mathbf{H}\mathbf{H}^T\right)\\
&=\frac{1}{2}\log\left((1 + \snr)(1 + \epsilon^2 \snr) + \left(1 + \sqrt{\epsilon}\right)^2\snr \right).
\end{align*}

Finally, let us compare the three rates in the setting where $\snr \rightarrow \infty,$
and where the parameter $\epsilon$ in our channel model tends to zero according\footnote{Recall that $f(\snr) \sim g(\snr)$ implies that $\lim_{\snr \rightarrow \infty} \frac{f(\snr)}{g(\snr)} = 1$. } to
$\epsilon \sim \frac{1}{\sqrt{\snr}}.$ In that special case, we can observe that
\begin{align*}
R_\text{ZF} \sim 1 \qquad R_\text{IF} \sim \frac{1}{2} \log (\snr) \qquad R_\text{ML} \sim \frac{1}{2} \log (\snr) \ . 
\end{align*}
Hence, the gap between zero-forcing and integer-forcing becomes unbounded for this sequence of channels as $\snr \rightarrow \infty$. Furthermore, integer-forcing achieves the same rate scaling as joint ML decoding.

\subsection{Example 3: The Gap between Integer-Forcing and Joint ML Decoding}
\label{ex:4}
Finally, we illustrate the point that integer-forcing can sometimes be arbitrarily worse than joint ML decoding. To see this, we consider a $2 \times 2$ real-valued MIMO channel with channel matrix
 \begin{align*}
 \mathbf{H} = \begin{bmatrix}
  1 & 1 \\
  0 &  \epsilon \\
 \end{bmatrix}   
 \end{align*}
where $0 < \epsilon < 1$. From~\eqref{eq:rjoint}, the rate attainable via joint ML decoding is
\begin{align*}
R_\text{ML}(\mathbf{H}) &= \min\bigg(\log(1 + \snr), \  \log\big(1 + \snr (1 + \epsilon^2)\big)  , \\
&\qquad \qquad \frac{1}{2} \log\big((1 + \snr (2 + \epsilon^2) + \snr^2 \epsilon^2 \big) \bigg) \\
&\geq \log(\epsilon \  \snr  ) \ .
\end{align*}
We note that the inverse of the channel matrix is given by
 \begin{align*}
 \mathbf{H}^{-1} = \begin{bmatrix}
  1 & -\frac{1}{\epsilon} \\
  0 &  \frac{1}{\epsilon} \\
 \end{bmatrix} \ . 
 \end{align*}
From Corollary \ref{cor:ratesub}, the exact integer-forcing rate is 
\begin{align*}
&R_{\iftext,\text{exact}}(\mathbf{H}) \\
&= 2 \max_{\substack{\mathbf{A} \in \mathbb{Z}^{2 \times 2} \\ \mathrm{rank}(\mathbf{A}) = 2}}\min_{m = 1,2} \frac{1}{2} \log \left( \frac{\snr}{\big\|\mathbf{H}{^{-T}}\mathbf{a}_m \big\|^2} \right)\\
&=  2 \max_{\substack{\mathbf{A} \in \mathbb{Z}^{2 \times 2} \\ \mathrm{rank}(\mathbf{A}) = 2}}\min_{m = 1,2} \frac{1}{2} \log \left( \frac{\snr}{a_{m,1}^2 + (a_{m,2} - a_{m,1})^2\frac{1}{\epsilon^2}} \right)\\
&\leq \max_{\substack{a_{m,\ell} \in \mathbb{Z} \\ a_{m,2} \neq a_{m,1} }} \log \left( \frac{\snr}{a_{m,1}^2 + (a_{m,2} - a_{m,1})^2\frac{1}{\epsilon^2}} \right)\\
&\leq \log \left(\epsilon^2 \snr \right) \ .
\end{align*}
Let $\epsilon \sim \frac{1}{\sqrt{\snr}}$ and consider the regime $\snr \rightarrow \infty$. For this sequence of channel matrices, the gap between (optimal) joint ML decoding and integer-forcing can be arbitrarily large. However, as we will see in Section \ref{sec:outage}, the average behavior of integer-forcing is quite close to that of joint ML decoding under Rayleigh fading.

 \section{Performance for Slow Fading Channels}
\label{sec:outage}
\subsection{Model and Definitions}
We now demonstrate that integer-forcing nearly matches the performance of the joint ML decoder under a slow fading channel model. As argued in Section~\ref{s:specialcase}, the integer-forcing receiver can match the performance of conventional linear receivers as a special case. However, these architectures are often coupled with some form of SIC. We will show that integer-forcing can even outperform the following standard SIC architectures:
\begin{itemize}
\item V-BLAST I:  The receiver decodes and cancels the data streams in a predetermined order, irrespective of the channel realization. Each data stream has the same rate. See \eqref{eq:rmmsesic1} for the rate expression.
\item V-BLAST II:  The receiver selects the decoding order separately for each channel realization in such a way as to maximize the effective SNR for the data stream that sees the worst channel.  Each data stream has the same rate. See \eqref{eq:rmmsesic2} for the rate expression.
\item V-BLAST III: The receiver decodes and cancels the data streams in a predetermined order. The rate of each data stream is selected using the channel statistics to maximize the sum rate. The rate expression is given in Section \ref{subsec:rateallocation}.
\end{itemize}

In Sections \ref{subsec:outage} and \ref{subsec:dmt}, we compare these schemes through simulations as well as their diversity-multiplexing tradeoffs. For completeness, we also compare integer-forcing to an SIC architecture that allows for both variable decoding order and unequal rate allocation in Appendix \ref{ap:vblast4}.

We adopt the standard quasi-static Rayleigh fading model where each element of the complex channel matrix is i.i.d.~according to a circularly symmetric complex Gaussian distribution of unit variance. The transmitter is only aware of the channel statistics while the receiver knows the exact channel realization. As a result, we will have to cope with some outage probability $p_{\text{outage}}$.

\begin{definition}
\label{def:outage}
Consider a scheme that encodes each data stream at the same rate and can support sum rate $R_{\text{scheme}}(\mathbf{H})$ over channel matrix $\mathbf{H}$. For a target sum rate $R$, the \textit{outage probability} is defined as
\begin{align*}
p_{\text{outage}}(R) = \mathbb{P}\big(R_{\text{scheme}}(\mathbf{H}) < R\big)\ .
\end{align*}
For a fixed probability $\rho \in (0, 1]$, we define the \textit{outage rate} to be
\begin{align*}
R_{\text{outage}}(\rho) = \sup\big\{ R: p_{\text{outage}}(R) \leq \rho \big\} \ .
\end{align*}
\end{definition}

\subsection{Rate Allocation}
\label{subsec:rateallocation}
Until now, we have assumed that each data stream is encoded at the same rate. This is optimal for linear receivers under isotropic fading. However, rate allocation can be beneficial in an outage scenario when combined with SIC. To compare the performance of integer-forcing to SIC with rate allocation, we now introduce V-BLAST III. This receiver architecture performs SIC with a fixed decoding order and allows for rate allocation across the different data streams using knowledge of the channel statistics at the transmitter. Without loss of generality for Rayleigh fading, if we fix a decoding order, we may take it to be $\pi = (1, 2, \ldots, \mrt)$. The rate at which the $m^{\text{th}}$ data stream follows from~\eqref{eq:sicstream} and~\eqref{e:mmsesicprojection}, 
\begin{align}
R_{\text{V-BLAST III},m}(\mathbf{H}) &= R_{\text{SIC},m}(\mathbf{H},\mathbf{b}_{\text{MMSE-SIC},m}) \ . \label{e:vblastiii}
\end{align}

On average, data streams are decoded later will achieve higher rates as they face less interference. Thus, V-BLAST III allocates lower rates to earlier streams and higher rates to later streams. We now generalize our definition of outage probability and rate to include rate allocation.
\begin{definition} \label{d:outagevblastiii}
Consider a scheme that achieves rate $R_{\text{scheme},m}(\mathbf{H})$ for the $m^{\text{th}}$ data stream. For a target sum rate $R$, the \textit{outage probability} is defined as
\begin{align*}
p_{\text{outage}}(R) = \min_{\substack{R_1, \ldots,  R_{\mrt} \\ \sum_{m = 1}^{\mrt} R_m \leq R}} \mathbb{P} \left(\bigcup_{m = 1}^{\mrt} \left\{R_{\text{scheme},m}(\mathbf{H}) < R_m\right\} \right) \ .
\end{align*}
For a fixed probability $\rho \in (0, 1]$, we define the \textit{outage rate} to be
\begin{align*}
R_{\text{outage}}(\rho) = \sup \left\{R : p_{\text{outage}}(R) \leq \rho \right\} \ .
\end{align*}
\end{definition}

\subsection{Outage Behavior}
\label{subsec:outage}
We now compare the outage rates and probabilities for the receiver architectures discussed above. First, note that the zero-forcing receiver performs strictly worse than the linear MMSE receiver and the V-BLAST I receiver performs strictly worse than the V-BLAST II receiver. We have chosen to omit zero-forcing and V-BLAST I from the plots to avoid overcrowding. All simulations are evaluated with respect to complex-valued channels with i.i.d.~Rayleigh fading (whose realization is only known at the receiver). The plots compare the performance of the joint ML~\eqref{eq:rjoint}, integer-forcing~\eqref{eq:alternaterate}, V-BLAST III~\eqref{e:vblastiii}, V-BLAST II~\eqref{eq:rmmsesic2}, and linear MMSE~\eqref{e:rmmse} receivers. 

\begin{figure}[h!]
\centering
\includegraphics[width=3.6in]{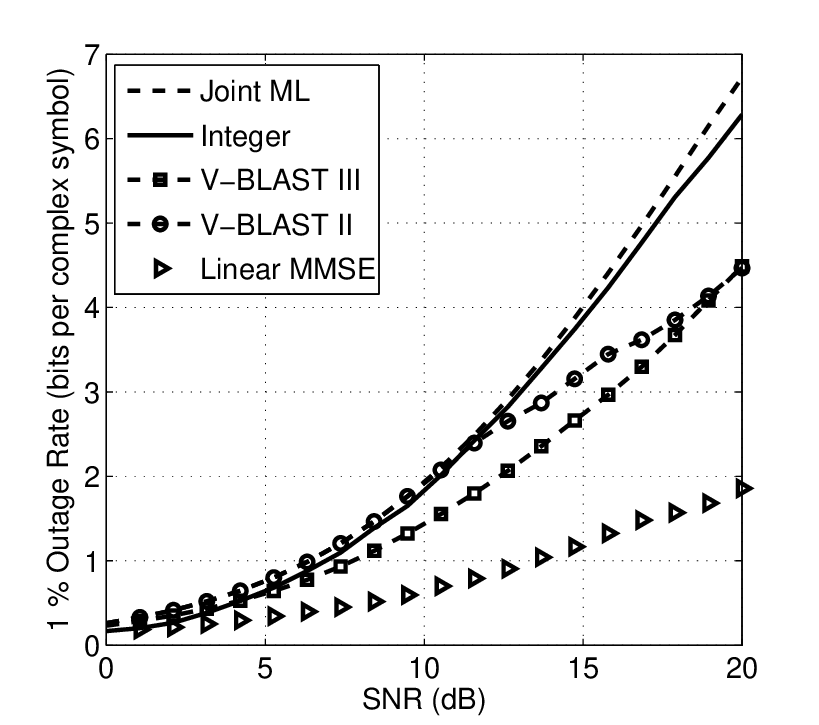}
\caption{1 percent outage rates for the $2 \times 2$ complex-valued MIMO channel under i.i.d.~Rayleigh fading.}
\label{f:out2p}
\end{figure}

\begin{figure}[h!]
\centering
\includegraphics[width=3.6in]{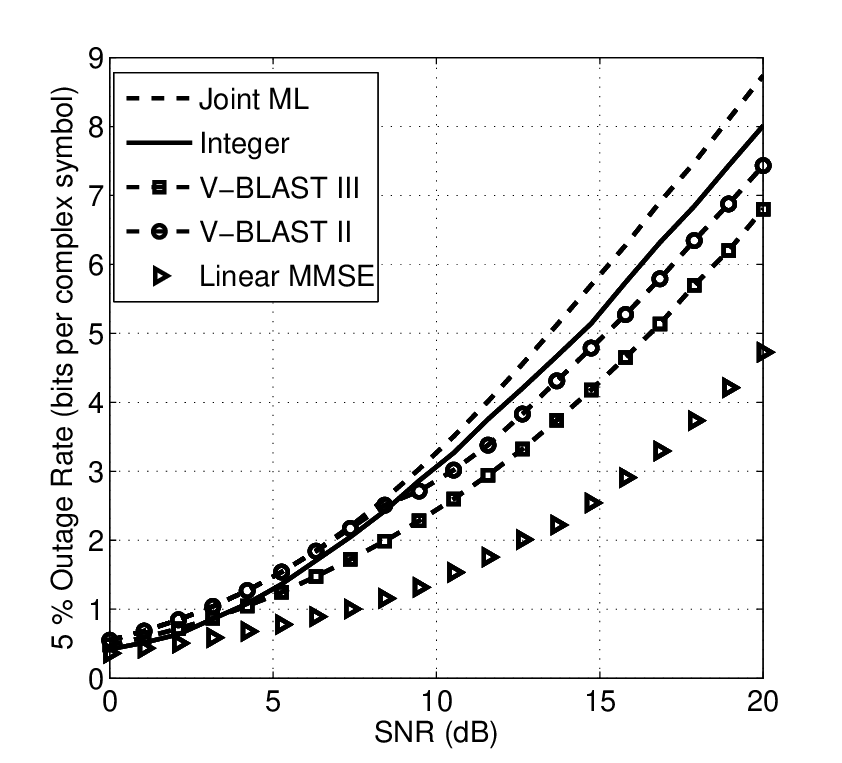}
\caption{5 percent outage rates for the $2 \times 2$ complex-valued MIMO channel under i.i.d.~Rayleigh fading.}
\label{f:out5p}
\end{figure}

\begin{figure}[h!]
\centering
\includegraphics[width=3.6in]{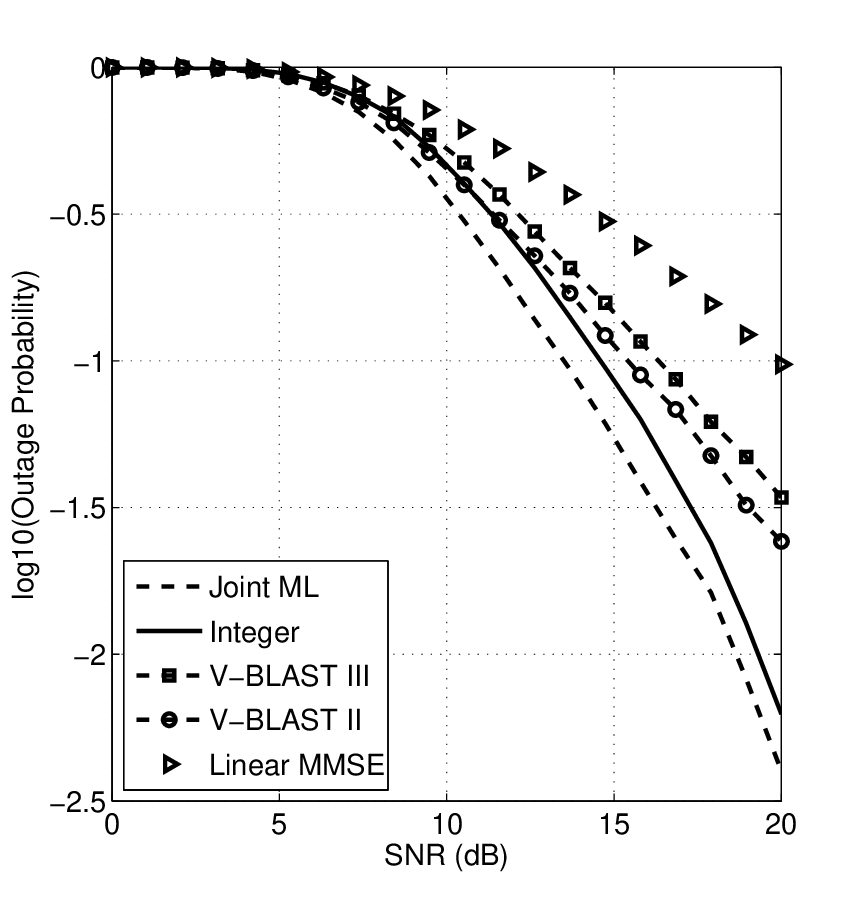}
\caption{Outage probability for a target sum rate of $R = 6$ for the $2 \times 2$ complex-valued MIMO channel under i.i.d.~Rayleigh fading.} 
\label{f:poutRo3}
\end{figure}

In Figures~\ref{f:out2p} and~\ref{f:out5p}, we have plotted the $1$ and $5$ percent outage rates, respectively. In both cases, the integer-forcing receiver nearly matches the rate of the joint ML receiver while the linear MMSE receiver achieves significantly lower performance. The SIC architectures with either an optimal decoding order, V-BLAST II, or an optimized rate allocation, V-BLAST III, improve upon the performance of the linear MMSE receiver considerably but still achieve lower rates than the integer-forcing receiver from medium SNR onwards. Our simulations suggest that the outage rate of the integer-forcing receiver remains within a small gap from the outage rate of the joint ML receiver. However, we recall from the example given in Section~\ref{ex:4} that it is not true that the integer-forcing receiver is uniformly near-optimal for all fading realizations. 

In Figure \ref{f:poutRo3}, we have plotted the outage probability for a target sum rate of $R = 6$. Note that integer-forcing achieves the same slope as joint ML decoding. In the next subsection, we investigate this behavior by deriving the DMT of the integer-forcing receiver.

\subsection{Diversity-Multiplexing Tradeoff}
\label{subsec:dmt}
The diversity-multiplexing tradeoff (DMT) provides a rough characterization of the performance of a MIMO transmission scheme at high SNR \cite{zt03}.
\begin{definition} Consider a family of coding schemes, indexed by $\snr$, that achieves sum rate $R(\snr)$ with probability of error $p_{\text{error}}(\snr)$ at a given $\snr$ value. This family of coding schemes is said to achieve \textit{spatial multiplexing gain} $r$ and \textit{diversity gain} $d$ if 
\begin{align*}
&\lim_{\snr \rightarrow \infty} \frac{ R(\snr)}{\log \snr} = r \\
&\lim_{\snr \rightarrow \infty}  \frac{ \log p_{\text{error}}(\snr)}{\log \snr} = -d.
\end{align*}
\end{definition}

Since the diversity and multiplexing gains are defined with respect to $\log{\snr}$, it is more natural to state the tradeoffs in terms of the number of transmit antennas $\mcr$ and receive antennas $\mct$ in the complex-valued representation.

In the case where each transmit antenna encodes an independent data stream\footnote{If joint encoding across the antennas is permitted, then a better DMT is achievable. See \cite{zt03} for more details.}, the optimal DMT is
\begin{align*}
d_{\text{ML}}(r) = \mcr \left(1 - \frac{r}{\mct} \right)
\end{align*} where $r \in [0,\mct]$ and can be achieved by joint ML decoding \cite{zt03}. If $\mcr \geq \mct$, the zero-forcing and linear MMSE receiver attain the same DMT~\cite{kcm09}, 
\begin{align*}
d_{\text{ZF}}(r) = d_{\text{MMSE}}(r)= (\mcr-\mct + 1)\left(1 - \frac{r}{\mct} \right) \ .
\end{align*}
Surprisingly, allowing the receiver to perform SIC does not change the DMT, even if the order is optimized based on the channel realization~\cite{jvl11},
\begin{align*}
d_{\text{V-BLAST I}}(r) = d_{\text{V-BLAST II}}(r) = (\mcr-\mct + 1)\left(1 - \frac{r}{\mct} \right) .
\end{align*} However, allowing for rate allocation at the transmitter can improve the DMT. For the special case of $\mcr = \mct$ (and a fixed decoding order), the DMT is~\cite{zt03}
\begin{align*}
d_{\text{V-BLAST III}}(r)&= \text{piecewise linear curve connecting the points} \\
&~~~(r_\ell, \mct - \ell) \text{~for~}1 \leq \ell \leq \mct  \text{~where}\\
 &~~~r_\ell = \begin{cases}
  0 & \ell = 0\ , \\
  \displaystyle\sum_{i = 0}^{\ell-1} \frac{\ell-i}{\mct-i} & 1 \leq \ell \leq \mct \  .
  \end{cases}
\end{align*}
The zero-forcing receiver chooses the matrix $\mathbf{B}$ to cancel the interference from the other data streams. As a result, the noise is heavily amplified when the channel matrix is near singular and the performance is limited by the minimum singular value of the channel matrix. In the integer-forcing linear receiver, the effective channel matrix $\mathbf{A}$ is not limited to the identity matrix but can be any full-rank integer matrix. This additional freedom is sufficient to recover the same DMT as the joint ML decoder.

The theorem below establishes that the integer-forcing receiver attains the optimal DMT when the number of receive antennas is greater than or equal to the number of transmit antennas. In other words, SISO decoding with equal rate allocation can attain the optimal DMT.

\begin{theorem}
\label{thm:dmt}
For a MIMO channel with $\mct$ transmit, $\mcr \geq \mct$ receive antennas, and i.i.d.~Rayleigh fading, the achievable diversity-multiplexing tradeoff for the integer-forcing receiver is \begin{equation}
d_{\text{IF}}(r) = \mcr \left(1 - \frac{r}{\mct} \right) \nonumber
\end{equation}
where $r \in [0,\mct]$.
\end{theorem}
The proof of Theorem \ref{thm:dmt} is given in Appendix \ref{ap:dmt}. It builds on a result due to Taherzadeh, Mobasher, and Khandani which showed that uncoded signaling coupled with lattice reduction can achieve the full diversity (with a multiplexing gain of zero) \cite{tmk07lll}. 

\begin{figure}[ht]
\begin{center}
\includegraphics[width=3.8in]{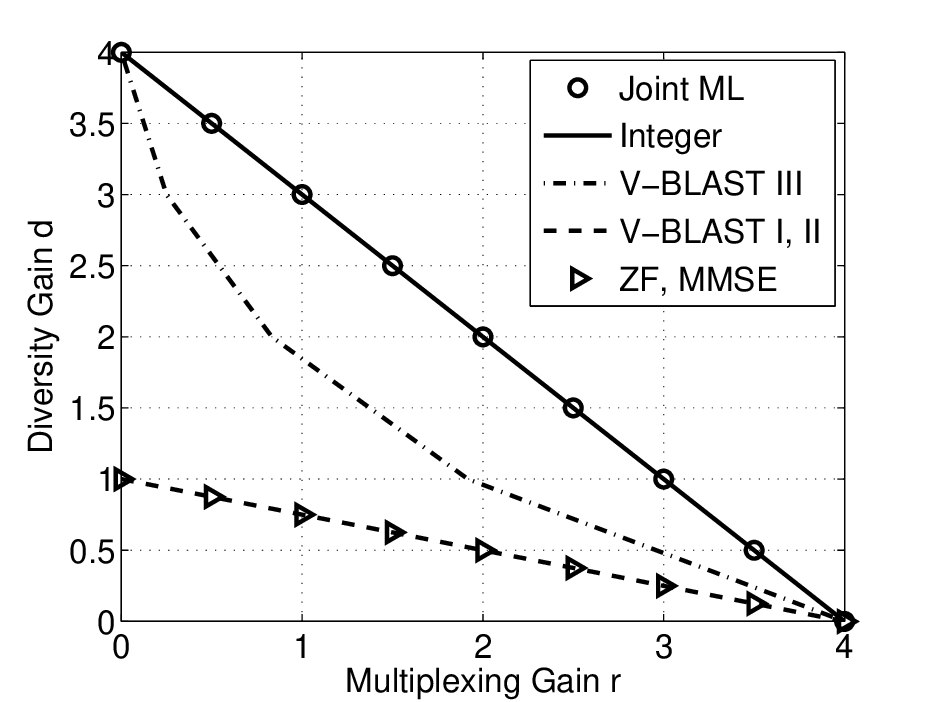}
\caption{Diversity-multiplexing tradeoffs for the complex-valued $4 \times 4$ MIMO channel (with independent data streams) under i.i.d.~Rayleigh fading.}
\label{fig:dmt}
\end{center}
\end{figure}

Figure \ref{fig:dmt} illustrates the DMT for a $4 \times 4$ MIMO channel under i.i.d.~Rayleigh fading. The integer-forcing receiver achieves the maximum diversity of $4$ while the zero-forcing, linear MMSE, V-BLAST I, and V-BLAST II receivers attain at most diversity of $1$. V-BLAST III achieves the optimal diversity at the point $r = 0$ but is suboptimal for all $r > 0$.

\subsection{Discussion}
As noted earlier, classical linear receivers are subject to a significant rate penalty when the channel matrix is ill-conditioned. Integer-forcing circumvents this issue by allowing the receiver to first decode integer combinations whose coefficients are matched with those of the channel. The outage plots in Section~\ref{subsec:outage} show that the integer-forcing receiver considerably outperforms the basic linear MMSE receiver. Moreover, integer-forcing can outperform more sophisticated SIC-based V-BLAST architectures, even if these are permitted to optimize their rate allocation while integer-forcing is not. We note that it is possible to develop integer-forcing schemes that permit unequal rate allocations~\cite{ng11IT} as well as a form of interference cancellation~\cite{znoeg10,oen13} but this is beyond the scope of the present paper.

Integer-forcing also attains the optimal DMT. Earlier work developed lattice-based schemes that attain the full DMT~\cite{ecd04,je10} but, to the best of our knowledge, ours is the first that decouples spatial decoding from temporal decoding. The caveat is that the DMT result presented in this paper is for the case when there is no spatial coding across transmit antennas, whereas the DMT results of~\cite{ecd04,je10} apply in general. Of course, one can include a space-time coding block after generating the coded data streams at the transmitter. Very recent work has examined this possibility and shown that integer-forcing continues to follow the performance of the joint ML decoder at finite SNR~\cite{de12}, attains the full DMT~\cite{oe13}, and can operate within a constant gap of the capacity of any MIMO channel~\cite{oe13}.

\section{Oblivious Interference Mitigation}
\label{sec:interference}
\subsection{MIMO Channel with Interference}
\label{sec:channelmodel}
So far, we have studied the performance of integer-forcing under the standard MIMO channel model and found that it achieves outage rates close those of joint ML decoding as well as the same DMT. In this section, we show that integer-forcing architectures are also successful at dealing with a different kind of channel disturbance, namely external interference. As a motivating example, consider neighboring cell-sites in a cellular deployment \cite{ghhss10}. In this setting, each receiver will see a noisy linear combination of the signals from its desired users as well as interfering users from neighboring cells.

We will assume that the interfering signal is low-dimensional (compared to the number of receive antennas), and will focus on the case where the variance of this interfering signal increases (at a certain rate) with the transmit power. We show that the integer-forcing architecture can be used to perform ``oblivious'' interference mitigation. By oblivious, we mean that the transmitter and receiver are unaware of the codebook of the interferer (if there is one). However, the receiver knows which subspace is occupied by the interference. This is quite reasonable in the context of our motivating cellular example, as the receiver can estimate the interferers' effective channels via their pilot symbols. By selecting integer vectors in a direction that depends both on the interference space and on the channel matrix, the integer-forcing receiver reduces the impact of interference beyond what is possible using traditional linear receivers. We will characterize the generalized degrees-of-freedom and show that it matches that of the joint ML decoder.

\begin{remark}
Oblivious receivers have been thoroughly studied in the context of cellular systems \cite{ssps09} and distributed MIMO \cite{sss09}.
\end{remark}

We now extend our channel model from~\eqref{eq:observation} to include external interference. For ease of notation and tractability, we will assume an equal number of transmit and receive antennas, $\mrr = \mrt = \mri$. The real-valued representation of the generalized model has channel output
\begin{align}
\mathbf{Y} = \mathbf{H}\mathbf{X} + \mathbf{J}\mathbf{V} + \mathbf{Z} \nonumber
\end{align} where $\mathbf{H} \in \mathbb{R}^{\mri \times \mri}$ is the channel matrix, $\mathbf{X} \in \mathbb{R}^{\mri \times n}$ is the channel input, $\mathbf{J} \in \mathbb{R}^{\mri \times K}$ is the $K$-dimensional interference subspace, $\mathbf{V} \in \mathbb{R}^{K \times n} $ is i.i.d.~Gaussian interference with mean zero and variance $\inr$, and and $\mathbf{Z} \in \mathbb{R}^{\mri \times n}$ is i.i.d.~Gaussian noise with mean zero and variance one. We assume that $\mathbf{H}$ and $\mathbf{J}$ are fixed during the whole transmission block and known only to the receiver.

The definition for messages, rates, encoders, and decoders follow along similar lines as those for the standard MIMO channel (see Definitions \ref{def:messages}, \ref{def:encoders}, \ref{def:decoder}, and \ref{def:rate} in Section \ref{sec:problem}).

\subsection{Conventional Receiver Architectures}
\label{sec:tradLR}

As before, the best performance is given by joint ML decoding. Assuming the use of i.i.d.~Gaussian codebooks at the transmitters, the following rate is achievable,
\begin{align}\label{eq:mlint}
&R_{\text{ML}}(\mathbf{H},\mathbf{J}) =\\
& \mri \cdot \min_{\mathcal{S} \subseteq \{1,\ldots,\mri\}}\frac{1}{2|\mathcal{S}|} \log\left(\frac{ \big| \mathbf{I} + \inr \ \mathbf{J}\mathbf{J}^T + \snr\  \mathbf{H}_\mathcal{S}\mathbf{H}^T_{\mathcal{S}}  \big| }{\big| \mathbf{I} + \inr \ \mathbf{J}\mathbf{J}^T \big|} \right) \nonumber
\end{align} where $\mathbf{H}_{\mathcal{S}}$ denotes the submatrix of $\mathbf{H}$ formed by taking the columns with indices in the subset $\mathcal{S} \subseteq \{1,2,\ldots,\mri\}$. 

As in the case without interference, conventional linear receivers process the channel output $\mathbf{Y}$ by multiplying it by a matrix $\mathbf{B} \in \mathbb{R}^{\mri \times \mri}$ to arrive at the effective output
\begin{align*}
\mathbf{\tilde{Y}} = \mathbf{B}\mathbf{Y}
\end{align*}
and recover the message $\mathbf{w}_m$ using only the $m^{\text{th}}$ row of the matrix $\mathbf{\tilde{Y}}.$
By analogy to \eqref{eq:rlinear},
the achievable sum rate (using i.i.d.~Gaussian codebooks) can be expressed as
 \begin{align}
\label{eq:linearint}
&R_{\text{linear}}(\mathbf{H}, \mathbf{J}, \mathbf{B}) =  \min_{m=1,\ldots,\mri} R_{\text{linear},m}(\mathbf{H}, \mathbf{J}, \mathbf{B}) \\
&R_{\text{linear},m}(\mathbf{H}, \mathbf{J}, \mathbf{B}) = \nonumber \\ & \frac{1}{2}\log \Bigg(1+\frac{\snr \big(\mathbf{b}^T_m \mathbf{h}_m \big)^2}{ \| \mathbf{b}_m \|^2 + \inr \| \mathbf{J}^T \mathbf{b}_m \|^2 +  \snr \sum\limits_{i \neq m} \big( \mathbf{b}_m^T \mathbf{h}_i \big)^2} \Bigg) \ . \nonumber
\end{align}

We now consider several choices for the matrix  $\mathbf{B}.$ Assuming $\mathbf{H}$ is full rank, the zero-forcing receiver, $\mathbf{B}_{\text{ZF}} = \mathbf{H}^{\dag}$, removes the interference between data streams but does not cancel the external interference (except in the very special case where the subspace spanned by $\mathbf{J}$ is orthogonal to the subspace spanned by $\mathbf{H}^{\dag}$). Alternatively, if we choose $\mathbf{B}_{\text{null}} = \mathbf{J}^{\perp},$ where $\mathbf{J}^{\perp}$ is a matrix whose rowspace is orthogonal to the columnspace of $\mathbf{J}$, then the external interference is nulled. This scheme works well in high $\inr$ regimes but is ineffective in high $\snr$ regimes since the interference between data streams is mostly unresolved. The linear MMSE receiver is optimal and sets $\mathbf{B}_{\text{MMSE}} = \snr\  \mathbf{H}\left( \mathbf{I} +  \inr\  \mathbf{J}\mathbf{J}^T + \snr\  \mathbf{H}\mathbf{H}^T    \right)^{-1}$. In general, it is not possible to eliminate the interference between $\mri$ data streams and the $K$-dimensional external interference using only $\mri$ receive antennas. One workaround is to reduce the number of transmitted streams to $\mri - K$ so that both forms of interference can be completely nulled.

As in Section~\ref{sec:sic}, we can enhance performance of a linear receiver via SIC. Optimizing over all decoding orders $\Pi$, we obtain the following achievable rate for V-BLAST II:
\begin{align}\label{eq:vblastint}
&R_{\text{V-BLAST II}}(\mathbf{H},\mathbf{J})= \\
&\mri \cdot \max_{\pi \in \Pi} \min_{m=1,\ldots,\mri}  R_{\text{SIC},\pi(m)}(\mathbf{H},\mathbf{J},\mathbf{b}_{\text{MMSE-SIC},m}) \nonumber \\
&R_{\text{SIC},\pi(m)}(\mathbf{H},\mathbf{J}, \mathbf{b}_m)= \nonumber \\&\frac{1}{2} \log \Bigg( 1 + \frac{\snr
\big(\mathbf{b}^T_m \mathbf{h}_{\pi(m)} \big)^2}{\|\mathbf{b}_m\|^2 + \inr \| \mathbf{J}^T \mathbf{b}_m \|^2 + \snr
\sum\limits_{i > m}{\big( \mathbf{b}_m^T\mathbf{h}_{\pi(i)} \big)^2}}\Bigg)  \nonumber 
\end{align}
where the MMSE equalization vector for the $m^{\text{th}}$ stream is $$\mathbf{b}_{\text{MMSE-SIC},m}^T = \snr \  \mathbf{h}_{\pi(m)}^T \big(\mathbf{I} + \inr \ \mathbf{JJ}^T +  \snr \ \mathbf{H}_{\pi_m}\mathbf{H}_{\pi_m}^T\big)^{-1}\ .$$

\subsection{Integer-Forcing Linear Receiver}
\label{sec:intLR}
We now apply the integer-forcing linear receiver proposed in Section \ref{sec:architecture} to the problem of mitigating interference. The channel output matrix $\mathbf{Y}$ is first multiplied by $\mathbf{B}$ to obtain the effective channel output $\mathbf{\tilde{Y}} = \mathbf{BY}$ whose $m^{\text{th}}$ row is the signal fed into the $m^{\text{th}}$ decoder. Each such row can be expressed as
\begin{align*}
\mathbf{\tilde{y}}_m^T &= \mathbf{b}^T_m \mathbf{H} \mathbf{X} + \mathbf{b}^T_m \mathbf{J} \mathbf{V} + \mathbf{b}^T_m \mathbf{Z} \\
&= \mathbf{a}^T_m \mathbf{X} + \big( \mathbf{b}^T_m \mathbf{H} - \mathbf{a}^T_m\big) \mathbf{X} + \mathbf{b}^T_m \mathbf{J} \mathbf{V} + \mathbf{b}^T_m \mathbf{Z}
\end{align*} where $\mathbf{b}^T_m$ is the $m^{\text{th}}$ row of $\mathbf{B}$ and $\mathbf{a}_m^T$ is the $m^{\text{th}}$ row of $\mathbf{A} \in \mathbb{Z}^{\mri \times \mri}$, the matrix of desired integer coefficients. As discussed in Section \ref{decoding_equations}, $\mathbf{a}^T_m \mathbf{X}$ is an integer combination of lattice codewords and is therefore itself a codeword. Overall, each decoder recovers its integer combination and, if all decoders are successful, the integer combinations are solved to reveal the transmitted codewords. 

\begin{theorem}
\label{thm:rateinterference}
Under the integer-forcing architecture, the following sum rate is achievable:
\begin{align}
&R_{\iftext}(\mathbf{H},\mathbf{J}) = \mri \cdot \max_{\substack{ \mathbf{A} \in \mathbb{Z}^{\mri \times \mri} \\ \mathrm{rank}(\mathbf{A}) = \mri}}  \max_{\mathbf{B} \in \mathbb{R}^{\mri \times \mri}} R_{\text{comp}}(\mathbf{H},\mathbf{J},\mathbf{A},\mathbf{B}) \label{eq:ifint} \\
&R_{\text{comp}}(\mathbf{H},\mathbf{J},\mathbf{A},\mathbf{B}) = \min_{m = 1,\ldots,\mri} \frac{1}{2}\log^+\bigg(\frac{\snr}{\sigma_{\text{eff},m}^2}\bigg) \nonumber \\
&\sigma_{\text{eff},m}^2 = \| \mathbf{b}_m \|^2 + \inr \ \big\| \mathbf{J}^T\mathbf{b}_m \big\|^2 +  \snr\  \big\| \mathbf{H}^T \mathbf{b}_m -\mathbf{a}_m \big\|^2  \nonumber \ .
\end{align} The proof follows along similar lines to that of Theorem~\ref{thm:rate} and is omitted.
\end{theorem}
\begin{corollary} \label{c:exactJif} Assume that $\mathbf{H}$ is full rank. The rate achievable via exact integer-forcing with equalization matrix $\mathbf{B}_{\text{exact}} = \mathbf{A} \mathbf{H}^{-1}$ is
\begin{align}
\label{eq:prate}
&R_{\iftext,\text{exact}}(\mathbf{H},\mathbf{J})  = \mri \cdot \max_{\substack{ \mathbf{A} \in \mathbb{Z}^{\mri \times \mri} \\ \mathrm{rank}(\mathbf{A}) = \mri}}  \min_{m} \frac{1}{2}\log^+\bigg( \frac{\snr}{\sigma_{\text{exact},m}^2} \bigg) \\
&\sigma_{\text{exact},m}^2 = \big\|\mathbf{H}^{-T}\mathbf{a}_m\big\|^2 + \inr\ \big\|\mathbf{J}^T\mathbf{H}^{-T}\mathbf{a}_m\big\|^2  \ . \label{eq:exactjnoise}
\end{align} 
\end{corollary}
\begin{remark}\label{r:conventionalJ}
Following the same arguments as in Section~\ref{s:specialcase}, it can be shown that the performance of any conventional linear receiver can obtained via integer-forcing with $\mathbf{A} = \mathbf{I}$.
\end{remark}
\begin{remark}
The achievable rate in Theorem \ref{thm:rateinterference} is maximized by the equalization matrix\begin{align*}
\mathbf{B}_{\text{opt}} = \snr \ \mathbf{A}\mathbf{H}^T\left(\mathbf{I} + \inr \ \mathbf{J}\mathbf{J}^T + \snr \ \mathbf{H}\mathbf{H}^T \right)^{-1} \ .
\end{align*}
\end{remark}

\subsection{Geometric Interpretation}
As argued in Section~\ref{sec:equations}, in the case without interference, the integer vectors $\mathbf{a}_1, \ldots, \mathbf{a}_{\mri}$ should be chosen in the direction of the maximum eigenvector of $\mathbf{H}^T\mathbf{H}$ to minimize the effective noise. Here, we argue that, when $\inr$ is large, the integer vectors should instead be chosen as close to orthogonal to the effective interference as possible. Assume that $\mathbf{H}$ is full rank and consider the (suboptimal) rate expression in \eqref{eq:prate}. Let $\mathbf{\tilde{J}} = \mathbf{H}^{-1}\mathbf{J}$. The effective noise variance from~\eqref{eq:exactjnoise} is upper bounded by \begin{align*}
\sigma_{\text{exact},m}^2 \leq \lambda^2_\text{max}(\mathbf{H}^{-1}) \|\mathbf{a}_m\|^2 + \inr\ \big\|{\mathbf{\tilde{J}}}^T\mathbf{a}_m \big\|^2 \ .
\end{align*} 

\begin{figure}[h!]
\begin{center}
\psset{unit=0.6mm}
\begin{pspicture}(0,-75)(120,120)
\rput(0,0){
\rput(60,112){\textbf{Zero-Forcing}}
\rput(15,0){
\psdots[dotscale=0.5](0,30)(0,45)(0,60)(0,75)(0,90)(0,105)
}

\rput(15,0){
\psdots[dotscale=0.5](0,30)(0,45)(0,60)(0,75)(0,90)(0,105)
}

\rput(30,0){
\psdots[dotscale=0.5](0,30)(0,45)(0,60)(0,75)(0,90)(0,105)
}

\rput(45,0){
\psdots[dotscale=0.5](0,30)(0,45)(0,60)(0,75)(0,90)(0,105)
}

\rput(60,0){
\psdots[dotscale=0.5](0,30)(0,45)(0,60)(0,75)(0,90)(0,105)
}

\rput(75,0){
\psdots[dotscale=0.5](0,30)(0,45)(0,60)(0,75)(0,90)(0,105)
}

\rput(90,0){
\psdots[dotscale=0.5](0,30)(0,45)(0,60)(0,75)(0,90)(0,105)
}

\rput(105,0){
\psdots[dotscale=0.5](0,30)(0,45)(0,60)(0,75)(0,90)(0,105)
}


\psline[linewidth=1.2]{->}(60,45)(20,90)
\rput(24,96){$\mathbf{\tilde{J}}$}
\psline[linestyle=dashed,linewidth=0.6]{->}(60,45)(75,45)
\psline[linestyle=dashed,linewidth=0.6]{->}(60,45)(60,60)
\rput(60,65){$\mathbf{a}_1$}
\rput(81,43){$\mathbf{a}_2$}
}

\rput(0,-100){
\rput(60,112){\textbf{Integer-Forcing}}
\rput(15,0){
\psdots[dotscale=0.5](0,30)(0,45)(0,60)(0,75)(0,90)(0,105)
}

\rput(15,0){
\psdots[dotscale=0.5](0,30)(0,45)(0,60)(0,75)(0,90)(0,105)
}

\rput(30,0){
\psdots[dotscale=0.5](0,30)(0,45)(0,60)(0,75)(0,90)(0,105)
}

\rput(45,0){
\psdots[dotscale=0.5](0,30)(0,45)(0,60)(0,75)(0,90)(0,105)
}

\rput(60,0){
\psdots[dotscale=0.5](0,30)(0,45)(0,60)(0,75)(0,90)(0,105)
}

\rput(75,0){
\psdots[dotscale=0.5](0,30)(0,45)(0,60)(0,75)(0,90)(0,105)
}

\rput(90,0){
\psdots[dotscale=0.5](0,30)(0,45)(0,60)(0,75)(0,90)(0,105)
}

\rput(105,0){
\psdots[dotscale=0.5](0,30)(0,45)(0,60)(0,75)(0,90)(0,105)
}

\psline[linewidth=1.2]{->}(60,45)(20,90)
\rput(24,96){$\mathbf{\tilde{J}}$}
\psline[linestyle=dashed,linewidth=0.6]{->}(60,45)(105,75)
\psline[linestyle=dashed,linewidth=0.6]{->}(60,45)(75,60)
\rput(112,75){$\mathbf{a}_2$}
\rput(75,65){$\mathbf{a}_1$}
}

\end{pspicture}
\end{center}\caption{The zero-forcing linear receiver (top) is equivalent to an integer-forcing linear receiver with integer vectors fixed to $\mathbf{a}_1 = [1~~0]^T$ and $\mathbf{a}_2 = [0~~1]^T$, which partially overlap with the interference subspace $\mathbf{\tilde{J}} = \mathbf{H}^{-1} \mathbf{J}$. Higher rates are possible by optimizing over the choice of integer vectors and, in this case, choosing integer vectors that are nearly orthogonal to $\mathbf{\tilde{J}}$. }\label{fig:eqnsinf}
\end{figure}
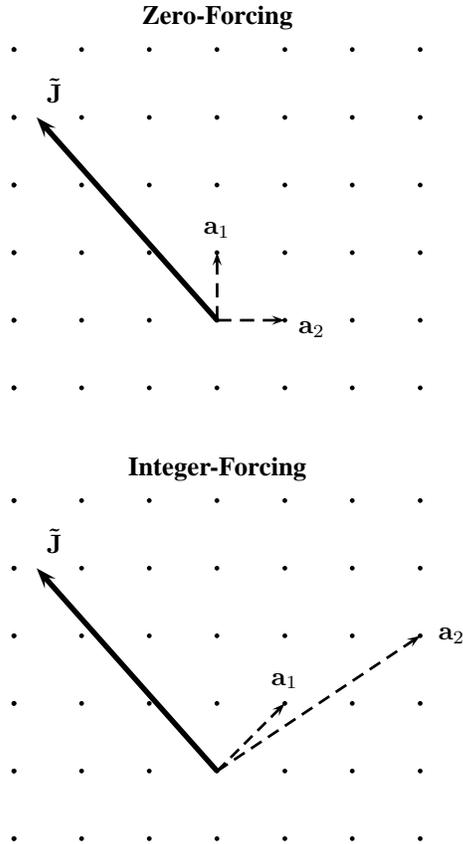

It follows that, in the high interference regime ($\inr \gg 1$), the integer vectors should be as orthogonal as possible to the effective interference space ${\mathbf{\tilde{J}}}$. This is illustrated in Figure~\ref{fig:eqnsinf} for one-dimensional interference. From Remark~\ref{r:conventionalJ}, the performance of a conventional linear receiver is equivalent to that of an integer-forcing linear receiver with $\mathbf{a}_1 = [1~0~\cdots~0]^T$, $\mathbf{a}_2 = [0~1~\cdots~0]^T$, $\ldots$, $\mathbf{a}_{\mri} = [0~0~\cdots~1]^T$. As a result, the interference space spanned by ${\mathbf{\tilde{J}}}$ has significant projections onto at least some of the decoding directions. By contrast, for the integer-forcing linear receiver, since $\mathbf{a}_1,\ldots , \mathbf{a}_{\mri}$ need only be linearly independent, we can choose each $\mathbf{a}_m$ to be close to orthogonal to ${\mathbf{\tilde{J}}}$.

\subsection{Outage Behavior}
\label{sec:interferenceoutage}

We now examine the outage performance of the receiver architectures discussed above. Consider a real-valued MIMO channel with $M = 2$ transmit and receive antennas as well as interference with dimension $K=1$. The elements of the channel matrix $\mathbf{H}$ are drawn i.i.d.~$\mathcal{N}(0,1)$ and the interference vector $\mathbf{J}$ is drawn uniformly over the $2$-dimensional unit sphere. The interference power is $\inr = \snr^{0.2}$. 

\begin{figure}[h!]
\centering
\includegraphics[width=3.6in]{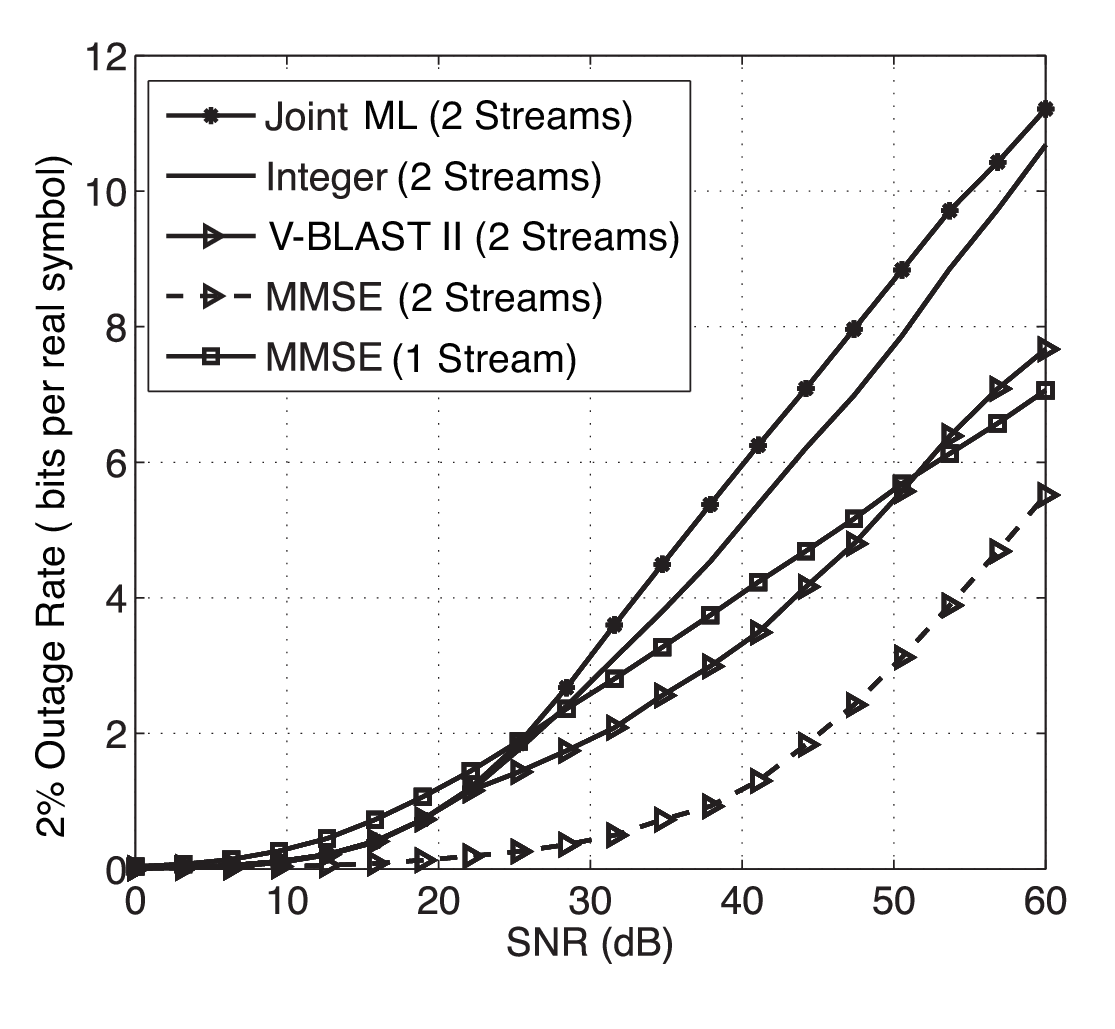}
\caption{2 percent outage rates for the $2 \times 2$ real-valued MIMO channel under i.i.d.~$\mathcal{N}(0,1)$ fading and $1$-dimensional interference with direction drawn uniformly over the unit sphere and strength $\alpha = 0.2$. }
\label{f:ifplot1}
\end{figure}

In Figure~\ref{f:ifplot1}, we have plotted the $2\%$ outage rates for joint ML decoding~\eqref{eq:mlint} as well as the integer-forcing~\eqref{eq:ifint}, V-BLAST II~\eqref{eq:vblastint}, and linear MMSE~\eqref{eq:linearint} receivers under the assumption that $2$ data streams are transmitted. We have also plotted the performance of the linear MMSE receiver with $1$ transmitted data stream.\footnote{Since there is only one data stream, the linear MMSE receiver is equivalent to joint ML decoding.} At low SNR, it is preferable to send only a single data stream, and thus there is no advantage to integer-forcing in this regime. However, beyond $25$dB, it is preferable to send $2$ data streams, and the integer-forcing receiver nearly matches the performance of joint ML decoding.

\subsection{Generalized Degrees-of-Freedom}
\label{sec:dof}
We now evaluate the generalized degrees-of-freedom as introduced in~\cite{etw07}. We specify the scaling of the interference-to-noise ratio through the parameter $0 \leq \alpha \leq 1$,
\begin{align*}
\inr = \snr^{\alpha} \ .
\end{align*} 
\begin{definition} Consider a family of coding schemes, indexed by $\snr$, that achieves sum rate $R_{\text{scheme}}(\mathbf{H},\mathbf{J},\snr)$ over a fixed channel matrix $\mathbf{H}$ and interference matrix $\mathbf{J}$. The \textit{generalized degrees-of-freedom (GDoF)} of this scheme is
\begin{equation*}
d_{\text{scheme}}(\mathbf{H},\mathbf{J}) = \lim_{\snr \rightarrow \infty} \frac{ R_{\text{scheme}}(\mathbf{H},\mathbf{J},\snr)}{\frac{1}{2}\log(1 + \snr)} \ .
\end{equation*}
\end{definition}

Assume that both the channel matrix $\mathbf{H}$ and the interference matrix $\mathbf{J}$ are full rank. A straightforward derivation shows that the joint ML decoder, linear MMSE receiver, and V-BLAST II receiver with $\mri$ i.i.d.~Gaussian data streams achieve
\begin{align*}
d_\text{ML}(\mathbf{H},\mathbf{J}) &= \mri - K\alpha \\
d_{\text{MMSE},\mri}(\mathbf{H},\mathbf{J}) &= \mri - \mri \alpha \\
d_{\text{V-BLAST II},\mri}(\mathbf{H},\mathbf{J}) &= \mri - \mri \alpha \ .
\end{align*} The linear MMSE and V-BLAST II receivers are suboptimal since they encounter interference in (some of) the effective channel outputs and the rate is determined by the worst data stream. This can be partially mitigated by reducing the number of data streams to $\mri - K$, which allows the receiver to employ $K$ of its antennas towards eliminating the interference before separating the data streams. Unfortunately, this still only yields a GDoF of \begin{align*}
d_{\text{MMSE},\mri-K}(\mathbf{H},\mathbf{J}) &= \mri - K \ .
\end{align*} 

In the next theorem, we show that the integer-forcing linear receiver achieves the same GDoF as the joint ML decoder (up to a set of channel and interference matrices of measure zero).
\begin{theorem}
\label{thm:dof}
For almost all full rank channel matrices $\mathbf{H} \in \mathbb{R}^{\mri \times \mri}$ and interference matrices $\mathbf{J} \in \mathbb{R}^{\mri \times K}$, the integer-forcing linear receiver achieves the GDoF \begin{align*}
d_{\text{IF}}(\mathbf{H},\mathbf{J}) = \mri - K\alpha \ .
\end{align*} 
\end{theorem} The proof is deferred to Appendix~\ref{ap:gdofif}.

\begin{figure}[h]
\begin{center}
\includegraphics[width=3.6in]{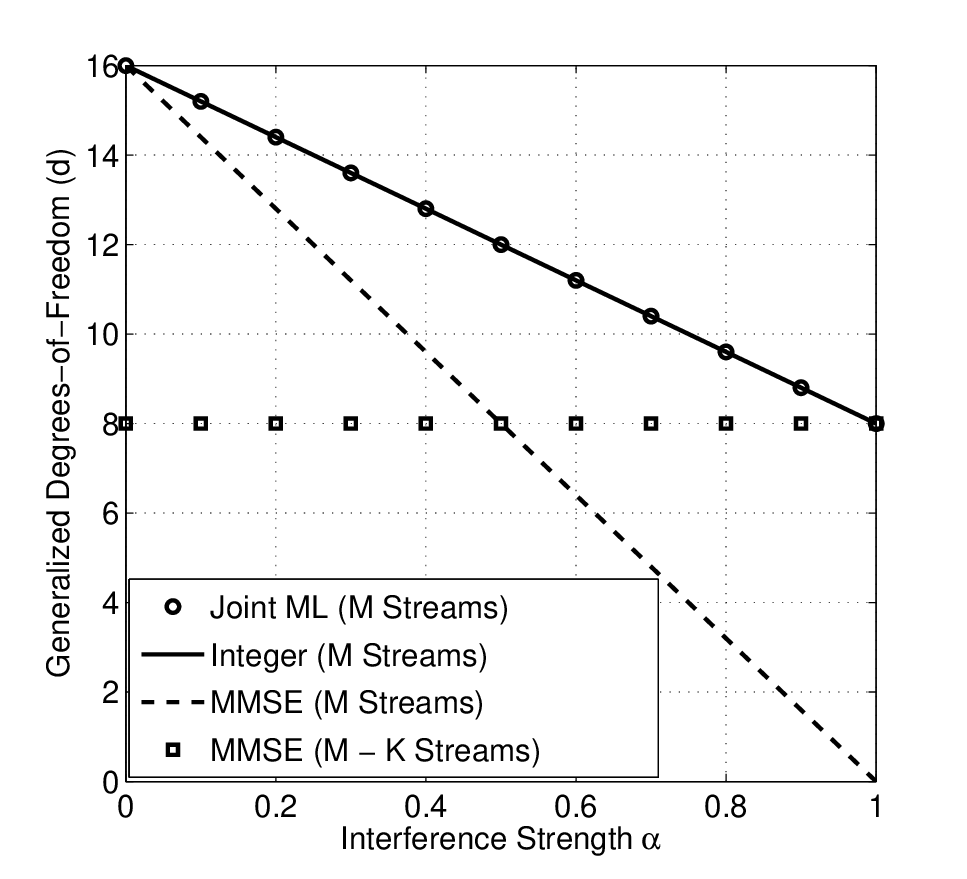}
\caption{Generalized degrees-of-freedom for the real-valued $16 \times 16$ MIMO channel with $8$-dimensional interference $(\mri = 16, K = 8).$}
\end{center}
\label{fig:dof}
\end{figure}

In Figure~\ref{fig:dof}, we have plotted the achievable GDoF for the joint ML decoder, integer-forcing, and the linear MMSE receiver with $\mri$ data streams and $\mri-K$ data streams. We have assumed that $\mri = 16$, $K = 8$, and that the channel and interference matrices are drawn from the set for which Theorem~\ref{thm:dof} applies. 

\section{Concluding Remarks and Extensions}

In this paper, we proposed a novel linear receiver architecture for MIMO channels that bridges the performance gap between conventional linear receivers and the optimal joint ML decoder. This integer-forcing linear receiver may be seen as an extension of the lattice reduction receiver to the case of coded transmission. It is well-suited to scenarios where a single receiver, equipped with multiple antennas, must recover multiple data streams (i.e., an uplink channel). Recent work has shown that the principles underlying integer-forcing can be applied in a broader context including downlink channels~\cite{hc12,hc12IT} and interference channels~\cite{ncnc13ISIT}. In another line of work, integer-forcing was applied to the intersymbol interference channel by adding the requirement that the codebook is cyclic~\cite{oe12}.

Although we focused on the setting where each transmit antenna encodes an independent data stream, integer-forcing can also be applied when there is space-time coding across the antennas. Specifically, after the data streams are mapped to codewords, the transmitter can apply a linear dispersion code \cite{hh02} $\mathbf{S}$ and transmit $\mathbf{SX}$. The achievable rates can be derived by simply replacing the channel matrix $\mathbf{H}$ with the effective channel $\mathbf{HS}$. The performance when the Golden code \cite{brv05} is used is investigated in \cite{de12} where it is shown that integer-forcing can operate quite close to the performance of the joint ML decoder at finite SNR under Rayleigh fading. Very recent work \cite{oe13} shows that integer-forcing, coupled with perfect linear dispersion space-time codes \cite{orbv06} not only achieves the optimal DMT for fading channels but also approaches the MIMO capacity to within a constant gap, regardless of the channel realization.

As discussed in Section~\ref{s:convlinear}, the performance of conventional linear receivers can be improved via successive interference cancellation. Furthermore, if the rates are  
chosen to correspond to one of the corner points of the associated multiple-access capacity region, then the V-BLAST II architecture can achieve the sum capacity~\cite{vg97}. Recent work has proposed a successive cancellation integer-forcing scheme and shown that it can attain the sum capacity~\cite{oen13}. Interestingly, this scheme can often operate at rate tuples that are much closer to the symmetric capacity than the corner points.

For complex-valued channels, it is possible to create compute-and-forward strategies from lattices over the Eisenstein integers. As a result, the receiver is able to decode linear combinations over the codewords, where the coefficients are taken from the Eisenstein (rather than Gaussian) integers. This can in turn improve the achievable rates for i.i.d.~Rayleigh fading. See~\cite{thbn14} for further details.

An interesting direction for future work is determining how closely the achievable rates derived here can be approached using modern channel codes (e.g., LDPC codes) and iterative decoding. Recent work on channel coding for compute-and-forward offers an excellent starting point \cite{fsk11,hc12IT,hn13,oe12,ozegn11,ylhyc12,hc11,bl12,tnp13,hnt14}.

\IEEEpeerreviewmaketitle

\appendices

\section{Uncoded Integer-Forcing as Lattice Reduction}
\label{ap:latticereduction}
We now connect integer-forcing to the class of symbol-level linear architectures known as {\em lattice-reduction} detectors \cite{yw02}. Consider a MIMO system where every transmit antenna sends an uncoded data stream using a QAM constellation. In this setting, a zero-forcing receiver first inverts the channel matrix $\mathbf{H}$ and feeds the equalized channel outputs into several slicers (i.e., detectors), each of which quantizes an entry of $\mathbf{X} + \mathbf{H}^{-1}\mathbf{Z}$ to the nearest constellation point. The goal of lattice reduction is to transform the received constellation into one that admits a lower probability of error, prior to feeding the output into a slicer. For any lattice-based constellation (such as QAM), any unimodular transformation $\mathbf{AX}$ will yield an effective constellation with (at least) the same minimum distance. A lattice-reduction receiver uses linear equalization to obtain $\mathbf{AX} + \mathbf{AH}^{-1} \mathbf{Z}$, employs slicers to quantize the entries to the nearest effective constellation points, and, finally, inverts the unimodular matrix $\mathbf{A}$ to recover estimates of the symbols $\mathbf{X}$. By optimizing over $\mathbf{A}$, the effective noise $\mathbf{AH}^{-1} \mathbf{Z}$ can be distributed more evenly across the data streams than in zero-forcing. It has been shown that lattice reduction can achieve the receive diversity \cite{tmk07lll}.

Clearly, lattice reduction is closely related to our proposed integer-forcing architecture. The key distinction is that integer-forcing works at the codeword level, whereas lattice reduction works at the symbol level. As a result, we can derive explicit rate expressions (e.g., Theorem~\ref{thm:alterrate}) that only depend on the channel matrix $\mathbf{H}$, the integer matrix $\mathbf{A}$, and the $\snr$.
Since lattice reduction does not directly permit channel coding, most studies have focused on its advantages in the high SNR regime. For instance, several works have proposed space-time codes that are amenable to lattice-reduction detectors and achieve the optimal diversity-multiplexing tradeoff \cite{ecd04,je10}.

Note that the integer-forcing architecture includes lattice reduction as a special case by setting the channel code blocklength to one, $n = 1$. Interestingly, integer-forcing does not require the effective channel matrix $\mathbf{A}$ to be unimodular: it can be any full-rank integer matrix. In the following example, we show that this restriction can sometimes result in an arbitrarily large performance gap.

We consider the $M \times M$ MIMO channel with channel matrix
 \begin{align}
 \mathbf{H} = \begin{bmatrix}
 1 & 0 & \cdots & 0 & 0\\
 0 & 1 & \cdots& 0 & 0\\
 \vdots & \vdots & \ddots & \vdots & \vdots\\
 0 & 0  & \cdots & 1&0 \\
 -1 & -1 & \cdots &-1 &2 \\
 \end{bmatrix} \label{e:fixedmatrixEx3} \ .
\end{align}
A simple calculation shows that the inverse of this channel matrix is
 \begin{align}
 \mathbf{H}^{-1} = \begin{bmatrix}
 1 & 0 & \cdots & 0 & 0\\
 0 & 1 & \cdots& 0 & 0\\
 \vdots  &  \vdots & \ddots & \vdots & \vdots\\
 0 & 0 &  \cdots & 1&0 \\
 \frac{1}{2}  & \frac{1}{2}  & \cdots &\frac{1}{2}  &\frac{1}{2}  \\
 \end{bmatrix} \label{e:fixedmatrixEx3inv} \ .
\end{align}
The integer matrix $\mathbf{A} = \mathbf{H}$ maximizes the achievable rate for the exact integer-forcing receiver from Corollary \ref{cor:ratesub}. Note that since $\mathbf{H}^{-1}$ has non-integer entries, $\mathbf{H}$ is not unimodular. The largest effective noise variance (as defined in~\eqref{eq:effectivenoise1}) is
\begin{align*}
\sigma^2_{\text{exact}} &= 1  \ .
\end{align*}

By contrast, for a lattice-reduction receiver, we must ensure that the effective channel matrix is unimodular. Using the fact that $\mathbf{H}^{-T}$ is a basis for the body-centered cubic lattice, it can be shown that the best choice of unimodular matrix is $\mathbf{A}_{\text{uni}} = \mathbf{I}$. It follows that the largest effective noise variance is 
\begin{align*}
\max_m \sigma^2_{\text{uni},m} &= \max \left\{ M/4, \ 1 \right\} \ .
\end{align*} Hence, as the number of antenna increases ($M \rightarrow \infty$), restricting the integer matrix to be unimodular can result in an arbitrarily large loss.

\section{Integer-Forcing vs. V-Blast IV}
\label{ap:vblast4}

Recall that, in V-BLAST II, the data streams have equal rates and the decoding order is optimized whereas,  in V-BLAST III, the rate allocation is optimized and the decoding order is fixed. In this appendix, we introduce V-BLAST IV, which allows for both rate allocation and an optimized decoding order. Let $\Pi$ denote the set of all possible permutations of $\{1,2,\ldots,\mrt\}$. Under V-BLAST IV, the data streams are decoded with respect to the ordering
\begin{align*}
\pi^{*} =  \argmax_{\pi \in \Pi} \min_m R_{\text{SIC},\pi(m)}(\mathbf{H},\mathbf{b}_{\text{MMSE-SIC},m}) 
\end{align*}
where $R_{\pi(m)}(\mathbf{H})$ and $\mathbf{b}_{\text{MMSE-SIC},m}$ are defined in~\eqref{eq:sicstream} and~\eqref{e:mmsesicprojection}, respectively. The outage probability and outage rate are calculated according to Definition~\ref{d:outagevblastiii}, which implicitly optimizes the rate allocation across data streams.

\begin{figure}[h!]
\centering
\includegraphics[width=3.7in]{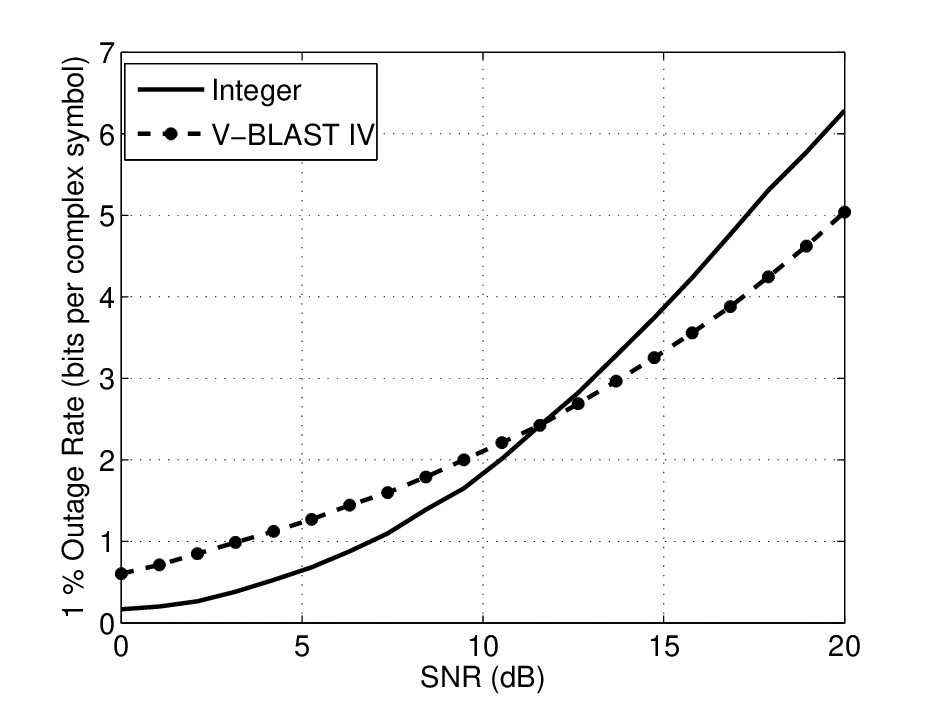}
\caption{1 percent outage rates for the $2 \times 2$ complex-valued MIMO channel under i.i.d.~Rayleigh fading.}
\label{f:outV41}
\end{figure}

\begin{figure}[h!]
\centering
\includegraphics[width=3.7in]{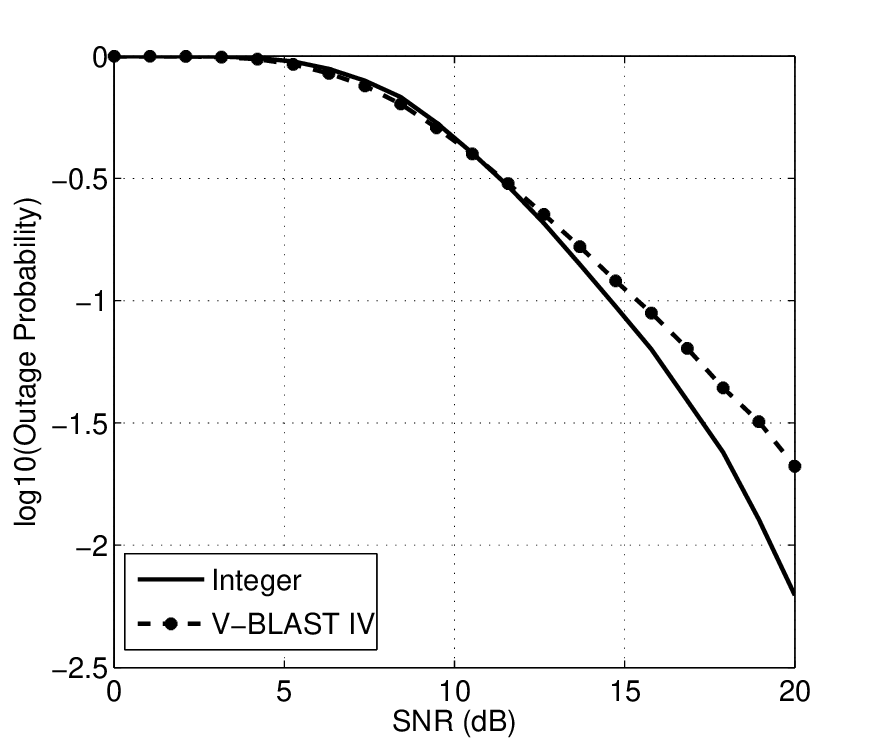}
\caption{Outage probability for the $2 \times 2$ complex-valued MIMO channel under i.i.d.~Rayleigh fading for a target sum rate of $R = 6$.}
\label{f:poutPV4}
\end{figure}

Consider a complex-valued MIMO channel with i.i.d.~Rayleigh fading and $\mct = \mcr = 2$ antennas. In Figure~\ref{f:outV41}, we have plotted the $1$ percent outage rate of V-BLAST IV and integer-forcing~\eqref{eq:alternaterate}. V-BLAST IV outperforms integer-forcing until approximately $12$dB and integer-forcing is superior from then onwards. In Figure~\ref{f:poutPV4}, we have compared the outage probability for a target sum rate of $R = 6$. The two curves are nearly identical until $12$dB, after which integer-forcing attains a smaller outage probability.

\section{Proof of Theorem \ref{thm:dmt}}
\label{ap:dmt}
In order to establish Theorem \ref{thm:dmt}, we need a few key facts about lattices, starting with the definition of dual lattices from \cite{lls90}.

\begin{definition}[Dual Lattice] Given a lattice $\Lambda \subset \mathbb{R}^{\mrt}$ with a rank-$L$ generator matrix $\mathbf{G} \in \mathbb{R}^{\mrt \times L}$,
\begin{align*}
\Lambda = \big\{ \mathbf{G} \mathbf{d} : \mathbf{d} \in \mathbb{Z}^L \big\} \ ,
\end{align*}
 the \textit{dual lattice} $\Lambda^*$ has generator matrix $\left(\mathbf{G}^T\right)^{\dag}$,
\begin{align*}
\Lambda^* = \left\{\left(\mathbf{G}^T\right)^{\dag} \mathbf{d}: \mathbf{d} \in \mathbb{Z}^{\mrt} \right\} \ .
\end{align*}
\end{definition}

To prove Theorem \ref{thm:dmt}, we will work with the successive minima for the involved lattices, a standard concept from the Diophantine approximation literature (see e.g. \cite{cassels,lang,ah}).

\begin{definition}[Successive Minima]
\label{def:smMSE}
Let $\mathcal{B} = \big\{\mathbf{x} \in \mathbb{R}^{\mrt}: \|\mathbf{x}\| \leq 1 \big\}$ be the unit ball. Given a lattice $\Lambda \subset \mathbb{R}^{\mrt}$ with a rank-$L$ generator matrix, the $m^{\text{th}}$ \textit{successive minimum} $\epsilon_m(\Lambda)$ is given by
\begin{align*}
\epsilon_m(\Lambda) &=  \inf \big\{\epsilon \in \mathbb{R}_+ : \exists~ m ~\text{linearly independent lattice} \\
&\qquad \qquad ~~~~~~~~~~~~~~~\text{points}~ \mathbf{v}_1, \ldots ,\mathbf{v}_m \in \Lambda \cap \epsilon \mathcal{B} \big\} \ .
\end{align*} Note that the successive minima are non-decreasing, $\epsilon_1(\Lambda) \leq \cdots \leq \epsilon_L(\Lambda)$.
\end{definition}

The following lemma links the successive minima of a lattice with those of its dual.

\begin{lemma}[{\cite[Proposition 3.3]{lls90}}]
\label{lm:smdual}
Let $\Lambda \subset \mathbb{R}^{\mrt}$ be an arbitrary lattice with a rank-$L$ generator matrix and $\Lambda^*$ be its dual lattice. The successive minima for $\Lambda$ and $\Lambda^*$ satisfy the following inequality:
\begin{equation*}
\epsilon^2_{m} (\Lambda^*)\epsilon^2_{1}(\Lambda) \leq \frac{m^2(m+3)}{4}~~~ \mbox{for} ~ m  = 1,2, \ldots,L  \ .
\end{equation*}
\end{lemma}

Finally, we need the following result of Taherzadeh, Mobasher, and Khandani~\cite{tmk07} concerning the first successive minimum of a lattice induced by an i.i.d.~Rayleigh channel matrix.

\begin{lemma}[{\cite[Lemma 3]{tmk07}}]
\label{lm:smfirst}
Let $\mathbf{H} \in \mathbb{R}^{\mrr\times \mrt}$ be the real-valued representation of a $\mcr \times \mct$ complex-valued matrix with i.i.d.~Rayleigh entries. Let $\Lambda=\left\{ \mathbf{H} \mathbf{d} : \mathbf{d} \in \mathbb{Z}^{\mrt} \right\}$ be the lattice generated by $\mathbf{H}$. Then, the first successive minimum of $\Lambda$ satisfies
\begin{align*}
&\mathbb{P} \big(\epsilon_1(\Lambda) \leq s\big) \\
&\leq
\begin{cases} \gamma s^{\mrr}, &\mrt < \mrr \ ,
\\
\gamma s^{\mrr} \max \left\{-(\ln s)^{1 + \mrr/2}, 1 \right\}, &\mrt = \mrr \ .
\end{cases}
\end{align*}
where $\gamma$ is a constant independent of $s$.
\end{lemma}

\begin{IEEEproof}[Proof of Theorem~\ref{thm:dmt}] First, condition on the event that $\mathbf{H}$ is full rank, which occurs with probability $1$ under i.i.d.~Rayleigh fading. Let $R = r \log \snr$ be the target rate where $r \in [0,\mct]$. For analytical convenience, we will work with the rate expression for exact integer-forcing from Corollary~\ref{cor:ratesub}. The outage probability is
\begin{align*}
&p_{\text{outage}}(R) \\
&= \mathbb{P}\big( R_{\iftext,\text{exact}}(\mathbf{H}) < r \log(\snr) \big) \\
&=\mathbb{P}\left( \max_{\substack{\mathbf{A} \in \mathbb{Z}^{\mrt \times \mrt} \\ \mathrm{rank}(\mathbf{A}) = \mrt}} R_{\text{comp}}(\mathbf{H},\mathbf{A},\mathbf{B}_{\text{exact}}) < \frac{2r}{\mrt} \log(\snr) \right) \\
&=\mathbb{P}\left( \max_{\substack{\mathbf{A} \in \mathbb{Z}^{\mrt \times \mrt} \\ \mathrm{rank}(\mathbf{A}) = \mrt}} \min_{m = 1,\ldots,\mrt} \frac{\snr}{\big\| \big(\mathbf{H}^T\big)^{\dag} \mathbf{a}_m \big\|^2} < \snr^{\frac{2r}{\mrt}} \right) \\
&=\mathbb{P}\left( \min_{\substack{\mathbf{A} \in \mathbb{Z}^{\mrt \times \mrt} \\ \mathrm{rank}(\mathbf{A}) = \mrt}} \max_{m = 1,\ldots,\mrt} \Big\| \big(\mathbf{H}^T\big)^{\dag} \mathbf{a}_m \Big\|^2 > \snr^{1 - \frac{2r}{\mrt}} \right) 
\end{align*}

Let $\Lambda_\text{channel}$ be the lattice generated by $\mathbf{H} \in \mathbb{R}^{\mrr \times \mrt}$ and $\Lambda_\text{dual}$ be the dual lattice generated by $\left(\mathbf{H}^T\right)^{\dag}$,
\begin{align*}
&\Lambda_\text{channel} = \left\{\mathbf{H} \mathbf{d} : \mathbf{d} \in \mathbb{Z}^{\mrt} \right\} \\
&\Lambda_\text{dual} = \left\{ \left(\mathbf{H}^T\right)^{\dag} \mathbf{d} : \mathbf{d} \in \mathbb{Z}^{\mrr} \right\}.
\end{align*}
From the definition of successive minima (Definition \ref{def:smMSE}), it follows that
\begin{align*}
\min_{\substack{\mathbf{A} \in \mathbb{Z}^{\mrt \times \mrt} \\ \mathrm{rank}(\mathbf{A}) = \mrt}} \max_{m = 1,\ldots,\mrt} \Big\| \big(\mathbf{H}^T\big)^{\dag} \mathbf{a}_m \Big\|^2 &= \epsilon_{\mrt}^2 (\Lambda_\text{dual})\  . 
\end{align*} Therefore, we can express the outage probability as 
\begin{align}
p_{\text{outage}} (r)&=\mathbb{P} \left( \epsilon^2_{\mrt}(\Lambda_\text{dual})  >  \snr^{1-\frac{2r}{\mrt}} \right)\label{eq:outage1}
\end{align}

Now, using Lemma \ref{lm:smdual}, we can bound the successive minima of $\Lambda_\text{dual}$ in terms of the successive minima of $\Lambda_\text{channel}$,
\begin{align}
\label{eq:succminima}
\epsilon^2_{\mrt}(\Lambda_\text{dual}) \leq \frac{2\mrt^3+3\mrt^2}{\epsilon^2_{1}(\Lambda_\text{channel})} \ .
\end{align}
Combining \eqref{eq:outage1} and \eqref{eq:succminima}, the outage probability is upper bounded by
\begin{align*}
p_{\text{outage}} (r) &\leq \mathbb{P} \left( \frac{2\mrt^3+3\mrt^2}{\epsilon^2_{1}(\Lambda_\text{channel})}  >  \snr^{1-\frac{2r}{\mrt}} \right)\\
&= \mathbb{P} \left( \epsilon^2_1(\Lambda_\text{channel})  < \frac{2\mrt^3 + 3\mrt^2}{\snr^{1-\frac{2r}{\mrt}}} \right)
\end{align*}
This probability can in turn be upper bounded using Lemma \ref{lm:smfirst}. For large $\snr$, we find that
\begin{align*}
p_{\text{outage}} (r)&\leq c \   \Big(\snr^{1-\frac{2r}{\mrt}}\Big)^{-\mrr/2}  \Big(\frac{1}{2}\ln\big( \snr^{1-\frac{2r}{\mrt}} \big)\Big)^{1+\mrr/2}
\end{align*}
where $c$ is a constant independent of $\snr$. The achievable diversity for multiplexing gain $r$ is thus
\begin{align*}
d_{\iftext}(r) &= \lim_{\snr \rightarrow \infty} \frac{-\log p_\text{outage}(r)}{\snr} \\
&\geq \lim_{\snr \rightarrow \infty} \frac{\frac{\mrr}{2}\left(1 - \frac{2r}{\mrt}\right)\snr}{\snr} - \frac{o(\snr)}{\snr}\\
&= \frac{\mrr}{2} \left(1 - \frac{2r}{\mrt}\right)\\
&= \mcr \left( 1 - \frac{r}{\mct}\right) \ . 
\end{align*}
\end{IEEEproof}

\section{Proof of Theorem~\ref{thm:dof}}
\label{ap:gdofif}

Our proof of Theorem \ref{thm:dof} uses the following lemma. 

\begin{lemma}
\label{lm:diophantine}
For almost all $\mathbf{T} \in \mathbb{R}^{K \times (\mri - K)}$, there exists a $Q' \in \mathbb{N}$ such that, for any $Q > Q'$, there exist $\mri$ linearly independent integer vectors $\mathbf{v}_1,\ldots,\mathbf{v}_\mri \in \mathbb{Z}^\mri$ of the form $[\mathbf{q}^T_m \  \mathbf{p}^T_m]^T \in \mathbb{Z}^{\mri - K} \times \mathbb{Z}^{K}$ for $m =1, 2, \ldots, \mri$ that satisfy
\begin{align*}
&\|\mathbf{q}_m \| \leq CQ (\log Q)^2 \\
&\big\|\mathbf{T}\mathbf{q}_m - \mathbf{p}_m \big\| \leq \frac{C (\log Q)^2}{Q^{(\mri - K)/K}}\ ,
\end{align*} where $C$ is a constant that is independent of $Q$.
\end{lemma}
The proof is given in Appendix \ref{ap:diophantine}. 

\begin{IEEEproof}[Proof of Theorem~\ref{thm:dof}]
We will work with the rate expression for exact integer-forcing from Corollary~\ref{c:exactJif}. Since $\inr = \snr^{\alpha}$, the largest effective noise variance is upper bounded by \begin{align}
&\max_{m} \sigma_{\text{exact},m}^2 \nonumber \\
 &= \min_{\substack{\mathbf{A} \in \mathbb{Z}^{\mri \times \mri} \\ \mathrm{rank}(\mathbf{A})= \mri }} \max_m
\big\|\mathbf{H}^{-T}\mathbf{a}_m\big\|^2 +
\snr^{\alpha}\big\|\mathbf{J}^T \mathbf{H}^{-T}\mathbf{a}_m\big\|^2 \nonumber \\
&\leq  \min_{\substack{\mathbf{A} \in \mathbb{Z}^{\mri \times \mri} \\ \mathrm{rank}(\mathbf{A})= \mri }} \max_m \lambda^2_\text{max} \big(\mathbf{H}^{-1}\big) \|\mathbf{a}_m\|^2 +
\snr^{\alpha} \big\|\mathbf{\tilde{J}}^T\mathbf{a}_m\big\|^2 \ ,
\label{eq-Interf-EffNoiseUpper}
\end{align}
where $\mathbf{\tilde{J}} = \mathbf{\mathbf{H}^{-1}\mathbf{J}}$. We now partition $\mathbf{\tilde{J}}^T$,
\begin{align*}
\mathbf{\tilde{J}}^T = \big[\mathbf{S}_1~ \mathbf{S}_2\big] \ ,
\end{align*}
where $\mathbf{S}_1 \in \mathbb{R}^{K\times (\mri- 2K)}$ and $\mathbf{S}_2
\in \mathbb{R}^{K \times K}$.
Since $\mathbf{\tilde{J}}$ has rank $K$, we can permute its columns so that the last $K$
columns are linearly independent. If we use the same permutation on the
coefficients of the vector $\mathbf{a}_m,$ the upper bound in \eqref{eq-Interf-EffNoiseUpper} will remain unchanged. Therefore, without loss of generality, we may assume
that $\mathbf{S}_2$ has rank $K$. Define $\mathbf{T} =
-\mathbf{S}_2^{-1}\mathbf{S}_1$. Then, we can write
\begin{align}
\mathbf{S}^{-1}_2 \mathbf{\tilde{J}}^T &= \big[\mathbf{S}^{-1}_2 \mathbf{S}_1 ~
\mathbf{S}^{-1}_2 \mathbf{S}_2\big] \nonumber \\
&= \big[-\mathbf{T} ~ \mathbf{I} \big]\ . \label{eq:tildeJ}
\end{align}

Let $\mathbf{q}_m$ denote the first $\mri - K$ entries of $\mathbf{a}_m$ and $\mathbf{p}_m$ denote the last $K$ entries, 
$$\mathbf{a}_m = \begin{bmatrix} \mathbf{p}_m \\ \mathbf{q}_m \end{bmatrix} \ . $$
We have that
\begin{align}
\big\|\mathbf{\tilde{J}}^T\mathbf{a}_m\big\|^2 &= \big\|\mathbf{S}_2 \mathbf{S}^{-1}_2
\mathbf{\tilde{J}}^T\mathbf{a}_m\big\|^2 \nonumber \\
&= \big\|\mathbf{S}_2 \big[-\mathbf{T}~ \mathbf{I}\big] \mathbf{a}_m\big\|^2 \nonumber \\
&\leq \lambda^2_{\text{max}}\big(\mathbf{S}_2\big) \big\|\big[-\mathbf{T}~
\mathbf{I}\big]\mathbf{a}_m\big\|^2 \nonumber\\
& = \lambda^2_\text{max}\big(\mathbf{S}_2\big) \big\|\mathbf{T}\mathbf{q}_m -
\mathbf{p}_m\big\|^2 \nonumber
\end{align} where the second line uses~\eqref{eq:tildeJ}. Combining this with~\eqref{eq-Interf-EffNoiseUpper} yields the upper bound \begin{align}
&\max_{m} \sigma_{\text{exact},m}^2 \label{eq:EffNoiseVar1} \leq \min_{\substack{\mathbf{A} \in \mathbb{Z}^{\mri \times \mri} \\ \mathrm{rank}(\mathbf{A})= \mri }} \max_m \ 
\lambda^2_\text{max} \big(\mathbf{H}^{-1}\big) \|\mathbf{a}_m\|^2 \\ 
& \qquad \qquad ~~~~~~~~~~~~~~~~~~+~
 \snr^{\alpha} \lambda^2_\text{max}\big(\mathbf{S}_2\big) \big\|\mathbf{T}\mathbf{q}_m -
\mathbf{p}_m\big\|^2 \ .\nonumber
\end{align}

We now proceed to upper bound $\| \mathbf{a}_m \|$ in terms of $\mathbf{q}_m$ and $\mathbf{p}_m$:
\begin{align*}
\|\mathbf{a}_m\| &\leq \|\mathbf{q}_m \| + \|\mathbf{p}_m\| \\
&= \|\mathbf{q}_m \| + \big\|\mathbf{{p}}_m + \mathbf{T}\mathbf{q}_m -
\mathbf{T}\mathbf{q}_m \big\| \\
&\leq \|\mathbf{q}_m \| + \big\|\mathbf{T}\mathbf{q}_m\big\|+
\big\|\mathbf{T}\mathbf{q}_m - \mathbf{p}_m \big\| \\
&\leq \big(1 + \lambda_\text{max}(\mathbf{T})\big)\|\mathbf{q}_m \| +
\big\|\mathbf{T}\mathbf{q}_m - \mathbf{p}_m \big\| 
\end{align*} This allows us to further upper bound~\eqref{eq:EffNoiseVar1} by
\begin{align}
\nonumber & \min_{\substack{\mathbf{A} \in \mathbb{Z}^{\mri \times \mri} \\ \mathrm{rank}(\mathbf{A})= \mri }} \max_m \ c_1  \Big(\| \mathbf{q}_m \|^2 +\|\mathbf{q}_m\| \big\| \mathbf{T} \mathbf{q}_m - \mathbf{p}_m \big\| \\ &\qquad \qquad~~~~~~~~~~~~~~+ \snr^\alpha \big\| \mathbf{T} \mathbf{q}_m - \mathbf{p}_m \big\|^2 \Big) \ , \label{eq:noise2}
\end{align} where $c_1$ is a constant that does not depend on $\snr$. 

Applying Lemma \ref{lm:diophantine}, it follows that, for almost all $\mathbf{T}$, there exists a $Q' \geq 1$ such that, for all $Q > Q'$,~\eqref{eq:noise2} is upper bounded by 
\begin{align}
&c_1 C^2 \big(\log Q\big)^4 \Big( Q^2 + Q^{1 - \frac{\mri-K}{K}} + \snr^\alpha Q^{-2 \frac{\mri-K}{K}} \Big) \\
&\leq c_2 \big(\log Q\big)^4 \Big( Q^2 +  \snr^\alpha Q^{-2 \frac{\mri-K}{K}} \Big) \ .
\end{align} where the inequality is due to the fact that $\mri \geq K$ and $c_2$ is a constant that does not depend on $\snr$. Now, set $Q^2 = \snr^{\gamma}$ to obtain
\begin{align*}
\max_{m} \sigma_{\text{exact},m}^2 \leq c_2 \big(\log \snr^\gamma\big)^4 \Big(\snr^{\gamma} + \snr^{\alpha -
\gamma \frac{\mri-K}{K}} \Big) \ .
\end{align*} We equalize the exponents by choosing $\gamma = \frac{K}{\mri} \alpha$ from which it follows that
\begin{align}
\label{eq:noise3}
\max_{m} \sigma_{\text{exact},m}^2 \leq c_3 \big(\log \snr \big)^4 \snr^{\alpha K/\mri} 
\end{align} where $c_3$ is a constant that does not depend on $\snr$. 

Finally, plugging the upper bound~\eqref{eq:noise3} into the rate expression~\eqref{eq:prate} for exact integer-forcing, we get a lower bound on the achievable rate (for almost all full rank $\mathbf{H}$ and $\mathbf{J}$),
\begin{align*}
&R_{\iftext,\text{exact}}(\mathbf{H},\mathbf{J}) \\
&\geq \mri \bigg( \frac{1}{2} \log\bigg(\frac{\snr}{\snr^{\alpha K/\mri}}\bigg) -  2\log\log(\snr) - c_4 \bigg) \\
&\geq \frac{\mri - \alpha K}{2} \log(\snr) - 2\mri \log \log(\snr) - c_4
\end{align*} where $c_4$ is a constant that does not depend on $\snr$. The desired GDoF result follows immediately.
\end{IEEEproof}

\section{Proof of Lemma \ref{lm:diophantine}}
\label{ap:diophantine}
In order to prove Lemma \ref{lm:diophantine}, we employ a technique introduced by Kratz in \cite{k81}. We first construct semi-norms $f: \mathbb{R}^{\mri} \rightarrow \mathbb{R}_+$ and $g: \mathbb{R}^{\mri} \rightarrow \mathbb{R}_+$ as well as a norm $h: \mathbb{R}^{\mri} \rightarrow \mathbb{R}_+$. We then apply Minkowski's Second Theorem to find $\mri$ linearly independent integer vectors that achieve the successive minima (with respect to the norm $h$). Afterwards, we will show that these integer vectors satisfy the conditions in Lemma \ref{lm:diophantine}. We will need the following definitions and theorems in the proof.
\begin{definition}[$h$-Unit Ball]
Let $h: \mathbb{R}^{\mri} \rightarrow \mathbb{R}_+$ be a norm. The \textit{$h$-unit ball} is
\begin{align*}
\mathcal{B}_h = \left\{\mathbf{x} \in \mathbb{R}^{\mri}: h(\mathbf{x}) \leq 1 \right\}\ .
\end{align*}
The volume of $\mathcal{B}_h$ is denoted by $V_h$.
\end{definition}
\begin{definition}[Successive $h$-Minima]
\label{def:smGen}
Let  $h: \mathbb{R}^{\mri} \rightarrow \mathbb{R}_+$ be a norm and $\mathcal{B}_h$ be the $h$-unit ball. For $m = 1,2,\ldots,\mri$, the $m^{\text{th}}$ \textit{successive $h$-minimum} $\epsilon_m$ is given by
\begin{align*}
\epsilon_m &=  \min \big\{\epsilon \geq 0 : \exists~ m ~\text{linearly independent integer points}~ \\
&\qquad ~~~~~~~~~~~~~~ \mathbf{v}_1, \ldots ,\mathbf{v}_m \in \mathbb{Z}^{\mri} \cap \epsilon \mathcal{B}_h \big \} \ .
\end{align*}
\end{definition}

\begin{definition}[Rational Independence]
We call a matrix $\mathbf{T}$ \textit{rationally independent} if, for all non-zero rational vectors $\mathbf{q}$ (i.e., vectors with rational entries), we have that $\mathbf{T}\mathbf{q} \neq \mathbf{0}$. Otherwise, we call $\mathbf{T}$ rationally dependent.
\end{definition}

\begin{theorem}[Minkowski's Second Theorem]
\label{thm:sm}
For any norm $h: \mathbb{R}^{\mri} \rightarrow \mathbb{R}_+$, the successive $h$-minima satisfy
\begin{align*}
 V_{h} \prod_{i=1}^{\mri} \epsilon_i \leq 2^{\mri} \ .
\end{align*}
\end{theorem}

\begin{theorem} [Dirichlet]
\label{thm:dirichlets}
For any $\mathbf{T} \in \mathbb{R}^{K \times (\mri-K)}$ and $Q >1$, there exists a $[\mathbf{q}^T\  \mathbf{p}^T]^T \in \mathbb{Z}^{\mri-K} \times \mathbb{Z}^{K} \setminus \{\mathbf{0}\}$ such that
\begin{align*}
&\|\mathbf{q}\|_{\infty} \leq Q\\
&\big\|\mathbf{T}\mathbf{q} - \mathbf{p}\big\|_{\infty} \leq \frac{1}{Q^{\frac{M-K}{K}}} \ .
\end{align*}
\end{theorem}

\begin{theorem}[Khintchine-Groshev]
\label{thm:kg}
Fix a function $\Psi: \mathbb{N} \rightarrow \mathbb{R}_{+}$. If
\begin{align*}
\sum_{q = 1}^{\infty} q^{\mri-K-1} \big(\Psi(q)\big)^{K} < \infty\ ,
\end{align*}
then, for almost all $\mathbf{T} \in \mathbb{R}^{K \times (\mri-K)}$, there are only finitely many solutions of the form $[\mathbf{q}^T \mathbf{p}^T]^T \in \mathbb{Z}^{\mri-K} \times \mathbb{Z}^{K} \setminus \{ \mathbf{0} \}$ to the inequality
\begin{align*}
\big\|\mathbf{T}\mathbf{q} - \mathbf{p}\big\|_{\infty} < \Psi\big(\|\mathbf{q}\|_{\infty}\big)\ .
\end{align*}
\end{theorem} Theorem~\ref{thm:sm} can be found in~\cite[Theorem V, p.156]{cassels}, Theorem~\ref{thm:dirichlets} can be found in~\cite[Theorem VI, p.13]{cassels}, and Theorem~\ref{thm:kg} can be found in~\cite[Section 1.3.4]{bernikdodson}.

\begin{IEEEproof}[Proof of Lemma \ref{lm:diophantine}] For any integer vector $\mathbf{v} \in \mathbb{Z}^{\mri}$, we denote the first $\mri-K$ components by $\mathbf{q}$ and the remaining $K$components by $\mathbf{p},$ and will thus write
\begin{align*}
 \mathbf{v} = \begin{bmatrix}
  \mathbf{q}  \\
  \mathbf{p}  \\
 \end{bmatrix}. 
 \end{align*}

Throughout the proof, we assume that the matrix $\mathbf{T} \in \mathbb{R}^{K \times (\mri - K)}$ is rationally independent. (Note that the set of rationally dependent matrices has Lebesgue measure zero.)

For a fixed $\mathbf{T}$, define the semi-norms $f$, $g$ as follows:
\begin{align*}
f(\mathbf{v}) &= \big\|\mathbf{T}\mathbf{q} - \mathbf{p}\big\| \\
g(\mathbf{v}) &= \|\mathbf{q}\|  \ .
\end{align*}
For a fixed $Q$, let $\lambda_1$ denote the minimum value of $f(\mathbf{v})$ under the constraint $g(\mathbf{v}) \leq Q$, \begin{align}
\lambda_1 &= \min_{\substack{\mathbf{v} \in \mathbb{Z}^{\mri} \setminus \{\mathbf{0}\} \\ g(\mathbf{v}) \leq Q}} f(\mathbf{v})\nonumber \\
& = \min_{\substack{\mathbf{q} \in \mathbb{Z}^{\mri-K}\\ \|\mathbf{q}\| \leq Q}} \min_{\substack{\mathbf{p} \in \mathbb{Z}^{K} \\ [\mathbf{q}^T  \mathbf{p}^T]^T \neq \mathbf{0}}} \big\|\mathbf{T}\mathbf{q} - \mathbf{p}\big\| \ ,
\label{eq:lambda1}
\end{align}
$\mathbf{q}_{1} \in \mathbb{Z}^{\mri - K}$ denote the integer vector that achieves $\lambda_1$,
\begin{align}
\label{eq:defq}
\mathbf{q}_1 = \argmin_{\substack{\mathbf{q} \in \mathbb{Z}^{\mri-K}\\\|\mathbf{q}\| \leq Q}} \min_{\substack{\mathbf{p} \in \mathbb{Z}^{K} \\ [\mathbf{q}^T \mathbf{p}^T]^T \neq \mathbf{0}}} \big\|\mathbf{T}\mathbf{q} - \mathbf{p}\big\| \ ,
\end{align}
and $\mu_1 = \| \mathbf{q}_1\| $ denote the length of $\mathbf{q}_1$.

Based on the seminorms $f$ and $g$, we define the function $h: \mathbb{R}^{\mri} \rightarrow \mathbb{R}_+$ as follows:
\begin{align}
\label{eq:h}
h(\mathbf{v}) &= \left(f^2(\mathbf{v}) + \frac{\lambda^2_1}{\mu^2_1} g^2(\mathbf{v}) \right)^{1/2}\\
&= \left(\big\|\mathbf{T}\mathbf{q} - \mathbf{p}\big\|^2+ \frac{\lambda^2_1}{\mu^2_1} \|\mathbf{q}\|^2 \right)^{1/2}. \label{eq:h2}
\end{align}

In the sequel, we show that $h$ is a norm for $Q > 1$. We define the $\mri \times \mri$ matrix $\mathbf{\Gamma}$ as follows:
\[ \mathbf{\Gamma} = \left[ \begin{array}{ccc}
\mathbf{T} & -\mathbf{I}_{K} \\
\frac{\lambda_1}{\mu_1}\mathbf{I}_{\mri-K} & \mathbf{0} \end{array} \right]\]
Note that we can rewrite the function $h$ using $\mathbf{\Gamma}$,
\begin{align*}
h(\mathbf{v}) = \big\|\mathbf{\Gamma}\mathbf{v}\big\| \  .
\end{align*} Since exchanging the rows of a matrix only affects the sign of its determinant, we have that
\begin{align*}
\big|\det(\mathbf{\Gamma}\big)\big| = \left| \det \left( \left[\begin{array}{ccc}
 \frac{\lambda_1}{\mu_1}\mathbf{I}_{\mri-K} & \mathbf{0} \\
\mathbf{T} & -\mathbf{I}_{K}
\end{array}\right] \right) \right| \ .  \end{align*} Now, using the fact that the determinant of a lower triangular matrix is just the product of its diagonal entries, we find that
\begin{align}
\big|\det\big(\mathbf{\Gamma}\big)\big| = \left(\frac{\lambda_1}{\mu_1} \right)^{\mri-K} .
\end{align}

Consider the case where $Q > 1$. Since $\mathbf{T}$ is rationally independent, it follows that $\lambda_1 > 0$. Since $\mu_1 \geq 0$ by definition, we have that $\frac{\lambda_1}{\mu_1} > 0$. Since $\mathbf{\Gamma}$ is full-rank and thus injective, $h$ is a norm.

Let $\mathbf{u} = \mathbf{T} \mathbf{v}$. It follows that the volume of the $h$-unit ball satisfies 
\begin{align}
V_{h} &= \int_{\left\{\mathbf{v}: \|\mathbf{\Gamma}\mathbf{v}\| \leq 1 \right\}} d\mathbf{v} \nonumber \\
&= \int_{\left\{\mathbf{u}: \|\mathbf{u}\| \leq 1 \right\}} \big|\det\big(\mathbf{\Gamma}^{-1}\big)\big| d\mathbf{u}\nonumber\\
&= \frac{1}{\big|\det\big(\mathbf{\Gamma}\big)\big|} \int_{\left\{\mathbf{u}: \|\mathbf{u}\| \leq 1 \right\}} d\mathbf{u} \nonumber\\
& = \frac{1}{|\det(\mathbf{\Gamma})|} V_{\mri}\nonumber\\
& = \left(\frac{\mu_1}{\lambda_1}\right)^{\mri-K}V_{\mri} \ ,
\label{eq:volume}
\end{align}
where $V_{\mri}$ denotes the volume of the unit ball in $\mathbb{R}^{\mri}$ with respect to the Euclidean norm.

Let $\epsilon_1, \ldots, \epsilon_{\mri}$ be the successive minima with respect to $h$ (see Definition \ref{def:smGen}). Let $\mathbf{v}_1, \ldots, \mathbf{v}_{\mri} \in \mathbb{Z}^{\mri}$ be the linearly independent integer points that achieve the successive minima, i.e., $h(\mathbf{v}_i) = \epsilon_i$. From Minkowski's Second Theorem (Theorem \ref{thm:sm}) and~\eqref{eq:volume}, we have that
\begin{align*}
 \left(\frac{\mu_1} {\lambda_1} \right)^{\mri-K} V_{\mri}   \prod_{i=1}^{\mri} \epsilon_i\leq 2^{\mri} .
\end{align*}
Rewriting the above, we get that
\begin{align}
\label{eq:m}
 \left(\frac{\mu_1} {\lambda_1} \right)^{\mri-K} \prod_{i=1}^{\mri} \epsilon_i \leq c ,
\end{align}
where $c$ is a constant that depends only on $\mri$. Rearranging \eqref{eq:m}, we arrive at
\begin{align}
&\epsilon_{\mri} \leq c \left(\frac{\lambda_1}{\epsilon_1} \cdots \frac{\lambda_1}{\epsilon_{\mri-K}}\right)\left(\frac{1}{\epsilon_{\mri-K+1}} \cdots \frac{1}{\epsilon_{\mri-1}}\right)\left(\frac{1}{\mu_1^{\mri-K}}\right) \nonumber \\
&= c \left(\frac{\lambda_1}{\epsilon_1} \cdots \frac{\lambda_1}{\epsilon_{M-1}}\right)\left(\frac{1}{\lambda_1^{K-1}\mu_1^{\mri-K}}\right) \  . \label{eq:e2M}
\end{align}

We now turn to show that $h(\mathbf{v}) \geq \lambda_1$ for all $\mathbf{v} \in \mathbb{Z}^{\mri} \setminus \{\mathbf{0}\}$. We consider the cases $\|\mathbf{q}\| < \mu_1$ and $\|\mathbf{q}\| \geq \mu_1$ separately. When $\|\mathbf{q}\| \geq \mu_1$, $h(\mathbf{v})$ from~\eqref{eq:h2} is lower bounded by
\begin{align*}
h(\mathbf{v}) &\geq \frac{\lambda_1}{\mu_1} \|\mathbf{q}\|\\
&\geq \lambda_1 .
\end{align*}
When $\|\mathbf{q}\| < \mu_1$, we begin by lower bounding~\eqref{eq:h2} by 
\begin{align*}
h(\mathbf{v}) &\geq \big\|\mathbf{T}\mathbf{q} - \mathbf{p}\big\| \  .
\end{align*}
Recall that $\mu_1 = \|\mathbf{q}_1\|$. From~\eqref{eq:defq}, $\mathbf{q}_1$ attains the minimum value $\lambda_1$ of $\big\| \mathbf{T} \mathbf{q} - \mathbf{p} \big\| $ across all non-zero integer vectors $[ \mathbf{q}^T \mathbf{p}^T]$ satisfying $\| \mathbf{q}\| \leq Q$. Since we have assumed $\| \mathbf{q} \| < \| \mathbf{q}_1\|$, $h(\mathbf{v}) \geq \lambda_1$ follows immediately. 

Since the successive minima can be written as $\epsilon_i = h(\mathbf{v}_i)$, we can lower bound them by $\epsilon_i \geq \lambda_1$ for some integer vectors $\mathbf{v}_1,\ldots,\mathbf{v}_M$. Combining this with \eqref{eq:e2M}, we obtain
\begin{align*}
h(\mathbf{v}_\mri) = \epsilon_{\mri} \leq c \frac{1}{\lambda_1^{K-1}\mu_1^{\mri-K}} \  .
\end{align*} Now, using the fact that $h(\mathbf{v}_i) \leq h(\mathbf{v}_\mri)$ for $i=1,2,\ldots,\mri$ and the definition of $h$ from~\eqref{eq:h}, we find that  
\begin{align}
\label{eq:f}
&f(\mathbf{v}_i) \leq h(\mathbf{v}_\mri) \leq c \frac{\lambda_1}{\lambda^{K}_1\mu_1^{\mri-K}}\\
&g(\mathbf{v}_i) \leq \frac{\mu_1}{\lambda_1} h(\mathbf{v}_\mri) \leq c \frac{\mu_1}{\lambda^{K}_1\mu_1^{\mri-K}} \ .
\end{align}

Recall that $\lambda_1$ and  $\mu_1$ are defined with respect to a fixed $Q$. We now show that for sufficiently large $Q$,
\begin{align}
\label{eq:lambda_mu}
&\lambda_1^{K}\mu_1^{\mri-K} \geq \frac{1}{\big(\log(\mu_1)\big)^2} \ .
\end{align}
We begin by defining the function 
\begin{align*}
\Psi(q) = \begin{cases}
 1 & q = 1 \ , \\
{\displaystyle \frac{1}{q^{(\mri-K)/K} \big(\log (q)\big)^{2/K}}} & q > 1 \  .
\end{cases}
\end{align*} and note that
\begin{align}
\sum_{q=1}^\infty q^{\mri-K-1} \big(\Psi(q)\big)^{K} < \infty  \ .
\end{align}
Thus, we can apply the Khintchine-Groshev Theorem (Theorem~\ref{thm:kg}) to establish that, for almost all $\mathbf{T} \in \mathbb{R}^{K \times (\mri - K)}$, there are only finitely many vectors $[\mathbf{q}^T \ \mathbf{p}^T]^T \in \mathbb{Z}^{\mri-K} \times \mathbb{Z}^{K}$ such that
\begin{align}
\label{eq:b}
\|\mathbf{q}\|_{\infty}^{\mri-K} \big\|\mathbf{T}\mathbf{q} - \mathbf{p}\big\|_{\infty}^{K} < \frac{1}{\big(\log\big( \|\mathbf{q}\|_{\infty}\big) \big)^2} \  .
\end{align}

Recall from \eqref{eq:defq} that $\mathbf{q}_1(Q)$ is the integer vector that achieves $\lambda_1(Q)$ for a given $Q$ (where we have made the dependence on $Q$ explicit). Clearly, $\left\{\|\mathbf{q}_1 (Q)\|_{\infty}\right\}_{Q=1}^{\infty}$ is a non-decreasing sequence. We now use the fact that $\mathbf{T}$ is rationally independent to argue that $\|\mathbf{q}_1 (Q)\|_{\infty}$ is unbounded as $Q \rightarrow \infty$. For the sake of a contradiction, assume that there exists some $B \in \mathbb{Z}_{+}$ such that $\|\mathbf{q}_1 (Q)\|_{\infty} \leq B$. This implies that $\mathbf{q}_1(Q)$ takes only a finite set of values. Hence, there exists a $D > 0$ such that
\begin{align}
\min_{\mathbf{p} \in \mathbb{Z}^{K}} \big\|\mathbf{T}\mathbf{q}_1(Q)- \mathbf{p}\big\|_{\infty} \geq D
\end{align} for all $Q \in \mathbb{N}$. However, by the definition of $\lambda_1(Q)$ in~\eqref{eq:lambda1} and Dirichlet's Theorem (Theorem \ref{thm:dirichlets}) we have that
\begin{align}
\label{eq:lambda1bound}
\min_{\mathbf{p} \in \mathbb{Z}^{K}} \big\|\mathbf{T}\mathbf{q}_1(Q) - \mathbf{p}\big\|_{\infty} & \leq \frac{1}{Q^{\frac{\mri-K}{K}}}
\end{align}
for all $Q \in \mathbb{N}$. This yields a contradiction with our assumption.

By Theorem \ref{thm:kg}, we know that there are only a finite number of solutions $\mathbf{q}_1(Q)$ that satisfy the condition in~\eqref{eq:b}. Let $Q'$ be the integer such that if $Q > Q'$, then $\mathbf{q}_1(Q)$ does not satisfy~\eqref{eq:b}. Since the $\ell_2$-norm is an upper bound on the $\ell_\infty$-norm, we have that, for $Q > Q'$,
\begin{align*}
\|\mathbf{q}_1(Q)\|^{\mri-K} \big\|\mathbf{T}\mathbf{q}_1(Q) - \mathbf{p}\big\|^{K} \geq \frac{1}{\big(\log\big( \|\mathbf{q}_1(Q)\big\|\big)^2} \ , 
\end{align*} which establishes~\eqref{eq:lambda_mu}. 
 
We now establish simple upper bounds on $\lambda_1$ and $\mu_1$. Using the definition of $\lambda_1$ from~\eqref{eq:lambda1} and the fact that $\| \mathbf{x} \| \leq \sqrt{K} \| \mathbf{x} \|_\infty$ for all $\mathbf{x} \in \mathbb{R}^K$, we get that \begin{align*}
\lambda_1 &\leq \min_{\substack{\mathbf{q} \in \mathbb{Z}^{\mri-K}\\ \sqrt{K} \|\mathbf{q}\|_\infty \leq Q}} \min_{\substack{\mathbf{p} \in \mathbb{Z}^{K} \\ [\mathbf{q}^T \ \mathbf{p}^T]^T \neq \mathbf{0}}} \sqrt{K} \big\|\mathbf{T}\mathbf{q} - \mathbf{p}\big\|_{\infty} \ .
\end{align*} Applying Dirichlet's Theorem (Theorem~\ref{thm:dirichlets}), we find that 
\begin{align}
\label{eq:d}
\lambda_1 \leq \sqrt{K} \left(\frac{\sqrt{K}}{Q}\right)^{(\mri-K)/K} \ .  
\end{align} Also, by definition, we have that
\begin{align}
\label{eq:e}
\mu_1 \leq Q \ .
\end{align}

Using \eqref{eq:f}, \eqref{eq:lambda_mu}, \eqref{eq:d},  and \eqref{eq:e}, and assuming that $Q$ is sufficiently large, we upper bound $f(\mathbf{v}_i)$ for $i=1,2,\ldots,\mri$ as follows: 
\begin{align*}
f(\mathbf{v}_i) &\leq c \frac{\lambda_1}{\lambda^{K}_1\mu^{\mri-K}}\\
&\leq c \lambda_1 (\log\mu_1)^2 \\
&\leq C \frac{(\log Q)^2}{Q^{(\mri-K)/K}},
\end{align*}
where $C$ is a constant that does not depend on $Q$. Similarly, we can upper bound $g(\mathbf{v}_i)$ for $i=1,2,\ldots,\mri$ as follows:
\begin{align}
g(\mathbf{v}_i) \leq c \mu_1 (\log(\mu_1))^2 \leq C Q (\log Q)^2 .
\end{align}
which concludes the proof of Lemma \ref{lm:diophantine}.
\end{IEEEproof}

\section*{Acknowledgment}
The authors are grateful to Or Ordentlich for pointing out the suboptimality of unimodular integer forcing, Giuseppe Caire for corrections in the proof of Theorem \ref{thm:dof}, and Seyong Park for many helpful discussions.

\ifCLASSOPTIONcaptionsoff
  \newpage
\fi

\bibliographystyle{IEEEtran}


\end{document}